\documentclass[11pt,a4paper,notitlepage]{article}
\usepackage[utf8]{inputenc}
\usepackage[english]{babel}
\usepackage[T1]{fontenc}
\usepackage{hyperref}
\usepackage{amsmath}
\usepackage{amssymb}
\usepackage{amsfonts}
\usepackage{slashed}
\usepackage{color} 
\usepackage{caption}
\usepackage{soul}
\usepackage{cite}
\usepackage{array,multirow,makecell}
\setcellgapes{1pt}
\makegapedcells
\newcolumntype{R}[1]{>{\raggedleft\arraybackslash }b{#1}}
\newcolumntype{L}[1]{>{\raggedright\arraybackslash }b{#1}}
\newcolumntype{C}[1]{>{\centering\arraybackslash }b{#1}}
\newcommand{\Tr}{\mathrm{Tr}}

\newcommand{\heq}{\mathrel{\hat{\approx}}}
    
\newcommand{\acts}{\triangleright}
\newcommand{\p}{\partial}
\newcommand{\f}{\frac}

\usepackage{enumerate}
\usepackage{subfig}
\usepackage{tikz-cd}

\newcommand{\cA}{{\mathcal A}}

\newcommand{\cC}{{\mathcal C}}
\newcommand{\cD}{{\mathcal D}}
\newcommand{\cE}{{\mathcal E}}
\newcommand{\cF}{{\mathcal F}}
\newcommand{\cG}{{\mathcal G}}

\newcommand{\cL}{{\mathcal L}}

\newcommand{\cN}{{\mathcal N}}

\newcommand{\cP}{{\mathcal P}}
\newcommand{\cQ}{{\mathcal Q}}
\newcommand{\cR}{{\mathcal R}}
\newcommand{\cS}{{\mathcal S}}
\newcommand{\cT}{{\mathcal T}}

\newcommand{\cZ}{{\mathcal Z}}

\newcommand{\fX}{{\mathcal X}}
\newcommand{\fY}{{\mathcal Y}}

\definecolor{mygray}{gray}{0.3}

\newcommand\beq{\begin{equation}}
\newcommand\eeq{\end{equation}}
\newcommand{\bes}{\begin{eqnarray}}
\newcommand{\ees}{\end{eqnarray}}
\def\nn{{\nonumber}}
\newcommand{\one}{\mbox{$1 \hspace{-1.0mm}  {\bf l}$}}

\def\vphi{{\varphi}}

\newcommand\restr[2]{{
  \left.\kern-\nulldelimiterspace 
  #1 
  \vphantom{\big|} 
  \right|_{#2} 
  }}

\def\extd{\mathrm {d}}
\newcommand{\U}{\mathrm{U}}

\usepackage{multicol}
\usepackage{amssymb}
\usepackage{graphicx}
\usepackage[left=2cm,right=2cm,top=2.5cm,bottom=2.5cm]{geometry}

\usepackage{nomencl}

\renewcommand{\nompreamble}{For the reader's convenience, we collect here our main field space notations; they will be of particular relevance in Secs.~\ref{sec:cov_phase_space} to \ref{sec:examples}.}
\makenomenclature

\usepackage{etoolbox}
\renewcommand\nomgroup[1]{%
  \item[\bfseries
  \ifstrequal{#1}{F}{Field space structures}{%
  \ifstrequal{#1}{G}{Weak equalities}{%
  \ifstrequal{#1}{S}{Gauge transformations and symmetries}{%
  \ifstrequal{#1}{C}{Spacetime structures}{%
  \ifstrequal{#1}{P}{Presymplectic structure}{}
  }}}}
]}

\begin{document}
\begin{center}
\textbf{\Large{Edge modes as reference frames and boundary actions from post-selection}}
\vspace{15pt}

{\large Sylvain Carrozza$^{a,}$\footnote{\url{s.carrozza@hef.ru.nl}} and Philipp A.\ H\"ohn$^{b,c,}$\footnote{\url{philipp.hoehn@oist.jp}}}.

\vspace{10pt}

$^{a}${\sl Institute for Mathematics, Astrophysics and Particle Physics, Radboud University,
Heyendaalseweg 135, 6525 AJ Nijmegen, The Netherlands.
}
\vspace{3pt}

$^{b}${\sl Okinawa Institute of Science and Technology Graduate University, Onna, Okinawa 904 0495, Japan.}

\vspace{3pt}

$^{c}${\sl Department of Physics and Astronomy, University College London, Gower Street, London, WC1E 6BT, United Kingdom.}

\end{center}

\vspace{5pt}

\begin{abstract}
\noindent We introduce a general framework realizing edge modes in (classical) gauge field theory as dynamical reference frames, an often suggested interpretation that we make entirely explicit. We focus on a bounded region $M$ with a co-dimension one time-like boundary $\Gamma$, which we embed in a global spacetime. Taking as input a variational principle at the global level, we develop a systematic formalism inducing consistent variational principles (and in particular, boundary actions) for the subregion $M$. This relies on a post-selection procedure on $\Gamma$, which isolates the subsector of the global theory compatible with a general choice of gauge-invariant boundary conditions for the dynamics in $M$. Crucially, the latter relate the configuration fields on $\Gamma$ to a dynamical frame field carrying information about the spacetime complement of $M$; as such, they may be equivalently interpreted as frame-dressed or relational observables. Generically, the external frame field keeps an imprint on the ensuing dynamics for subregion $M$, where it materializes itself as a local field on the time-like boundary $\Gamma$; in other words, an edge mode. We identify boundary symmetries as frame reorientations and show that they divide into three types, depending on the boundary conditions, that affect the physical status of the edge modes. Our construction relies on the covariant phase space formalism, and is in principle applicable to any gauge (field) theory. We illustrate it on three standard examples: Maxwell, Abelian Chern-Simons and non-Abelian Yang-Mills theories. In complement, we also analyze a mechanical toy-model to connect our work with recent efforts on (quantum) reference frames.
\end{abstract}

\setcounter{tocdepth}{2}
\tableofcontents

\section{Introduction}\label{sec:intro}

It is well known that gauge field theories defined on manifolds with boundaries can support the emergence of dynamical edge modes. This is unquestionable in the quantum theory, and in condensed matter physics in particular, where such emergent degrees of freedom can be related to a wealth of interesting phenomena (famously, chiral edge states are known to arise from the topological order underlying the quantum hall effect \cite{halperin1982quantized, wen1992theory}). In a broader context, and already at the classical level, the emergence of edge modes was argued to be a necessary consequence of gauge symmetry in the presence of boundaries by Donnelly and Freidel \cite{Donnelly:2016auv}. This has led to a series of efforts to better understand the notion of edge modes in gauge theories \cite{Donnelly:2014fua,Geiller:2017xad,Gomes:2018dxs, Gomes:2019xto, Riello:2020zbk, Riello:2021lfl,Blommaert:2018oue,Geiller:2019bti,Wong:2017pdm, Mathieu:2019lgi, Mathieu:2020fwg, Francois:2020tom}, gravity \cite{Freidel:2015gpa, Donnelly:2016rvo,Geiller:2017xad, Speranza:2017gxd, Geiller:2017whh,Freidel:2019ees,Freidel:2020xyx,Freidel:2020svx, Freidel:2020ayo, Dupuis:2020ndx, Geiller:2020xze, Donnelly:2020xgu, Livine:2021qwx, Freidel:2021cbc,Wieland:2017cmf,Wieland:2017zkf,Wieland:2020gno,Wieland:2021vef,Takayanagi:2019tvn,DePaoli:2018erh,Geiller:2020okp,Freidel:2021ajp,Kirklin:2019ror} and string theory \cite{Donnelly:2016jet,Donnelly:2020teo,Jiang:2020cqo}. However, the genericity of this particular argument for edge modes, its interpretation and physical relevance, all remain debated in the literature. In particular, it is not a priori obvious how the two notions of edge modes just mentioned --- physical edge degrees of freedom on the one hand, and Donnelly--Freidel edge modes on the other hand --- are related to one another. 

To various degrees, it was already suggested in the literature that Donnelly--Freidel edge modes might be best understood as some kind of dynamical reference frames (see e.g.~\cite{Wieland:2020gno, Riello:2020zbk} and also \cite{Donnelly:2016rvo}). Part of the purpose of the present work is to realize this idea in an entirely explicit manner. This will allow us to rederive the structure of Donnelly--Freidel edge modes from first principles, while making their interpretation as reference frames completely transparent; in particular, we will show that they are dynamical reference frames in the same sense as they appear in the recent quantum reference frame literature \cite{Giacomini:2017zju,Vanrietvelde:2018pgb,Vanrietvelde:2018dit,Hohn:2018iwn,Hohn:2018toe,Hohn:2019cfk,Hoehn:2020epv,Krumm:2020fws,Hoehn:2021flk,Hoehn:2021wet,delaHamette:2020dyi,Castro-Ruiz:2019nnl,Giacomini:2020ahk,Ballesteros:2020lgl}. By investigating their interplay with generic boundary conditions, we will also elucidate how they can sometimes (but not always) support a non-trivial algebra of boundary symmetries acting on the physical phase space. At least in the classical set-up we are considering, this is a necessary condition for Donnelly--Freidel edge modes to support the emergence of physical degrees of freedom, and therefore, to reveal themselves in an experiment. We will show that different types of boundary conditions lead to a refinement of the notion of boundary symmetry into three distinct types. Since our primary focus will be on Donnelly--Freidel edge modes, from now on we will simply refer to them as ``edge modes''.

\medskip

Beyond the question of edge modes \emph{per se}, our work establishes new connections between subjects that were previously discussed in the literature, which we believe to be of broader interest. 

From a technical standpoint, we will employ covariant phase space methods~\cite{Witten_Crnkovic, Lee:1990nz, ashtekar1991covariant, Henneaux:1992ig, Wald:1993nt, Jacobson:1993vj, Wald:1999wa,Iyer:1994ys,Iyer:1995kg} (see~\cite{Khavkine:2014kya, Gieres:2021ekc} for reviews) to describe the dynamics of fields in a (bounded) spacetime region $M$. For definiteness, we will take $M$ to have the topology of a cylinder, with time-like boundary $\Gamma$ (but generalizing the construction to the case of a null boundary should be reasonably straightforward). This problem has received renewed interest in recent years~\cite{Harlow:2019yfa, Geiller:2019bti, Margalef-Bentabol:2020teu, Chandrasekaran:2020wwn}, in part because of its relevance to holography, e.g.\ see~\cite{Compere:2008us,Faulkner:2013ica,Faulkner:2017tkh,Jafferis:2015del,Belin:2018bpg,Belin:2018fxe,Kirklin:2019ror,Kirklin:2019xug,Kirklin:2020zic,Pedraza:2021fgp,Pedraza:2021mkh}, gravitational observables \cite{Harlow:2021dfp}, and more generally, to the characterization of asymptotic symmetries in gauge and gravitational subsystems, see e.g.~\cite{Compere:2018aar,Barnich:2001jy,Barnich:2007bf,Barnich:2009se,Barnich:2011mi,Compere:2011ve,Geiller:2020okp,Geiller:2020edh,Chandrasekaran:2020wwn,Freidel:2021fxf,Freidel:2021cbc}. Our work revisits such constructions, from the vantage point of a global field space of solutions in a spacetime $M\cup \bar M$, where $\bar M$ is the complementary spacetime region of $M$. In that, it provides a direct link between the global methods of~\cite{Lee:1990nz} and the regional analysis of~\cite{Harlow:2019yfa,Geiller:2019bti}. In more concrete terms: given a consistent variational principle for $M\cup \bar M$, we will provide a general algorithm inducing consistent variational principles, and specifically boundary actions, for fields in $M$.

By identifying edge modes as reference frames, our work establishes a bridge between field theory and the recent literature on (quantum) reference frames. This permits us to identify the above mentioned boundary symmetries as \emph{edge frame reorientations} and to prove that frame-dressed observables\,---\,a systematic generalization of the examples considered in~\cite{Donnelly:2016auv}\,---\,can be equivalently understood as relational observables in the sense of~\cite{Rovelli:1990ph,Rovelli:1990pi,Rovelli:2001bz,rovelliQuantumGravity2004,Rovelli:1989jn,Rovelli:1990jm,Dittrich:2004cb,Dittrich:2005kc}. It furthermore enables us to consider different systems of edge frame fields and to show how to translate from the description relative to one frame field to that relative to another, i.e.\ to establish the transformation between the relational observables relative to different edge frames. Given that edge modes are only defined on the boundary $\Gamma$ of $M$, rather than its full bulk, they provide a particularly promising avenue for extending the efforts on quantum frame covariance of physical properties \cite{Giacomini:2017zju,Vanrietvelde:2018pgb,Vanrietvelde:2018dit,Hohn:2018iwn,Hohn:2018toe,Hohn:2019cfk,Hoehn:2020epv,Krumm:2020fws,Hoehn:2021flk,Hoehn:2021wet,delaHamette:2020dyi,Castro-Ruiz:2019nnl,Giacomini:2020ahk,Ballesteros:2020lgl} into the quantum field theory setting. Conversely, the distinction between (i) gauge transformations and symmetries (i.e.\ frame reorientations) or, equivalently, constraints and charges, as originally proposed in \cite{Donnelly:2016auv}, and (ii) different types of symmetries depending on the boundary conditions as established below, has so far not been explored in the quantum reference frame literature and warrants further investigation there.

\medskip

A basic idea we will invoke in our construction is that of \emph{post-selection} of a certain subspace of field configurations in a global space of solutions. We view post-selection as a unifying concept allowing one to incorporate operational constraints into a given overarching theory to select the subsector of that theory consistent with the constraints. These are associated to physical events of a contingent nature. Such constraints may encode the boundary conditions describing the context of a controlled experiment (e.g.~a scattering experiment), or more broadly speaking, the state of the world we happen to observe as agents. This broad understanding seems particularly relevant in the gravitational context, where controlled experiments are more the exception than the norm. In (quantum) general relativity, our control/engineering abilities are essentially non-existent, but what we observe around us is nonetheless post-selected from within the set of all possibilities admitted by the global theory. 

In this paper, it is post-selection of field configurations on $\Gamma$ that will provide a link between the global covariant phase space for $M\cup \bar M$, e.g.\ in the sense of \cite{Lee:1990nz}, and the local one for $M$, e.g.\ in the sense of \cite{Harlow:2019yfa,Geiller:2019bti,Chandrasekaran:2020wwn}. The two additional guiding principles we will rely on in this endeavour are: 1) that post-selection should operate on \emph{physical} (as opposed to gauge) degrees of freedom, and thus, on gauge-invariant functionals in the global solution space for $M \cup \bar M$; 2) that it should lead to a consistent factorization of the global variational problem into two independent ones, for $M$ and its complement. Up to more minor and technical assumptions that we will make along the way, this will be sufficient to derive the structure of edge modes on $\Gamma$, in a self-contained and conceptually illuminating way. In particular,  edge modes assume the role of ``internalized'' external dynamical frames for subregion $M$ and do not have to be added to the theory; they are always part of the global theory and induced onto the regional one, encoding how the subregion $M$ of interest relates to its spacetime complement $\bar M$. Furthermore, the fact that edge modes can ultimately restore the invariance of the regional presymplectic structure under field-dependent gauge transformations becomes secondary in our approach: it is a direct consequence of our construction, not a main postulate (as it was in~\cite{Donnelly:2016auv}).

Post-selection naturally prompts us to examine the interplay between edge modes and different boundary conditions on $\Gamma$. Incidentally, we will investigate  general boundary conditions, that can be obtained by application of a linear canonical transformation of some set of gauge-invariant local canonical Darboux coordinates for the presymplectic current on $\Gamma$. This will allow us to observe that, within the same global theory space, the physical role of edge modes for the post-selected dynamics in subregion $M$ can in fact be contingent on the choice of boundary conditions. For instance, in Maxwell theory, edge modes drop out of the regional presymplectic structure (and hence the regional physical phase space) if one decides to impose Neumann boundary conditions, i.e.\  boundary conditions on the flux $\restr{\star F}{\Gamma}$. This suggests that, in this specific example, the concept of edge mode could be dispensed with altogether. This is consistent with the interpretation of edge modes as external frames for subregion $M$ (originating in the complement $\bar M$) and the fact that Neumann boundary conditions correspond to reflecting boundary conditions and thus an essentially independent dynamics of $M$ and $\bar M$. Indeed, it was demonstrated explicitly in~\cite{Gomes:2018dxs, Gomes:2019xto, Riello:2020zbk, Riello:2021lfl} that edge mode degrees of freedom are not strictly necessary in order to construct a consistent phase space for Maxwell theory with boundary condition on the flux. However, our work also demonstrates that they do have physical significance for other (e.g., Dirichlet and Robin) boundary conditions in Maxwell, Chern-Simons and Yang-Mills theories.\footnote{For Yang-Mills (and in particular Maxwell) theories, the imprint of edge modes on the reduced phase space remains arguably mild: once boundary conditions have been imposed, at most a finite number of global charges survive at the physical level. This is dramatically less than in Chern-Simons theory, which generates an algebra of symmetries parametrized by a space of functions on $\Gamma$ (hence, infinite-dimensional, see Sec.~\ref{sec:examples}).} 

This ultimately leads to a refinement of the notion of boundary symmetries, as proposed in \cite{Donnelly:2016auv}, into three types depending on the boundary conditions, that we call \emph{symmetries, meta-symmetries} and \emph{boundary gauge symmetries}. Symmetries are physical transformations in the regional phase space, while meta-symmetries are symplectomorphisms between different regional phase spaces (i.e.\ different subregion theories), and boundary gauge symmetries arise when the charges turn into additional first-class constraints on the boundary (which will always be the case for Neumann boundary conditions).

\medskip 

Even though we are mainly motivated by (quantum) gravity, our focus in the present article will be on non-gravitational gauge theories, in which the metric is non-dynamical (though can be curved). We however anticipate our interpretation of edge modes as dynamical reference frames to extend (with suitable adaptations) to gravitational theories, a problem we plan to return to in the future. In the long run, our hope is that this viewpoint will help to further establish the notion of gravitational edge mode as a valuable concept for quantum gravity, in line with recent efforts in this direction~\cite{Freidel:2020xyx, Freidel:2020svx, Freidel:2020ayo, Dupuis:2020ndx, Geiller:2020xze, Donnelly:2020xgu, Livine:2021qwx, Freidel:2021cbc,Wieland:2020gno,Wieland:2021vef,Wieland:2017cmf,Wieland:2017zkf}. Given the holographic nature of gravitational systems (see for instance~\cite{Marolf:2008mf}), and the fact that the quantum reference frame literature provides a promising framework in which to define quantum generalizations of the equivalence principle (see e.g.~\cite{Giacomini:2020ahk,Giacomini:2021aof} for initial explorations), we believe this hope is warranted.

\

\noindent{\bf Organization of the paper.} We start in Sec.~\ref{sec:mechanical} by analyzing a Newtonian mechanical model featuring a local gauge symmetry generated by a single constraint. This serves as a simple illustration of the general formalism to be developed later, in a context where conceptual connections with the existing literature on dynamical reference frames are not obscured by field theory technicalities. Readers who are unfamiliar with the covariant phase space formalism will also find in this section a gentle introduction to its philosophy. A more substantial introduction to this formalism is provided in Sec.~\ref{sec:cov_phase_space}, this time in the context of field theory. This is where we fix some of our basic notations for the global field space, and introduce additional background material on the relation between covariant and canonical methods. In Sec.~\ref{sec_gaugesplit}, we start by formalizing the notion of dynamical reference frame in the context of field theory. This allows us, in a second step, to explicitly realize gauge field theory edge modes as dynamical reference frames. We conclude that section by analyzing two types of transformations on edge frames: boundary symmetries (following the nomenclature of~\cite{Donnelly:2016auv}), which we identify as physical frame reorientations; and changes of frames, which can be formalized as field-dependent generalizations of the former. Along the way, we establish that the frame-dressed observables entering the construction of Donnelly and Freidel can be generalized and understood as a covariant incarnation of the relational observables originally defined in a canonical set-up by Dittrich \cite{Dittrich:2004cb,Dittrich:2005kc} (extending earlier work by Rovelli \cite{Rovelli:1990ph,Rovelli:1990pi,Rovelli:2001bz,rovelliQuantumGravity2004,Rovelli:1989jn,Rovelli:1990jm}). We finally proceed with post-selection itself, which for didactic reasons, we describe in two steps. In Sec.~\ref{sec:geometry}, we work exclusively on-shell of the bulk equations of motion, which already allows us to induce a conserved presymplectic structure for the subregion $M$. Moving on to an off-shell description of the same procedure in Sec.~\ref{sec:splitting}, we construct a general and systematic algorithm inducing consistent variational principles for the subregion $M$, that is, to deduce an appropriate boundary action imposing both the bulk equations of motion and the boundary conditions as dynamical equations on $\Gamma$. Finally, we illustrate our results on standard examples of field theories in Sec.~\ref{sec:examples}: scalar field theory, which does not necessitate the introduction of edge modes, but in which the general algorithm of Sec.~\ref{sec:splitting} is still operational; Maxwell theory, which we take as a main basic example to illustrate our formalism throughout the paper, and only work out in full detail in that section; Abelian Chern-Simons theory, a particularly interesting example in which the edge modes construction gives rise to an infinite-dimensional algebra of boundary symmetries; and finally, Yang-Mills theory, as a way to illustrate non-Abelian features of the general formalism.


\addcontentsline{toc}{section}{Main notations}\hypertarget{thelist}{}

\nomenclature[C,0]{\(M\)}{Spacetime region with the topology of $S^{d-1} \times \mathbb{R}$}
\nomenclature[C,1]{\(\bar M\)}{Spacetime complement of $M$}
\nomenclature[C,2]{\(\mathring{M}\)}{Interior of region $M$}
\nomenclature[C,3]{\(\Gamma\)}{Time-like boundary of $M$}
\nomenclature[C,4]{\(\Sigma \cup \bar\Sigma\)}{Space-like Cauchy surface in $M\cup\bar M$, such that $\Sigma \subset M$ and $\bar\Sigma \subset \bar M$}
\nomenclature[C,5]{\(\epsilon\)}{Spacetime volume form}
\nomenclature[C,6]{\(\epsilon_\Gamma \)}{Volume form on $\Gamma$ induced by $\epsilon$}
\nomenclature[C,7]{\(\extd\)}{Spacetime exterior derivative}
\nomenclature[C,8]{\(\wedge\)}{Spacetime wedge product}

\nomenclature[C,9]{\(\restr{}{\Gamma}\)}{Pullback to $\Gamma$}

\nomenclature[F,1]{\(\cF\)}{Space of global field configurations}
\nomenclature[F,2]{\(\cS\)}{Subspace of solutions ($\cS \subset \cF$)}
\nomenclature[F,3]{\(\cP\)}{Physical phase space of gauge orbits ($\cP = \cS/ \cG$, where $\cG$ is the gauge group)}
\nomenclature[F,4]{\(\cF_M \,, \cS_M \,, \cP_M\)}{Analogously defined field spaces for the spacetime subregion $M$}
\nomenclature[F,4]{\(\cS_M^{X_0}\)}{Subspace of solutions in $S_M$ obeying the boundary condition $X=X_0$}
\nomenclature[F,5]{\(\delta\)}{Field-space exterior derivative}
\nomenclature[F,6]{\( \)}{Field-space wedge product (kept implicit)}
\nomenclature[F,7]{\( \cdot \)}{Field-space interior product}

\nomenclature[G]{\(=\)}{Equality on $\cF$ (resp. $\cF_M$)}
\nomenclature[G]{\(\approx\)}{Equality on $\cS$ (resp. $\cS_M$)}
\nomenclature[G]{\(\heq\)}{Equality on $\cS_M^{X_0}$}

\nomenclature[P,1]{\(\Theta\)}{Presymplectic potential (defined on $\cF$)}
\nomenclature[P,2]{\(\omega\)}{Presymplectic current $\omega := \delta \Theta$}
\nomenclature[P,3]{\(\Omega\)}{Presymplectic form on $\cS$ ($\Omega := \int_{\Sigma \cup \bar\Sigma} \omega$)}
\nomenclature[P,4]{\(\omega_\partial\)}{Boundary presymplectic current ($(2,d-2)$-form on $\Gamma$)}
\nomenclature[P,5]{\(\Omega_M\)}{Presymplectic form on $\cS_M^{X_0}$ ($\Omega_M := \int_{\Sigma} \omega + \int_\Gamma \omega_\partial$)}

\nomenclature[S,1]{\(\fX_\alpha\)}{Field-space vector field generating the gauge transformation $\delta_\alpha := \fX_\alpha \cdot \delta$}
\nomenclature[S,2]{\(C[\alpha]\)}{Gauge constraint associated to $\fX_\alpha$ ($\delta C[\alpha]=\fX_\alpha \cdot \Omega_M$)}
\nomenclature[S,3]{\(\fY_\rho\)}{Vector field generating the symmetry (i.e. frame reorientation) $\Delta_\rho := \fY_\rho \cdot \delta$}
\nomenclature[S,4]{\(Q[\rho]\)}{Charge associated to $\fY_\rho$ ($\delta Q[\rho]=\fY_\rho \cdot \Omega_M$)}

\printnomenclature

\section{Edge modes and post-selection in a mechanical system*}\label{sec:mechanical}

*\emph{This section may be skipped by a quick reader only interested in our field theory constructions. It may serve as an introduction to the covariant phase space formalism for the uninitiated.}

\medskip

As a warm-up exercise, and for the sake of conceptual clarity, we shall illustrate a few illuminating observations in a mechanical toy model before later exploiting them in the field theory context. This model mimics finite region gauge theories through groups of particles subject to translation invariance.  Specifically, this model serves to show 
\begin{itemize}
    \item that edge modes constitute ``internalized'' external reference frames for a subregion and are dynamical frames in the same sense in which they have been discussed in the recent quantum reference frame literature \cite{Giacomini:2017zju,Vanrietvelde:2018pgb,Vanrietvelde:2018dit,Hoehn:2021wet,Hohn:2019cfk,Hoehn:2020epv,Hohn:2018iwn,Hohn:2018toe,Castro-Ruiz:2019nnl,delaHamette:2020dyi,Krumm:2020fws,Hoehn:2021flk,Giacomini:2020ahk,Ballesteros:2020lgl}; 
  
    \item how to change from the description of the subregion relative to one choice of edge mode frame to another one;
    \item how to obtain a subregion phase space structure from a boundary condition induced foliation of the global space of solutions  through a splitting post-selection procedure;
     \item that gauge transformations and symmetries need to be distinguished \cite{Donnelly:2016auv}, where we identify the latter as frame reorientations. These frame reorientations have to be further distinguished into  three types: symmetries, meta-symmetries and boundary gauge symmetries. Symmetries leave a subregion theory defined by given boundary conditions invariant, while a meta-symmetry maps one such theory into another one defined by different boundary conditions. Boundary gauge symmetries arise in addition to the already present gauge symmetry and only for Neumann boundary conditions.

\end{itemize}

While the mechanical toy model captures many of the qualitative features we shall later encounter in gauge field theory, it also leads to some differences which we try to highlight. For example, owing to the discrete nature of the degrees of freedom, it does not give rise to corner terms. Furthermore, gauge constraints arise as identities already off-shell and, due to the simplicity of the model, are non-local in contrast to our field theory examples.

\subsection{Covariant phase space in mechanics}

Suppose we are given a group of $N$ particles in 1D Newtonian space, collectively called $M$, subject to an action $S_M=\int_{t_1}^{t_2}L_M$ with Lagrangian form
\beq 
L_M=\cL(q_i,\dot q_i)\extd t\,.\nn
\eeq 
Our configuration space is $\cQ=\mathbb{R}^N$ which is the mechanical analog of the space of instantaneous field configurations, i.e.\ space of field configurations on some Cauchy slice. By contrast, the mechanical analog of the space of field configurations here is the space of spatiotemporal configurations of $M$, in other words, the space of \emph{histories} for the time interval $[t_1,t_2]$, which we take to be $\cF_{\mathring{M}}:=\{c:[t_1,t_2]\rightarrow\cQ\,|\,\text{$c$ is a $C^2$ curve}\}$ and for simplicity often refer to as ``field space''. $\mathring{M}$ denotes here internal or ``bulk'' degrees of freedom in distinction of the situation when we include external frame particles below.  The action is a function $S_M:\cF_{\mathring{M}}\rightarrow\mathbb{R}$, while the Lagrangian can be thought of as a functional $\cL[c,t]=\cL(q_i(t),\dot q_i(t))$ on $\cF_{\mathring{M}}\times[t_1,t_2]$ because a history $c=\{q_i(t),t\in[t_1,t_2]\}$ and an instant $t$   determine the positions and velocities $(q_i(t),\dot q_i(t))$.  For the moment it is irrelevant whether the Lagrangian $\cL$ features a gauge symmetry. 

$\cF_{\mathring{M}}$ is itself a smooth infinite-dimensional manifold, allowing us to define standard differential geometric notions on it. In what follows, we will be using the covariant phase space formalism, constructing a phase space from the subspace $\cS_{\mathring{M}}\subset\cF_{\mathring{M}}$ of solutions to the equations of motion. In line with standard convention for the covariant phase space in field theory, which we adopt throughout this work, we denote the exterior derivative on field space $\cF_{\mathring{M}}$ by $\delta$, while reserving the notation $\extd$ for the exterior derivative of spacetime forms. In the mechanical case, these are forms on the time manifold $[t_1,t_2]$ underlying the action $S_M$. As such, we will be dealing with $(r,s)$-forms $\mu$, which means $\mu$ is an $r$-form on $\cF_{\mathring{M}}$ and an $s$-form on spacetime. For example, the Lagrangian form $L_M$ is a $(0,1)$-form. The two exterior derivatives commute with each other ($\left[ \extd , \delta \right] = 0$), and both square to zero ($\extd^2 = 0$ and $\delta^2 = 0$).

The variation of the action on field space $\delta S_M=\int_{t_1}^{t_2}\delta L_M$ is determined through the exterior derivative of the Lagrangian $(0,1)$-form
\beq\label{eq:L_M-var}
\delta L_M = \sum_{i=1}^N\left(\frac{\partial\cL}{\p q_i}-\f{\extd }{\extd t}\f{\p\cL}{\p\dot q_i}\right)\delta q_i\extd t+\extd\Theta_{\mathring{M}}\,,
\eeq 
where the Lagrangian symplectic potential (a $(1,0)$ form) takes the form
\beq 
\Theta_{\mathring{M}}=\sum_{i=1}^N\f{\p\cL}{\p\dot q_i}\delta q_i\,
\eeq 
and gives rise to the Lagrangian presymplectic structure (a $(2,0)$ form)
\beq 
\Omega_{\mathring{M}} =\sum_{i,j=1}^N \delta\Theta_{\mathring{M}}=\sum_{i,j=1}^N\f{\p^2\cL}{\p\dot q_i\p q_j}\delta q_j\delta q_i+\sum_{i,j=1}^N\f{\p^2\cL}{\p\dot q_i\p\dot q_j}\delta\dot q_j\delta q_i\,.
\eeq 
Again in line with standard covariant phase space convention, we keep the field space wedge product between field space one-forms $\delta q_j$ and $\delta\dot q_i$ implicit, keeping in mind that $\delta q_j\delta \dot q_i=-\delta \dot q_i\delta q_j$; the notation $\wedge$ is reserved for the spacetime wedge product.

The presymplectic structure is conserved on-shell, which follows from 
\beq 
\extd\Omega_{\mathring{M}} =\frac{\extd\Omega_{\mathring{M}}}{\extd t}\extd t\approx0\,,\label{omcons}
\eeq 
where $\approx$ denotes evaluation on $\cS_{\mathring{M}}\subset\cF_{\mathring{M}}$, i.e.\ on solutions to the Euler-Lagrange equations of motion $\frac{\partial\cL}{\p q_i}-\f{\extd }{\extd t}\f{\p\cL}{\p\dot q_i} = 0$. Indeed, this follows immediately from applying $\delta$ to $\delta L_M \approx \extd \Theta_{\mathring{M}}$, which is implied by equation~\eqref{eq:L_M-var}.

\emph{Defining}\footnote{We are not performing a Legendre transformation and so $p_i$ is here simply a function on $\cF_{\mathring{M}}\times[t_1,t_2]$.} $p_i:=\f{\p\cL}{\p\dot q_i}$, we find the potential and presymplectic structure in the simple forms
\beq 
\Theta_{\mathring{M}} =\sum_{i=1}^N p_i\delta q_i\qquad\text{and}\qquad\Omega_{\mathring{M}}=\sum_{i=1}^N \delta p_i\delta q_i\,,
\eeq 
which will become useful shortly.

\subsection{Edge modes as ``internalized'' external reference frames}

In this subsection, we will illustrate the mechanical analog of edge modes. To this end, we have to introduce a gauge symmetry. As the simplest possibility, suppose the Lagrangian is of the form 
\beq 
\cL(q_i,\dot q_j) = \cL\left(\{q_i-q_j,\dot q_i-\dot q_j\}_{i,j=1}^N\right)\,,\label{transllag}
\eeq 
so that it features an invariance under translations
$q_i(t)\to q_i(t)+\alpha(t)$ and $\dot q_i(t)\to\dot q_i(t)+\dot\alpha(t)$, for an arbitrary function of time $\alpha(t)$.\footnote{For example, the Lagrangian could read $\frac{1}{2N}\sum_{i, j=1}^N(\dot q_i-\dot q_j)^2-V\left(\{q_i-q_{i+1}\}\right)$, which amounts to subtracting the center-of-mass kinetic energy from the total kinetic energy and a translation-invariant potential.} In the mechanical context, the time manifold $\mathbb{R}$ assumes the role of spacetime and since $\alpha$ depends on $t$, we can view this as a gauge-transformation of the corresponding action.\footnote{ For simplicity, we choose this spatially global gauge symmetry. In order to better mimic the local gauge symmetries in field theory, one could for instance choose a Lagrangian of the form
${\cL=\frac{1}{2}\sum_{i}(\dot q_{2i}-\dot q_{2 i + 1})^2- V\left(\{q_{2i+1} -  q_{2i} - q_{2i -1 } + q_{2i -2 } \}\right)}
$.
This would lead to a ``local'' gauge symmetry $(q_{2i}(t), q_{2i+1}(t)) \to (q_{2i}(t)+ \alpha_i (t), q_{2i+1} (t) + \alpha_i (t))$, in term of a gauge parameter $\alpha_i (t)$ which can be explicitly ``space'' dependent. The exposition would become somewhat more convoluted, however. } The gauge group is thus the translation group $(\mathbb{R},+)$.

We denote by $\fX_\alpha$ the vector field on $\cF_{\mathring{M}}$ which generates the above translation. It satisfies $\fX_\alpha(q_i)=\alpha=\fX_\alpha\cdot\delta q_i$ and $\fX_\alpha(\dot q_i)=\dot\alpha=\fX_\alpha\cdot\delta\dot q_i$, where $\fX_\alpha\cdot\mu$ denotes the interior product of the vector field $\fX_\alpha$ and the form $\mu$ on $\cF_{\mathring{M}}\times[t_1,t_2]$.
Let $g(\alpha)\in\mathbb{R}$ denote the translation group element amounting to translation by $\alpha$. Let $f(q_i(t),\dot q_i(t))$ be an arbitrary functional on $\cF_{\mathring{M}}\times[t_1,t_2]$ which depends analytically on positions and velocities. It transforms under this group action as
\beq\label{gaugeactsmech}
g(\alpha)\acts f(q_i,\dot q_i)=e^{\fX_\alpha}f=f(q_i+\alpha,\dot q_i+\dot\alpha)\,,
\eeq 
which clearly satisfies $g(\alpha+\beta)\acts f=g(\alpha)\acts g(\beta)\acts f=g(\beta)\acts g(\alpha)\acts f$.

\subsubsection{Internal description of a group of particles}\label{ssec_intpart}

We can select any of the particles of $M$, say particle $N$, as a dynamical \emph{internal} reference frame for the translation group $(\mathbb{R},+)$, relative to which we describe the remaining ones. Indeed, we can build gauge-invariant quantities, such as relative distances $q_i-q_N$, using the frame degrees of freedom. More formally, note that the frame position $q_N$ constitutes a (translation) group-valued dynamical reference frame $U=q_N$  on field space, which transforms by group multiplication (here, the addition) under translations, $g(\alpha)\acts U=g(\alpha)+U=q_N+\alpha$.  The group being Abelian, left and right actions of the group are of course the same. Using this frame, we can turn any functional $f(q_i(t),\dot q_i(t))$ into a gauge-invariant one by subjecting it to a frame-dressed gauge transformation defined by $\alpha=-q_N+x$, for an arbitrary real function $x(t)$:
\beq 
O_{f,U}(x):=g(-q_N+x)\acts f=\left(U-x\right)^{-1}\acts f:=\restr{g(\alpha)\acts f}{\alpha=-U+x}\,,\label{relobs0}
\eeq 
where $(U-x)^{-1}$ denotes the inverse group element.\footnote{For the translation group, we of course have $(U-x)^{-1}=-(U-x)$. The reason we also use the notation $g^{-1}$ for the inverse group element here is for facilitating the comparison with the corresponding construction of gauge-invariant observables for general groups in gauge field theory, which will be the focus of  Sec.~\ref{sec_gaugesplit}.} In coordinates, we thus have 
\beq 
O_{f,U}(x,q_i,\dot q_i) = f(q_i-q_N+x,\dot q_i-\dot q_N+\dot x)\,,\label{relobs}
\eeq 
which clearly is translation-invariant. This frame-dressed observable is the (Lagrangian analog) of a \emph{relational} Dirac observable. They constitute so-called gauge-invariant extensions of gauge-fixed quantities \cite{Henneaux:1992ig,Dittrich:2004cb,Hohn:2019cfk,Chataignier:2019kof} and here encode the question ``what is the value of $f$ when the reference frame (particle $N$) is in orientation (position) $x$?'' \cite{Dittrich:2004cb,Rovelli:1989jn,Rovelli:1990pi,Rovelli:1990jm,Rovelli:1990ph,rovelliQuantumGravity2004}. This is particularly transparent when setting $f=q_i$, giving the relative distance
\beq \label{mechconfigrelobs}
Q_{i|N}(x):=O_{q_i,U}(x)=q_i-q_N+x\,.
\eeq 
Since the value of $O_{q_i,U_N}(x)$ is constant on any gauge orbit, this value is the same as on the gauge-fixing surface $q_N=x$, in this sense measuring the position of particle $i$ when frame particle $N$ is in position $x$. 

Next, we observe that $\fX_\alpha$ constitutes a null direction of the presymplectic structure. To this end, we first note that, owing to our assumption on the Lagrangian in equation~\eqref{transllag}, $p_i=\f{\p\cL}{\p\dot q_i}$ is translation-invariant, so that $\fX_\alpha(p_i)=\fX_\alpha\cdot\delta p_i=0$. Furthermore, the total momentum vanishes identically on \emph{all} of $\cF_{\mathring{M}}\times[t_1,t_2]$\footnote{This is because $P_M$ constitutes a primary constraint (upon Legendre transformation) in the language of constrained systems. This is yet another difference to the field theory examples later, where the gauge constraints are \emph{not} identities on field space, but only vanish on the subspace of solutions. The reason is that they are secondary constraints.}
\beq 
P_M=\sum_{i=1}^N p_i =0\,.\label{Mmom}
\eeq 
This yields 
\beq 
\fX_\alpha\cdot\Omega_{\mathring{M}} = -\alpha\delta P_M=0\,.
\eeq 
In particular, $\fX_\alpha$ is also off-shell a degenerate direction of the presymplectic structure. By contrast, in the field theory examples later, the constraints will be related to the equations of motion in such a way that gauge directions in field space will only be on-shell null directions of the presymplectic structure.

We can use the relational observables to split the symplectic potential and presymplectic structure into gauge-invariant and gauge parts, that is $
\Theta_{\mathring{M}}=\mathring{\Theta}_{\rm inv}+\mathring{\Theta}_{\rm gauge}$ and $\Omega_{\mathring{M}}=\mathring{\Omega}_{\rm inv}+\mathring{\Omega}_{\rm gauge}$, respectively,
where 
\begin{align}\label{mechsymp}
\mathring{\Theta}_{\rm inv}&:= \sum_{j=1}^{N-1}p_j\delta Q_{j|N}(x)\qquad \mathring{\Theta}_{\rm gauge}:= P_M\delta q_N=0\,,\\
 \mathring{\Omega}_{\rm inv}&:=\sum_{j=1}^{N-1}\delta p_j\delta Q_{j|N}(x)\qquad \mathring{\Omega}_{\rm gauge}:=\delta P_M\delta q_N=0\,.
\end{align}
For distinction from the case with external frame particles below, we have equipped the forms with a $\mathring{}$. A characteristic feature of the gauge contributions $\mathring{\Theta}_{\rm gauge}$ and $\mathring{\Omega}_{\rm gauge}$ (here, but also in field theory) is that they involve the constraints.
However, in contrast to the field theory case later, the gauge part vanishes identically here.

Let us now consider the construction of a phase space for the particle group $M$. Owing to the degeneracy of the presymplectic form $\Omega_{\mathring{M}}$, the field space $\cF_{\mathring{M}}$ is not a phase space. In order to construct a phase space $\cP_{\mathring{M}}$ from it, we have to factor out the null directions of $\Omega_{\mathring{M}}$: $\cP_{\mathring{M}}:=\cF_{\mathring{M}}/\!\sim$, where $c\sim c'$ if the histories $c,c'$ differ by such a null direction. Clearly, $\fX_\alpha$ is a null direction, but there are more: recalling that $\Omega_{\mathring{M}}$ is a $(2,0)$-form, we have to fix a time $t\in[t_1,t_2]$ in order to turn it into a genuine two-form $\Omega_{\mathring{M}}(t)$ on $\cF$. Note that outside of the space of solutions $\cS_{\mathring{M}}\subset\cF_{\mathring{M}}$, $\Omega_{\mathring{M}}(t)$ will depend on the choice of $t$. It is clear that any non-zero vector field $X$ which vanishes at $t$ will be a null direction of $\Omega_{\mathring{M}}(t)$, simply because  it only changes  $q_i(t'),\dot q_i(t')$ at times $t' \neq t$. Taking both types of degeneracies together, the equivalence class $[c]$ consists of all histories that have the same gauge-invariant $(Q_{j|N}(x,t),\dot Q_{j|N}(x,t))_{j=1}^{N-1}$ at $t$ as $c$. The phase space $\cP_{\mathring{M}}$ can thus be parametrized by these gauge-invariant variables and is isomorphic to $\mathbb{R}^{2N-2}$, hence finite-dimensional in contrast to $\cF_{\mathring{M}}$. We can, however, equivalently use $(Q_{j|N},p_j)_{j=1}^{N-1}$ as a coordinate system for $\cP_{\mathring{M}}$. To see this, note that we can write $p_j=\frac{\p\cL}{\p\dot Q_{j|N}}$ since $\frac{\p\dot Q_{j|N}}{\p\dot q_j}=1$ and  $\cL\left(\{q_i-q_j,\dot q_i-\dot q_j\}_{i,j=1}^N\right)=\cL\left(\{Q_{j|N},\dot Q_{j|N}\}_{j=1}^{N-1}\right)$, so $p_j$ is determined by $(Q_{j|N},\dot Q_{j|N})_{j=1}^{N-1}$. Conversely, $(Q_{j|N},p_j)_{j=1}^{N-1}$ determine $(Q_{j|N},\dot Q_{j|N})_{j=1}^{N-1}$ by the implicit function theorem because $\frac{\p p_j}{\p\dot Q_{j'|N}}=\frac{\p^2\cL}{\p\dot Q_{j|N}\p\dot Q_{j'|N}}$ is non-degenerate owing to the gauge-invariance of the involved variables. Hence, $\cP_{\mathring{M}}$ is also equipped with the non-degenerate symplectic form $\mathring{\Omega}_{\rm inv}(t)$.

In the present class of models, it turns out that the so constructed phase space, in fact, does not depend on $t$. The reason is that it coincides with the phase space constructed directly from the space of solutions $\cS_{\mathring{M}}\subset\cF_{\mathring{M}}$ on which $\Omega_{\mathring{M}}(t)$ is independent of $t$ thanks to equation \eqref{omcons}. To see this, note that $\fX_\alpha$ is also a degenerate direction of the pullback of $\Omega_{\mathring{M}}(t)$ to $\cS_{\mathring{M}}$, and that the set of $\fX_{\alpha}$ contains any non-zero vector field $X$ tangential to $\cS_{\mathring{M}}$ which vanishes at $t$. The ensuing equivalence relation $s\sim s'$ of solutions $s,s'\in\cS_{\mathring{M}}$ is thus simply the restriction of the field space equivalence relation $c\sim c'$ to the subspace of solutions. First consider the null directions defined by non-zero vector fields which vanish at $t$ (which are comprised of gauge directions $\fX_\alpha$ for which $\alpha,\dot\alpha$ vanish at $t$). Factoring  out these null directions from $\cS_{\mathring{M}}$ yields a finite-dimensional space isomorphic to $T\cQ\simeq\mathbb{R}^{2N}$, parametrized by initial data $(q_i(t),\dot q_i(t))$ at some time $t\in[t_1,t_2]$. Further factoring out all remaining gauge directions $\fX_\alpha$, which amounts to identifying the initial data sets $(q_i(t),\dot q_i(t))$ and $(q_i(t)+\alpha(t),\dot q_i(t)+\dot\alpha(t))$ for arbitrary $\alpha(t)$, produces a space which is conveniently parametrized by the relational observables $(Q_{j|N},\dot Q_{j|N})_{j=1}^{N-1}$. This is, of course, the same phase space as above, i.e.\ altogether
\beq \label{pkin=c}
\cP_{\mathring{M}}=\cS_{\mathring{M}}/\!\sim \,\,=\cF_{\mathring{M}}/\!\sim\,.
\eeq

This is also equivalent to the reduced phase space one would obtain if, instead of the covariant phase space method, one were to first perform a Legendre transformation, followed by a constraint reduction. Indeed, here the Legendre transformation, in coordinates defined by $p_i=\frac{\partial\cL}{\partial\dot q_i}$ (viewing $\cL$ at fixed $t$ as a function $\cL:T\cQ\to\mathbb{R}$), would map the space $T\cQ$ of positions and velocities $(q_i,\dot q_i)$ to the kinematical phase space $\cP_{\rm kin}:=T^*\cQ$. On $\cP_{\rm kin}$, $P_M=0$  no longer is an identity, but a non-trivial primary constraint defining a constraint submanifold $\cC\subset\cP_{\rm kin}$. Factoring out the gauge orbits of $P_M$ from $\cC$ then yields the reduced phase space $\cP_{\mathring{M}}$, coinciding with equation~\eqref{pkin=c}. However, within the covariant phase space formulation, the Legendre transformation is not a natural map to consider as it relies on fixed $t$.

The ultimate reason for the coincidence of the two phase spaces, constructed from the field space and space of solutions, respectively, is that $\fX_\alpha$ constitutes a degenerate direction of $\Omega_{\mathring{M}}$ \emph{both} on- and off-shell. This will be different in Maxwell, Chern-Simons and Yang-Mills theory later where constraints will only hold on-shell (they are secondary), such that gauge directions $\fX_\alpha$ are only degenerate directions of the pullback of $\Omega_{\mathring{M}}$ to $\cS_{\mathring{M}}$. In \emph{those} cases, the equivalence relation $\sim$ does not encompass gauge directions and so $\cP_{\rm kin}:=\cF_{\mathring{M}}/\!\!\sim,\cC:=\cS_{\mathring{M}}/\!\!\sim$ and $\cP:=\left(\cS_{\mathring{M}}/\!\sim\right)/\cG$, where $\cG$ denotes the group of spacetime gauge transformations, are distinct sets, constituting what are usually known as the kinematical phase space, constraint surface and reduced or physical phase space, respectively; see e.g.\  \cite{Lee:1990nz} (as well as Sec.~\ref{sec:covariant_to_standard}). 
A field theory example where the kinematical and physical phase spaces coincide as here is parametrized field theory \cite{Lee:1990nz}.

Altogether, this constitutes a purely \emph{internal} description of the group of particles $M$. At times it will be interesting though to explore how a subsystem relates to its environment.

\subsubsection{External description of the particle group: ``extending the Heisenberg cut''}\label{ssec_extHeis}

While the internal description of the particle group $M$, e.g.\ relative to particle $N$, thus lives in the physical phase space $\cP_{\mathring{M}}=\cS_{\mathring{M}}/\!\sim$, i.e.\ on the set of equivalence classes of solutions, the kinematical phase space $\cP_{\rm kin}$, as we shall now elucidate, can be associated with the description of $M$ relative to an \emph{external} reference frame. As just seen, the kinematical phase space is not a natural object to consider in the covariant phase space formulation of the present mechanical model because it is here not obtained through the natural reduction methods of the covariant formulation as in equation~\eqref{pkin=c}, but through a Legendre transformation. 
 We nevertheless discuss this physical interpretation of $\cP_{\rm kin}$ here since it is useful for understanding edge modes conceptually and as we are ultimately interested in field theory where no Legendre transformation is needed in order to construct $\cP_{\rm kin}$ (see Sec.~\ref{sec:covariant_to_standard}). From the point of view of the covariant method, the following interpretation of $\cP_{\rm kin}$ will thus be more natural to entertain in field theory.

$\cP_{\mathring{M}}$ comprises the physical states of $M$ that are distinguishable relative to an internal frame of $M$ at any time $t\in[t_1,t_2]$, while $\cP_{\rm kin}$ can, in a sense to be made more precise below, be viewed as comprising those states that are distinguishable relative to a frame external to $M$. This frame could be fictitious or a physical one that we have thus far ignored. In the sequel, we consider the latter situation and  ``internalize'' such an external frame through a field space and corresponding phase space extension. The set of physical states of the ensuing internal description of $M$ relative to the new reference frame will then be equivalent to $\cP_{\rm kin}$. 
Although we are not in the quantum theory and neither considering measurement interactions, this is somewhat reminiscent of the extension of the Heisenberg cut. In the quantum theory, the below extension is related to what is known as the `paradox of the third particle' \cite{Krumm:2020fws,angelo2011physics} (see also \cite{Hoehn:2021flk} for a summary).

Suppose the group of particles $M$ is not alone in the world and there is a further particle that we call $R_1$ (see Fig.~\ref{fig:gauge_vs_sym}). It will assume the role of the ``internalized'' external reference frame and will be the mechanical analog of an edge mode for $M$. To accommodate it, we will have to extend the translation-invariant Lagrangian in equation~\eqref{transllag} as well as the underlying field space accordingly. In line with the subsequent field theory discussion, let us denote this field space by $\cF_M$, which is defined in complete analogy to $\cF_{\mathring{M}}$ above, incorporating all histories of the now $N+1$ particles (group $M$ and frame $R_1$). For this $(N+1)$-body problem, we can now repeat the same exercise as before. Specifically, the symplectic potential and presymplectic structure on $\cF_M$ (which depend on $t$) read 
\beq 
\Theta_{\Sigma_t}=\sum_{i=1}^N p_i \delta q_i  + p_{R_1}\delta q_{R_1}\qquad\text{and}\qquad\Omega_{M}=\sum_{i=1}^N\delta p_i\delta q_i+\delta p_{R_1}\delta q_{R_1}\,,
\eeq 
where, in preparation for the field theory case, $\Sigma_t$ here highlights that the potential depends on the ``time slice'' $t$ both on- and off-shell, in contrast to $\Omega_M$ which on $\cS_M$ will be $t$-independent due to equation~\eqref{omcons}. In contrast to equation~\eqref{Mmom}, we now have the identity 
\beq 
P_{MR_1}:=P_M+p_{R_1}=0\label{Mmom2}
\eeq
on all of $\cF_M\times[t_1,t_2]$, so $P_M$ need no longer vanish.

All the relational observables of the previous subsection remain invariant observables on $\cF_M$. However, thanks to the additional frame particle $R_1$ we can now turn \emph{all} the kinematical degrees of freedom associated with $M$ into gauge-invariant ones. Specifically, we can now also relate the old internal frame,  particle $N$, to the ``edge mode'' $R_1$, yielding new relational observables, such as
\beq 
Q_{N|R_1}(y):=O_{q_N,U_{R_1}}(y)=q_N-q_{R_1}+y\,.\label{NR1}
\eeq 
This permits us to decompose the symplectic potential and symplectic structure once more into gauge-invariant and gauge parts, $\Theta_{\Sigma_t}=\Theta_{\rm inv}+\Theta_{\rm gauge}$ and $\Omega_{M}=\Omega_{\rm inv}+\Omega_{\rm gauge}$, taking equation \eqref{mechsymp} into account, where: 
\bes  
\Theta_{\rm inv}&:=&\sum_{j=1}^{N-1}p_j\delta Q_{j|N}(x)+P_M\delta Q_{N|R_1}(y)\,,\qquad \Theta_{\rm gauge}:=P_{MR_1}\delta q_{R_1}\nn\\
\Omega_{\rm inv}&:=&\sum_{j=1}^{N-1}\delta p_j\delta Q_{j|N}(x)+\delta P_M\delta Q_{N|R_1}(y)\,,\qquad\Omega_{\rm gauge}:=\delta P_{MR_1}\delta q_{R_1}\,.
\ees 
In particular, observe that only the last term in the invariant parts are dressed by the edge mode $R_1$. These are the mechanical analogs of the radiative contributions to the presymplectic potential and presymplectic structure on the boundary  that we shall later see in field theory, while the first terms are the mechanical analog of the bulk contribution in field theory. In contrast to the field theory case, no corner term can arise in this discrete model and the gauge parts vanish identically.

It is clear that the physical phase space
\beq 
\cP_M:=\cS_M/\!\sim\,=\cF_M/\!\sim 
\eeq 
of $M$ and $R_1$ together is isomorphic to $\mathbb{R}^{2N}$, equipped with the non-degenerate symplectic form $\Omega_{\rm inv}$, and parametrized by the pairs $(Q_{j|N},p_j;Q_{N|R_1},P_M)_{j=1}^{N-1}$. Specifically, we can describe $\cP_{M}$ equivalently through a gauge-fixing since, e.g., $q_{R_1}=y$ for some $y\in\mathbb{R}$ is a globally valid gauge fixing. On the corresponding gauge-fixing surface we have $Q_{N|R_1}(y)=q_N$ such that the coordinates on $\cP_M$ become $(Q_{j|N},p_j;q_N,P_M)_{j=1}^{N-1}$. This is equivalent to the Legendre transformation induced natural coordinates $(q_i,p_i)_{i=1}^N$ on the kinematical phase space $\cP_{\rm kin}:=T^*\cQ\simeq\mathbb{R}^{2N}$; in particular, the invariant part of the symplectic structure can (in this gauge-fixing) be written equivalently as $\Omega_{\rm inv}=\sum_{i=1}^N\delta p_i\delta q_i$. In this sense, the set of gauge-invariant states of $M$ relative to $R_1$ is equivalent (even symplectomorphic) to the kinematical phase space associated with $M$ only, i.e.\ $\cP_M\simeq\cP_{\rm kin}$. This equivalence is rooted in the above observation that we can turn \emph{all} kinematical degrees of freedom associated with $M$ into gauge-invariant ones by relating them to the edge mode $R_1$. 

\subsection{Symmetries as frame reorientations vs.\ gauge transformations}\label{ssec_symvsgaugemech}

\begin{figure}[htb]
    \centering
    \includegraphics[scale=.7]{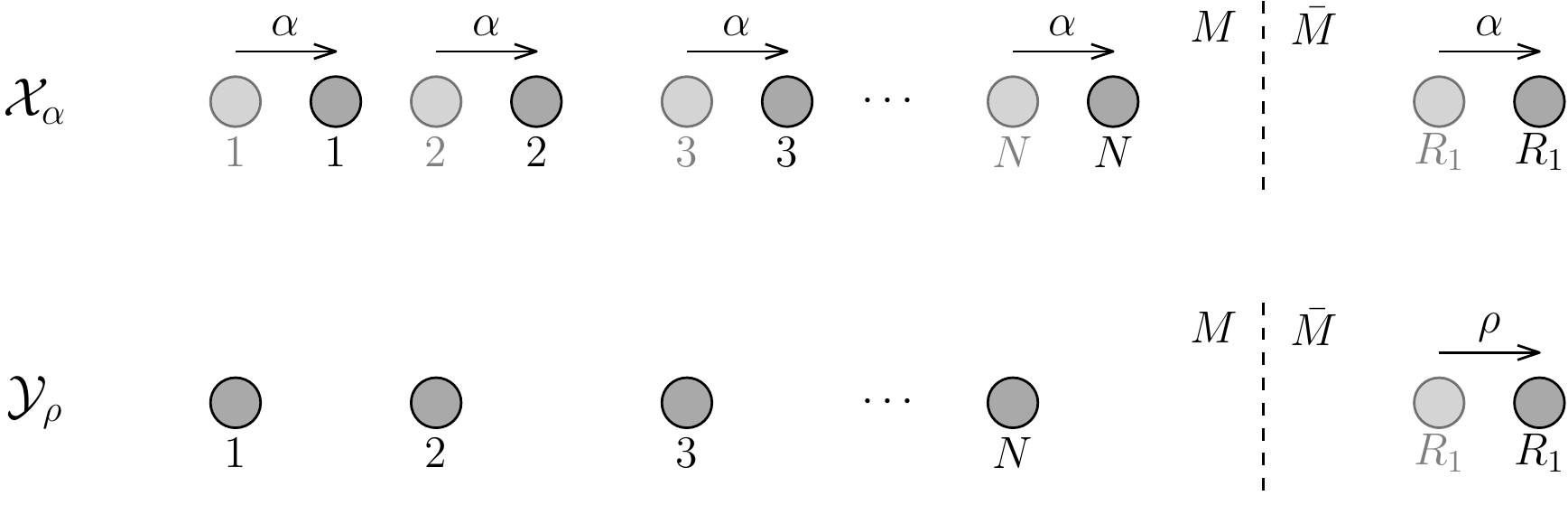}
    \caption{Two types of transformations which are indistinguishable from their action on particles in region $M$, but distinguishable relative to the dynamical frame $R_1$. The vector field $\fX_\alpha$ (top) moves both the particles in $M$ and the frame $R_1$ by the same distance, therefore does not affect relative distances and corresponds to a gauge direction. The symmetry (frame reorientation) generator $\fY_\rho$ (bottom) only acts on $R_1$,  thereby affects relative distances to the reference frame: it generates a physical transformation.}
    \label{fig:gauge_vs_sym}
\end{figure}

Given the internalized external frame $R_1$, we now have to distinguish two types of transformations which are indistinguishable in a purely internal description of $M$ on $\cP_{\mathring{M}}$, as they leave all the relations within $M$ invariant (see Fig.~\ref{fig:gauge_vs_sym}):
\begin{description}
\item[Gauge transformations:] A global translation 
\beq 
q_i\to q_i+\alpha\,,\quad i=1,\ldots,N\,,\quad \qquad\text{and}\qquad q_{R_1}\to q_{R_1}+\alpha 
\eeq
of $M$ and $R_1$ leaves the relation between the new frame $R_1$ and $M$ invariant and thus describes the same physical situation. Denote the vector field on $\cF_M$ generating this transformation once more by $\fX_\alpha$, where $\alpha$ may be ``field-dependent'', i.e.\ depend on where it is evaluated on field space $\cF_M$. It constitutes a null-direction of the presymplectic structure
\beq 
\fX_\alpha\cdot\Omega_{M}=\delta C[\alpha]=0\,,
\eeq 
with ``constraint''\footnote{We write it in quotation marks as it vanishes identically everywhere on $\cF_M\times[t_1,t_2]$.}
\beq 
C[\alpha]=-\alpha P_{MR_1}=0\,.
\eeq
\item[Symmetries (frame reorientations):] A relative translation of $M$ and $R_1$ by a distance $\rho$. While we could equivalently formulate such a transformation as a transformation of only $M$, we shall write it as a transformation purely of the new frame $R_1$:
\beq \label{framere}
q_i\to q_i\,,\quad i=1,\ldots,N\,,\quad \qquad\text{and}\qquad q_{R_1}\to q_{R_1}+\rho\,.
\eeq
This transformation does \emph{not} leave the relation between $R_1$ and $M$ invariant and thereby changes the physical situation. Since the position $q_{R_1}$ constitutes the ``orientation'' of the frame, this transformation amounts to a \emph{frame reorientation}. Denote the vector field on $\cF_M$ generating this transformation by $\fY_\rho$, where $\rho$ must now be field-space-\emph{independent}.
\\
For concreteness, let us assume that the only dependence of the  Lagrangian for $M,R_1$ on the velocities is of the form $\cL_{\rm kin}=\frac{1}{4(N+1)}\sum_{I,J=1}^{N+1}(\dot q_a-\dot q_b)^2$, where $I,J$ runs over all particles in $M$ and the edge mode $R_1$, which is tantamount to assuming unit mass for the particles and subtracting the center-of-mass kinetic energy from the total kinetic energy, e.g.\ see \cite{Vanrietvelde:2018pgb,Vanrietvelde:2018dit}. This yields $p_I=\dot q_I-\frac{1}{N+1}\sum_J\,\dot q_J$ and therefore the variations
\beq 
\Delta_\rho Q_{j|N}=\fY_\rho\cdot\delta Q_{j|N}=0, \quad\Delta_\rho Q_{N|R_1}=\fY_\rho\cdot\delta Q_{N|R_1}=-\rho, \quad\text{and}\quad\Delta_\rho p_i:=\fY_\rho\cdot\delta p_i=-\frac{\dot\rho}{N+1}\,,\nn
\eeq 
where $i=1,\ldots, N$ runs over the particles of $M$. Contraction with the presymplectic structure gives
\beq 
\fY_\rho\cdot\Omega_{M}=\fY_\rho\cdot\Omega_{\rm inv}=\delta Q[\rho]\,,
\eeq 
with gauge-invariant \emph{charge}
\beq \label{mechcharge}
Q[\rho]=\rho P_M-\frac{1}{N+1}\dot\rho\sum_{j=1}^{N-1} Q_{j|N}-\frac{N}{N+1}\dot\rho \,Q_{N|R_1}\,.
\eeq 
which does not in general vanish even on-shell.
Specifically, for an edge-mode-independent functional on $\cF_M$ of the form $f(q_i,\dot q_i)$, we have, thanks to the analog of equation~\eqref{relobs} for when $R_1$ is used as frame, 
\beq \label{mechrelabel}
O_{f,U_{R_1}+\rho}(y)=O_{f,U_{R_1}}(y-\rho)
\eeq 
and
\beq 
\rho\,\frac{\p O_{f,U_{R_1}}(y)}{\p y}+\dot\rho\,\frac{\p O_{f,U_{R_1}}(y)}{\p\dot y}=\big\{Q[\rho],O_{f,U_{R_1}}(y)\big\}=\cZ_O \cdot \fY_\rho\cdot\Omega_{\rm inv}\,,
\eeq 
where $\cZ_O$ is the Hamiltonian vector field on $\cP_M$ associated with the relational observable $O_{f,U_{R_1}}(y)$. In line with the interpretation of frame reorientations, we thus see that the charge generates changes in the frame orientation label $y$ of the edge mode for such observables. 
\\
Just like the group action $\acts$ in equation~\eqref{gaugeactsmech} corresponds to gauge transformations acting on all dynamical variables, we can now define the group action $\odot$ corresponding to symmetries, i.e.\  
\beq\label{symactsmech}
g(\rho)\odot f(q_i,q_{R_1},\dot q_i,\dot{q}_{R_1})=e^{\fY_\rho}f=f(q_i,q_{R_1}+\rho,\dot q_i,\dot{q}_{R_1}+\dot\rho)\,.
\eeq
This permits us to rewrite equation~\eqref{mechrelabel} as
\beq \label{obssymtransmech}
g(\rho)\odot O_{f,U_{R_1}}(y)=O_{f,g(\rho)\odot U_{R_1}}(y)=O_{f,U_{R_1}}(y-\rho)\,.
\eeq
This will become useful to describe frame changes shortly.
\end{description}

This distinction between gauge transformations and symmetries is the mechanical analog of the distinction in gauge theories put forth in \cite{Donnelly:2016auv}. What is new here is the connection with dynamical reference frames and especially the identification of symmetries as frame reorientations of the ``edge mode''.  We shall see  in Sec.~\ref{ssec_3types}, that we have to introduce a further distinction of symmetries into three  types.

We finally note that an infinitesimal frame reorientation $\fY_\rho$, which we have so far defined on $\cF_M$, does not necessarily translate into an actual symmetry of the solution space $\cS_M$. The translation parameter $\rho$ needs to verify some non-trivial set of differential equations in order for $\fY_\rho$ to act tangentially to $\cS_M$, that is, in order for $\Delta_\rho$ to map solutions to solutions. In the present model, requiring those equations to have non-trivial solutions turns out to impose severe constraints on the shape of the translation-invariant potential $V$. As a result, for the purpose of illustrating the physical action of symmetries on solutions, it will be convenient to simply assume a vanishing potential $V=0$. In that case, it is clear that $\fY_\rho$ leaves the space of solutions invariant (in particular $\dot{p}_{R_1} = 0$, where $p_{R_1} = \dot{q}_{R_1} - \frac{1}{N+1}(\sum_{i=1}^N \dot{q}_i + \dot{q}_{R_1})$), if and only if $\ddot{\rho}(t)=0$. That is, we must require $\rho(t)$ to be a linear function of time.

\subsection{Reference frame changes}\label{ssec_rfc}

The choice of dynamical reference frame is not unique. For instance, we have already used the internal frame particle $N$ and the ``edge mode'' $R_1$. As in the field theory context later, there could be additional external frames that we can internalize as further edge modes. It is therefore natural to inquire about a further type of transformation we have so far not considered: changes of reference frame. How can we change from the description relative to one dynamical frame to that relative to another? Plenty of recent efforts in the context of quantum reference frames have focused on developing a framework for dynamical frame covariance \cite{Giacomini:2017zju,Vanrietvelde:2018dit,Vanrietvelde:2018pgb,Hohn:2018iwn,Hohn:2018toe,Hoehn:2021wet,Hohn:2019cfk,Hoehn:2020epv,Krumm:2020fws,Hoehn:2021flk,delaHamette:2020dyi,Castro-Ruiz:2019nnl,Giacomini:2020ahk,Ballesteros:2020lgl}. For example, the classical discussion of \cite{Vanrietvelde:2018dit,Vanrietvelde:2018pgb} could be directly applied to the present mechanical scenario. Here, we shall restrict our attention to the transformation of relational observables under changes of dynamical frame, i.e.\ we consider an active change from the relational observables relative to one frame to those relative to another. The following discussion can thereby be viewed as an extension of the exploration in some of the above references where relational observables relative to different frame choices where considered (e.g., see \cite{Hoehn:2021wet}), but the explicit transformations between them were not given.\footnote{The frame changes in these references rather encompass how one and the same invariant observable can be described relative to different internal frame choices. This involves an additional reduction using a gauge fixing procedure compared to the present treatment.} It also bears some resemblance to the ``clock transformations'' in \cite[Sec.\ 5]{Dittrich:2007jx}.

\begin{figure}[htb]
    \centering
    \includegraphics[scale=.7]{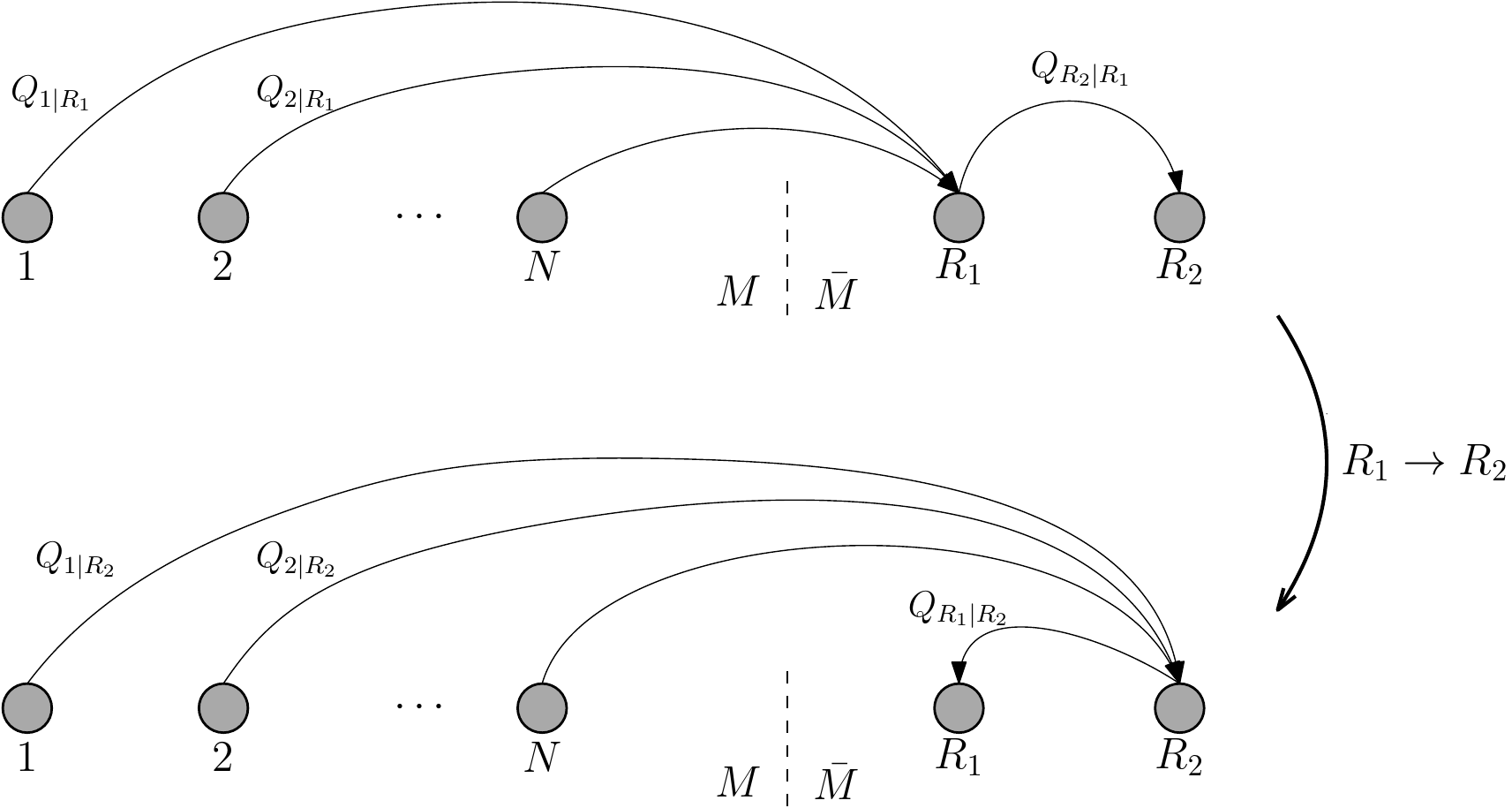}
    \caption{Pictorial illustration of the frame change mapping the relational observables $\{ Q_{i\vert R_1} \}$ to $\{ Q_{i\vert R_2} \}$. It is implemented by a \emph{field-dependent} reference frame transformation, namely a translation by the relative distance $U_{R_2} - U_{R_1}$.}
    \label{fig:change_of_frame}
\end{figure}

Suppose we are given a second ``edge mode'' $R_2$ in addition to $R_1$ and the original group of particles $M$.  $R_2$ can be accommodated by extending accordingly the translation-invariant Lagrangian, as well as the field space (and thus physical phase space), in the same manner in which these structures were extended in the previous subsection to internalize the first ``edge mode'' $R_1$. We denote the group-valued orientation variables of the two dynamical external frames to $M$ by $U_{R_k}=q_{R_k}$, $k=1,2$. See Fig.~\ref{fig:change_of_frame}.

Consider an arbitrary functional $f(q_I(t),\dot q_I(t))$ which depends analytically on positions and velocities, where $I$ runs over the $N$ particles in $M$, as well as $R_1,R_2$. Invoking equation~\eqref{relobs0}, the corresponding relational observable relative to frame $R_k$ reads
\beq 
O_{f,U_{R_k}}(y_k)=\left(U_{R_k}-y_k\right)^{-1}\acts f:=\restr{g(\alpha)\acts f}{\alpha=-U_{R_k}+y_k}\,.\label{relobs1}
\eeq 
This implies that the transformation from $O_{f,U_{R_1}}$ to $O_{f,U_{R_2}}$
is given by a \emph{relation-conditional translation} by the gauge-invariant relative distance between the two frames (which is a dynamical element of the translation group $(\mathbb{R},+)$):
\beq 
g_{21}(U_{R_1},U_{R_2}):=Q_{{R_1}|{R_2}}(0):=O_{U_{R_1},U_{R_2}}(0) = U_{R_1}-U_{R_2}\,.
\eeq 
This implies that the transformation from the description relative to frame $R_1$ to the one relative to frame $R_2$ is given by the \emph{relation-conditional translation} $\odot$ of the frame $R_1$:
\bes
U_{R_1}\quad&\longmapsto&\quad U_{R_2}=g_{21}^{-1}\odot U_{R_1}:=U_{R_1}-g_{21}\,,\label{mechobsrfc}\\
O_{f,U_{R_1}}(y_1)\quad&\longmapsto&\quad O_{f,U_{R_2}}(y_2)=\left(-y_1+y_2+g_{21}\right)^{-1}\odot O_{f,U_{R_1}}(y_1):=O_{f,(-y_1+y_2+g_{21})^{-1}\odot U_{R_1}}(y_1)\,.\nn
\ees
Note that this constitutes an extension of the definition of $\odot$ in equation~\eqref{symactsmech} to field-dependent frame translations, in complete analogy to how equation~\eqref{relobs0} extends field-independent to field-dependent gauge transformations. In particular, just like  equation~\eqref{relobs0} gives rise to observables invariant under gauge transformations, equation~\eqref{mechobsrfc} generates observables invariant under symmetries (reorientations of the frame $R_1$).
Indeed, 
\bes
O_{f,U_{R_2}}(y_2)&=&\big((U_{R_1}-y_1)-(U_{R_1}-y_1)+(U_{R_2}-y_2)\big)^{-1}\acts f \nn\\
&=& \Big(\big(-g_{21}-y_2+y_1\big)^{-1}\acts\big(U_{R_1}-y_1\big)^{-1}\Big)\acts f\nn\\
&=& \big(-g_{21}-y_2+y_1\big)^{-1}\odot O_{f,U_{R_1}}(y_1)\label{mechtrafo3}\\
&=&\restr{g(\beta)\acts f}{\beta=(-y_1+y_2+g_{21})^{-1}\odot \alpha(U_{R_1})}\,,\nn
\ees
where $(-y_1+y_2+g_{21})^{-1}\odot \alpha(U_{R_1})=-U_{R_2}+y_2$ and $\alpha(U_{R_1})=-U_{R_1}+y_1$ is the field-dependent group parameter used in equation \eqref{relobs1}.\footnote{Note that we cannot write the third line in \eqref{mechtrafo3} in terms of the global translation group action as ${\big(g_{21}-y_2+y_1\big)^{-1}\acts O_{f,U_{R_1}}(y_1)}$ since $g\acts O_{f,U_{R_1}}(y_1)=O_{f,U_{R_1}}(y_1)$ for all $g\in\mathbb{R}$.}
We can therefore interpret the frame change transformation as a relation-conditional frame reorientation. Specifically, it is equivalent to the relation-conditional translation of the frame orientation label
\beq \label{clab}
y_1\mapsto y_1+(-y_1+y_2+g_{21})=Q_{R_1|R_2}(y_2)
\eeq 
i.e.
\beq \label{mechlabeltrans}
\big(-g_{21}-y_2+y_1\big)^{-1}\odot O_{f,U_{R_1}}(y_1)=O_{f,U_{R_1}}\big(Q_{R_1|R_2}(y_2)\big)\,.
\eeq 
This is especially transparent in the coordinate expression, cf.\ equation~\eqref{relobs}, for relational observables $O_{f,U_{R_1}}(y_1)= f(q_i-U_{R_1}+y_1,\dot q_i-\dot U_{R_1}+\dot y_1)$.

It is  instructive to illustrate this on a generating set of observables. Since the momenta are translation-invariant and therefore trivially relational observables relative to both $R_1,R_2$, we only need to look at the relational observables associated with the positions $f = q_I$:
\beq 
Q_{I|{R_1}}(y_1) = q_I-U_{R_1}+y_1\quad\longmapsto\quad q_I-U_{R_1}+Q_{R_1|R_2}(y_2)=Q_{I|{R_2}}(y_2)\,.
\eeq 
This includes the tautological observables (e.g., encoding ``what is the position of $R_1$ when $R_1$ is in position $y_1$?'')
\bes
Q_{R_1|R_1}(y_1)=y_1 &\longmapsto&  Q_{{R_1}|{R_2}}(y_2)\nn\\
Q_{{R_2}|{R_1}}(y_1)=U_{R_2}-U_{R_1}-y_1&\longmapsto& y_2=Q_{{R_2}|{R_2}}(y_2)\,.\nn
\ees 
Including the tautological observables and the identity $P_M+p_{R_1}+p_{R_2}=0$, the change of frame thus amounts to a canonical transformation on the physical phase space
\beq 
\left(Q_{i|R_1},p_i; Q_{R_2|R_1},p_{R_2}\right)_{i=1}^N\quad\longmapsto\quad\left(Q_{i|R_2},p_i;Q_{R_1|R_2},p_{R_1}\right)_{i=1}^N\,.
\eeq 
This can also be viewed as a ``field redefinition'' on the physical phase space.

It is important to note that, despite the change of frame transformation admitting the interpretation of a relation-conditional (and hence configuration-dependent) frame reorientation, it is \emph{not} a symmetry; as seen in the previous subsection, symmetries are configuration-\emph{independent} translations of the frame $R_1$. In particular, the change of frame transformation is \emph{not} generated by the charges $Q[\rho]$. Note also that, in contrast to a symmetry, it does not change the relations between the $N+2$ particles, only the description of them.

\subsection{Subsystem phase space from post-selection on the global space of solutions}\label{ssec_mechpost}

Thus far, we have considered the covariant phase space structures associated with the group of particles $M$ with and without the two ``edge modes'' $R_1,R_2$. Accommodating the edge modes required us to extend the field and phase spaces, embedding a setup with fewer into a setup with more particles. 
Let us now consider the opposite process: how can we recover the covariant phase space structure of fewer particles from that with more particles? This serves as a toy scenario for recovering local subregion phase space structures from those associated with a global spacetime region in field theory, which we shall elaborate on later. This can be achieved through a post-selection on the global space of solutions.

Suppose that $R_1,R_2$ are part of a second group $\bar M$ of $\bar N$ particles, the complement of $M$. We shall now illustrate how to recover the phase space structure of the subsystem $M$ from the global one for $M\cup\bar M$.
To this end, let us assume for concreteness and simplicity that we are dealing with free particle dynamics subject to a translation-invariant Lagrangian of the form ${\cL_{M\cup\bar M}=\frac{1}{4(N+\bar N)}\sum_{I,J=1}^{N+\bar N}(\dot q_I-\dot q_J)^2}$, where $I,J$ now runs over all particles in $M\cup\bar M$. Then $p_I=\dot q_I-\frac{1}{N+\bar N}\sum_{J}\,\dot q_J$, so that $p_i=\dot q_i-\f{1}{N+\bar N}\sum_{j=1}^N\dot q_j-\f{\bar N}{N+\bar N}\dot q_{\bar C}$, where $i=1,\ldots,N$ runs over the particles of $M$. Here, $q_{\bar C}:=\f{1}{\bar N}\sum_{\bar j=N+1}^{N+\bar N}q_{\bar j}$, where $\bar{j}$ labels the $\bar N$ particles in $\bar M$, is the center-of-mass of $\bar M$. In other words, the momenta in $M$ now only depend on $\bar M$ through its center-of-mass. We denote the global field space on which the Lagrangian is defined by $\cF$ and the subspace of solutions by $\cS$. Let us choose particle $N$ as the internal frame for group $M$, as before, while we now select the center-of-mass $\bar C$ of $\bar M$ as an external frame for $M$ and an internal frame for $\bar M$. We could in principle choose any of the configuration degrees of freedom in $\bar M$ as the edge mode, however, to ensure a suitable dynamical decoupling, it will be convenient to choose $q_{\bar C}$ as the edge mode. Repeating the previous exercises on $\cF$, we find the presymplectic structure in the form
\beq \label{globpres}
\Omega = \sum_{j=1}^{N-1}\delta p_j\delta Q_{j|N}+\delta P_M\delta Q_{N|\bar C}+\left(\delta P_M+\delta P_{\bar M}\right)\delta q_{\bar C}+\sum_{\bar{j}=N+1}^{N+\bar N}\delta p_{\bar{j}}\delta Q_{\bar{j}|\bar C}\,,
\eeq 
 where $j$ labels the $N-1$ particles in $M$, except particle number $N$, and $P_{\bar M}=P_{\bar C}=\sum_{\bar{j}}p_{\bar{j}}$ is the total momentum of $\bar M$. The relational observables $Q_{N|\bar C},Q_{\bar j|\bar C}$ are constructed as before according to equation~\eqref{mechconfigrelobs}.\footnote{There is a redundancy in the last term of equation~\eqref{globpres} as $\sum_{\bar j=1}^{\bar N}Q_{\bar j|\bar C}=0$, but this will be of no relevance below.} We now have the identity
\beq 
P_M+P_{\bar M}=0\,.
\eeq 
The third contribution to $\Omega$ is the pure gauge part and thus vanishes identically. The first and last contributions pertain solely to $M$ and $\bar M$, respectively, and are  mechanical analogs of the bulk contribution of a subregion. By contrast, the second term $\Omega_{\rm rad}:=\delta P_M \delta Q_{N|\bar C}$ is the mechanical analog of the edge-mode-dressed radiative contribution on the boundary between adjacent subregions $M$ and $\bar M$ in field theory. We will consider the particle $N$ as the boundary of $M$ and $(Q_{N|\bar C},P_M)$ as the ``boundary variables''. 

In \emph{splitting post-selection}, we shall now post-select on the global space of solutions $\cS$ those solutions that are compatible with a choice of \emph{gauge-invariant} boundary conditions on particle $N$. This will split the global solution space into subgroup solution spaces for $M$ and $\bar M$. The ``boundary contribution`` $\Omega_{\rm rad}$ will play an important role in this process. In order to encompass different types of boundary conditions (Dirichlet, Neumann, Robin and mixed), let us consider the linear combinations of the translation-invariant data
\beq \label{Xmech}
X=a\,Q_{N|\bar C}+b\, P_M\,,\qquad\qquad Y=c\,Q_{N|\bar C}+d\,P_M\,,
\eeq 
where $a,b,c,d$ are real functions of $t$ subject only to $ad-bc=1$, in which case they define a symplectic transformation
\beq 
\Omega_{\rm rad}=\delta P_M\delta Q_{N|\bar C} = \delta Y\delta X\,.
\eeq

The space of solutions $\cS$ can be parametrized by solutions ${\left(Q_{j|N}(t), Q_{N|\bar C}(t),q_{\bar C}(t),Q_{\bar{j}|\bar C}(t); t\in[t_1,t_2]\right)}$  and so also by ${\left(Q_{j|N}(t), X(t),q_{\bar C}(t),Q_{\bar{j}|\bar C}(t); t\in[t_1,t_2]\right)}$. In particular, we can thus use the  configuration variable $X$ to foliate $\cS$
\beq 
\cS = \underset{X_0}{\bigsqcup}\,  \cS_{X_0} \,, \qquad  \cS_{X_0} := \left\lbrace \text{all histories $c(t) \in \cS$} \vert X(t) = X_0(t) \right\rbrace  \,,
\eeq 
where $X_0(t)$ is some non-dynamical background function specifying the gauge-invariant boundary condition for particle $N$ (which, of course, have to be consistent with the equations of motion of the relevant Lagrangian). For example, we could have $a(t_1)=d(t_1)=1$ and $b(t_1)=c(t_1)=0$ initially and $a(t_2)=d(t_2)=0$ and $b(t_2)=1=-c(t_2)$ at the end, so that the boundary conditions are initially on $X(t_1)=Q_{N|\bar C}(t_1)$, hence Dirichlet, and finally on $X(t_2)=P_M(t_2)$, thus Neumann, and so altogether of mixed type. Clearly, different choices of the functions $a,b,c,d$ will lead to distinct foliations of $\cS$.

Post-selection now proceeds by selecting a leaf $\cS_{X_0}$ from the global solution space $\cS$, i.e.\ restricting to all solutions consistent with the history $X_0(t)$ of the variable $X$. Pulling back to any of these leaves, we thus have $\restr{\delta X}{\cS_{X_0}}=\delta X_0=0$ and so
\beq 
\restr{\Omega_{\rm rad}}{\cS_{X_0}}=0\,.
\eeq 
Henceforth, we also write the weak equality $\heq$ for identities that only hold on $\cS_{X_0}$ (i.e.\ after the boundary conditions have been imposed), so that $\Omega_{\rm rad}\heq0$.
In other words, post-selection guarantees that there is no (physical) symplectic flux between $M$ and $\bar M$ since $\Omega_{\rm rad}$ is the only gauge-invariant contribution to the global presymplectic structure $\Omega$ in \eqref{globpres} that involves degrees of freedom from $M$ and $\bar M$ that are generically non-constant (the momenta, which depend on \emph{all} particles, satisfy $\dot p_I\approx0$).

This also splits the phase space structures associated with $M$ and $\bar M$. In particular, we can consider the restriction of any leaf to $M$. As can be easily checked, the equations of motion for the $q_i$ and edge mode $q_{\bar C}$ decouple from the remaining ones for $\bar M$ and can thereby be solved independently; we can therefore perform post-selection on this subset. This suggests to define the restriction of the leaf $\cS_{X_0}$ to $M$ by including the edge mode in the form
\beq 
\cS_M^{X_0}:=\Big\{\restr{c(t)}{M}=\left(Q_{j|N}(t),Q_{N|\bar C}(t),q_{\bar C}(t)\right)\,,\,t\in[t_1,t_2]\,\Big\vert \,c\in\cS_{X_0}\Big\}\,.
\eeq 
Accordingly, we equip this ``regional'' space of solutions for $M$ with the presymplectic structure (cf.\ equation~\eqref{globpres})
\beq 
\sum_{j=1}^{N-1}\delta p_j\delta Q_{j|N}+\delta P_M\delta Q_{N|\bar C}+\left(\delta P_M+\delta P_{\bar M}\right)\delta q_{\bar C}=\Omega_{\mathring{M}}+\Omega_{\rm rad}\heq\Omega_{\mathring{M}}\,,
\eeq 
recalling that the second term on the left hand side vanishes on-shell of the boundary conditions, while the last term vanishes identically. On-shell, 
the surviving piece of the presymplectic structure for $M$ is thus $\Omega_{\mathring{M}}=\mathring{\Omega}_{\rm inv}=\sum_j\delta p_j\delta Q_{j|N}$ given in \eqref{mechsymp}, which is the ``bulk''  non-degenerate symplectic form for $M$ on the physical phase space $\cP_{\mathring{M}}$ discussed at the end of Sec.~\ref{ssec_intpart}. Indeed, using the equations of motion in the form $\dot p_I\approx0$ and noting that $\dot Q_{j|N}=p_j-p_N$ and $\dot Q_{N|\bar C}=p_N+\f{1}{\bar N}P_M$, it is straightforward to check that $\f{\extd}{\extd t}\left(\Omega_{\mathring{M}}+\Omega_{\rm rad}\right)\approx0$, i.e.\ the restricted presymplectic structure for $M$ is conserved on-shell. It is also gauge-invariant and any $\fX_\alpha$ constitutes a degenerate direction of it.\footnote{Neither conservation nor invariance would be true had we only included the generally non-vanishing $\delta P_M\delta q_{\bar C}$ for the edge mode contribution.}
Hence, upon factoring out the degenerate directions of the presymplectic structure from $\cS_M^{X_0}$, we recover the phase space $\cP_{\mathring{M}}$ of internally distinguishable states for the subsystem $M$ from splitting post-selection on the global solution space for $M\cup\bar M$. As discussed shortly, this does \emph{not} mean, however, that the dynamics on $\cP_{\mathring{M}}$ induced by $\cS_M^{X_0}$ will coincide with the \emph{isolated} dynamics for $M$ defined by the pure subsystem Lagrangian $\cL$ implicitly considered in Sec.~\ref{ssec_intpart}; depending on the boundary condition $X_0$, the Hamiltonian may be distinct.

Splitting post-selection can also be performed at the level of the action, however, we shall only discuss this in the field theory case later. This will lead to a systematic algorithm for generating boundary actions.

\subsection{Three types of symmetry transformations (frame reorientations)}\label{ssec_3types}

Each leaf $\cS_{X_0}$ in the global space of solutions $\cS$ can be regarded as (the space of solutions of) a specific subsystem theory for $M$ (resp.\ $\bar M$), defined by the boundary conditions $X_0$. Indeed, since the ``edge-mode-dressed'' variables $X,Y$ contain degrees of freedom from both $M$ and $\bar M$, these boundary conditions specify how $M$ couples to its complement $\bar M$. Different choices of $X_0$ for $X$ will lead to distinct subsystem Hamiltonians (e.g., different parameters or explicit time dependence) for the remaining $N-1$ particles in $M$. This will be discussed in more detail in \cite{CH2}.  That is to say, different ``subleaves'' $\cS_M^{X_0}$  yield the same physical phase space $\cP_{\mathring{M}}$, but will correspond to distinct subsystem dynamics and thereby contain different sets of solutions for $Q_{j|N}(t)$. We are therefore justified to think of the different $\cS_M^{X_0}$ as different theories for the subsystem $M$ and, consequently, of the global space of solutions $\cS$ as a ``meta-theory'' containing all the different subsystem theories.

In Sec.~\ref{ssec_extHeis}, we discussed the distinction between gauge and symmetry transformations, originally put forth in \cite{Donnelly:2016auv}, and interpreted the latter as frame reorientations. Since our subsystem theory is now subjected to boundary conditions when considered as a post-selection of the global theory, we have to explore the consequences for symmetries in the presence of these boundary conditions. In particular, we have to investigate whether the frame reorientations preserve the boundary conditions or whether they give rise to changes of leaves $\cS_{X_0}$ in the foliations of $\cS$. 

For concreteness, let us once more assum a free particle dynamics as in the previous subsection. The vector field $\fY_\rho$ on $\cS$ generating the frame reorientation \eqref{framere} (for the center-of-mass) then leads to the variations $\Delta_\rho p_i:=\fY_\rho\cdot\delta p_i=-\frac{\bar N}{N+\bar N}\dot\rho$, where $i=1,\ldots, N$ runs over the particles of $M$. This has two consequences: first, it leads to the following variation of the variable fixed through the boundary conditions
\beq\label{mechpresbdry} 
\Delta_\rho X = \fY_\rho\cdot\delta X = -\rho a-\frac{N \bar N}{N+\bar N}\dot\rho b\,.
\eeq 
Hence, only symmetries with $\rho$ such that $\Delta_\rho X =0$ will leave the boundary conditions invariant.
Second, restricting $\fY_\rho$ to the subleaf $\cS_M^{X_0}$, we can contract it with its presymplectic structure ${\Omega}_{\mathring{M}}+\Omega_{\rm rad}$. Note that, while the pullback of $\Omega_{\rm rad}$ to $\cS_M^{X_0}$ vanishes, the contraction with $\fY_\rho$, which may not be tangential to $\cS_{X_0}$ in $\cS$, will not in general vanish. This leads to 
\beq 
\fY_\rho\cdot\left({\Omega}_{\mathring{M}}+\Omega_{\rm rad}\right)=\delta Q[\rho]
\eeq 
with charges in the form (cf.\ \eqref{mechcharge})
\beq \label{Qmech}
Q[\rho]=\rho P_M-\frac{\bar N}{N+\bar N}\dot\rho\sum_{j=1}^{N-1} Q_{j|N}-\frac{N\bar N}{N+\bar N}\dot\rho \,Q_{N|\bar C}\,.
\eeq 
Given that they constitute generators of an Abelian transformation group (translations of $R_1$), they generate an Abelian Poisson algebra, $\{Q[\rho],Q[\sigma]\}=\fY_\sigma\cdot\left(\fY_\rho\cdot\left(\Omega_{\mathring{M}}+\Omega_{\rm rad}\right)\right)=0$.

This leads to a further distinction of three types of symmetry transformations (frame reorientations). 
\begin{description}
\item[Symmetries.] These are frame reorientations $\fY_\rho$ which preserve the boundary condition $X=X_0$, hence leave $\cS_{X_0}$ in $\cS$, and feature an \emph{unconstrained charge}. These are symmetries with \emph{time-dependent} $\rho(t)$ such that $\Delta_\rho X=0$ and $\delta Q[\rho]\not\heq 0$, by which we mean that the pullback of $\delta Q[\rho]$ to the leaf $\cS_M^{X_0}$ does not vanish (e.g., $Q[\rho]$ cannot be proportional to $X$). The above implies that this is only possible for $a,b\neq0$ and thus requires \textbf{Robin boundary conditions}. Solving $\Delta_\rho X=0$ in this case and recalling from the end of Sec.~\ref{ssec_symvsgaugemech} that for free particles frame reorientations must be of the form $\rho(t)=c_1t+c_2$ with $c_1,c_2=const$ to be determined, yields 
\beq 
\f{c_2}{c_1}=- t -\f{N+\bar N}{N\bar N} \f{b}{a}\,,\nn
\eeq 
which only has a solution for special choices of $a,b$. When it has a solution, only the ratio $c_2/c_1$ is specified, which means there will then be a one-parameter family of solutions. The Abelian Poisson algebra of
the non-vanishing charges on $\cS_M^{X_0}$,
\beq 
Q[\rho]=-\frac{\bar N\,c_1}{N+\bar N}\left(\frac{N}{a} X_0+\sum_{j=1}^{N-1} Q_{j|N}\right)\,,
\eeq
is then of dimension $\dim\mathbb{R}=1$, in line with generating a translation group symmetry.
\item[Meta-symmetries.] These are frame reorientations $\fY_\rho$ which do \emph{not} preserve the boundary condition, $\Delta_\rho X\neq0$, and so change leaf $\cS_{X_0}$ (or subsystem theory) in the foliation of the meta-theory $\cS$. For example, in the case of \textbf{Dirichlet boundary conditions}, i.e.\ $a=1$ and $b=0$ so that $X=Q_{N|\bar C}=Q_{N|\bar C}^0$, we have $\Delta_\rho Q_{N|\bar C}=-\rho$ so that \emph{any} frame reorientation will lead to a change of subsystem theory. In fact, Dirichlet boundary conditions are the only ones which imply that every frame reorientation is a meta-symmetry. By contrast, for some fixed Robin boundary conditions, hence $a,b\neq0$, there will typically exist some $\rho$ satisfying $\Delta_\rho X\neq0$ and others satisfying $\Delta_\rho X=0$, so Robin boundary conditions will generally admit meta-symmetries, as well as symmetries.
\\
While $\fY_\rho$ thus leads to a change of leaf $\cS_{X_0}$, note that it is a symplectomorphism in the meta-theory $\cS$. Indeed, using Cartan's magic formula $\cL_\fY\Omega =\delta(\fY\cdot\Omega)+\fY\cdot(\delta\Omega)$, we find
\beq 
\cL_{\fY_\rho}\left(\Omega_{\mathring{M}}+\Omega_{\rm rad}\right)=\delta\left(\delta Q[\rho]\right)=0\,,
\eeq 
which is the infinitesimal form of a symplectomorphism.
\item[Boundary gauge symmetries.] These are frame reorientations $\fY_\rho$ preserving the boundary conditions, $\Delta_\rho X=0$, and generating a \emph{constrained charge} $\delta Q[\rho]\heq0$. For instance, we could have $Q[\rho]=g\,X$  for some background function $g(t)$. In the present model, this is only possible for \textbf{Neumann boundary conditions}, i.e.\ for $a=0$ so that $X=b P_M=X_0$, and $\rho$ time-independent, which leads to the new ``boundary constraint'' 
\beq 
C_\partial[\rho]=\rho\left(P_M-\frac{1}{b}X_0\right)\heq 0\,.
\eeq 
 Since $\fY_\rho$ defines a null-direction, $\fY_\rho\cdot\left({\Omega}_{\mathring{M}}+\Omega_{\rm rad}\right)=\delta C_\partial[\rho]=0$ on this subsystem theory, it constitutes a gauge transformation. Indeed, all physical variables that are to survive on the subsystem phase space have to leave the conditions defining the theory invariant and must therefore be invariant under $\fY_\rho$. While $(Q_{j|N},p_j)$ are invariant (recall that $\dot\rho=0$), we have $\Delta_\rho Q_{N|\bar C}=-\rho$. Hence, while the edge-mode-dressed $Q_{N|\bar C}$ is gauge-invariant under global translations $\fX_\alpha$, it fails to be invariant under the boundary gauge symmetries appearing for Neumann conditions. However, since $\fY_\rho$ therefore only varies a term in $\Omega_{\rm rad}$ (through $Y$) which vanishes on-shell of the boundary conditions, it constitutes a direction that in any case is factored out from $\cS_M^{X_0}$ in the reduced phase space construction. It therefore does not lead to an additional reduction of the dimension of the reduced phase space, in contrast to the analogous situation in field theory later.
A special case of this constraint is $P_M\heq 0$ in which case one obtains precisely the purely internal subsystem theory on $\cP_{\mathring{M}}$ discussed in Sec.~\ref{ssec_intpart}, in which no edge mode appeared.
\end{description}

In conjunction with the previous subsection, this implies that the edge-mode-dressed configuration observables of $M$ drop out from the ensuing phase space $\cP_{\mathring{M}}$ for the subsystem $M$,\footnote{The momentum variables $p_i$, which depend on $\dot q_{\bar C}$, however, survive on $\cP_{\mathring{M}}$.} regardless of the boundary conditions. However, the edge mode $\bar C$ plays a crucial role in the formulation of the boundary conditions giving rise to the foliation of the global solution space into subsystem theories for $M$. Through the boundary conditions, the edge mode will also determine the  dynamics on $\cP_{\mathring{M}}$ induced by the global Lagrangian on $M\cup\bar M$ \cite{CH2}. Altogether, the edge mode is therefore relevant for describing in an invariant manner how a subsystem couples to its complement.

Below (especially in Sec.~\ref{sec:examples}), we shall elaborate on these observations in the field theory setting, where the situation is similar, but also more subtle.

\subsection{Reference frame changes as changes of foliation of the solution space}\label{ssec_mechrfcfol}

Symmetries as frame reorientations thus divide into those that are tangential or transversal to a leaf $\cS_{X_0}$ in a foliation of $\cS$, but they always preserve the foliation. What kind of transformations map between distinct foliations of $\cS$ and so between different classes of boundary conditions, i.e.\ subsystem theories?  

For a fixed external frame $R_1$, different foliations arise from different choices for the functions $a,b,c,d$ that define the ``boundary variables'' $X,Y$ in equation~\eqref{Xmech}. These different foliations are therefore related by the corresponding canonical transformations in $\cS$. 

Similarly, the second external frame $R_2$ will give rise do different edge-mode-dressed ``boundary variables'' $X',Y'$, not encompassed by  $X,Y$ in equation~\eqref{Xmech}. These lead to a distinct family of foliations of $\cS$. It is the reference frame transformations of Sec.~\ref{ssec_rfc} that link the family of foliations relative to frame $R_1$ with the one relative to frame $R_2$. 
This bears some analogy to spacetime, where different foliations are  related by (spacetime) reference frame transformations too.

\section{Covariant phase space method in field theory}\label{sec:cov_phase_space}

We now begin our systematic investigation of edge modes in (gauge) field theory. In the remainder of the paper, we will consider relativistic and Lagrangian field theories defined on $d$-dimensional Lorentzian manifolds of the form $M\cup \bar M$, where $M$ is the spacetime subregion of primary interest and $\bar M$ its complement. For definiteness, the interface between $M$ and $\bar M$, which we denote by $\Gamma$, is assumed to be time-like with the topology of $S^{d-2}\times \mathbb{R}$; see Fig.~\ref{fig:regions} from Sec.~\ref{sec_gaugesplit}. The main purpose of this section is twofold: first, to summarize the key ingredients of the covariant phase space formalism we will need at the global level, that is in $M \cup \bar M$; and second, to anticipate the main questions we will have to address in going from the global level to the regional covariant phase space for subregion $M$.

\subsection{Presymplectic structure of global field space}\label{sec:presympl_global}

We start from the  space of off-shell field configurations in $M\cup \bar{M}$, which we denote by $\cF$. Given a Lagrangian $L$, one can define a presymplectic potential $\Theta$ on $\cF$ via
\beq
\delta L = E_a \delta \Phi^a + \extd \Theta \,,
\eeq
where $\Phi^a$ are local coordinates on $\cF$, $E_a \approx 0$ are the bulk equations motion associated to $L$ and $\omega := \delta \Theta$ is the \emph{presymplectic current}. We note that $L,\Theta$ and $\omega$ are $(0,d)$-, $(1,d-1)$- and $(2,d-1)$-forms, respectively, where we recall that an $(r,s)$-form is an $r$-form on $\cF$ and an $s$-form on $M\cup\bar M$. By construction, $\omega$ is spacetime closed on-shell, since:
\beq\label{eq:closed_presymplectic}
\extd \omega = \extd \delta \Theta = \delta \extd \Theta =   
\delta \left( \delta L - E_a \delta \Phi^a \right) = - \delta E_a \delta \Phi^a \approx 0 \,.
\eeq 
We can then introduce a canonical presymplectic two-form 
\beq\label{eq:global_presymplectic}
\Omega = \int_{\Sigma \cup \bar \Sigma} \omega \, 
\eeq 
on the subspace of on-shell configurations $\cS\subset\cF$, where $\Sigma \cup \bar\Sigma$ is some arbitrary Cauchy surface in $M\cup\bar M$, such that $\Sigma \subset M$ and $\bar\Sigma \subset \bar M$. 
In case $M\cup \bar M$ has spatial boundaries (whether asymptotic or finite), we assume that suitable boundary (or fall-off) conditions have been imposed at the global level so that $\omega \approx 0$ there. Together with \eqref{eq:closed_presymplectic}, this in turn ensures that $\Omega$, seen as a two-form on the solution space $\cS$, is independent from the choice of Cauchy surface $\Sigma \cup \bar\Sigma$.

Vector fields $\fX_\alpha$ on $\cF$ which on-shell lie in the kernel of the presymplectic form,
\beq
\fX_\alpha \cdot \Omega \approx 0\,,
\eeq
are infinitesimal gauge transformations. They form a Lie algebra \cite{Lee:1990nz} and generate the gauge group $\cG$ (group of spacetime gauge transformations). Any two solutions related by an element of $\cG$ lie on the same gauge orbit and must be considered as physically equivalent. Ignoring technical subtleties (such as the Gribov problem), one obtains a symplectic form $\hat\Omega$ on the space of gauge orbits $\cP := \cS / \cG$ by quotienting $\Omega$ with respect to~$\cG$. We shall refer to $\cP$ as the \emph{physical phase space} or, equivalently, as the \emph{reduced phase space}.

Our goal in the rest of the paper will be to elucidate how to induce a consistent dynamics for fields in the subregion $M$ via a post-selection procedure on the solution space of the global dynamical system in $M \cup \bar{M}$. In the presence of gauge symmetries, post-selection can in principle be performed at two levels. As a first option, one can post-select in the physical phase space of gauge orbits equipped with the symplectic form $\hat\Omega$, to directly obtain a phase space $\cP_M$ for subregion $M$, equipped with a symplectic form $\hat\Omega_M$. Alternatively, one can implement post-selection prior to gauge reduction. Given $\cS$ and $\Omega$ as input, one then looks for a presymplectic form $\Omega_M$ for the post-selected solution space $\cS_M$ in region $M$. Technical obstructions aside, one might subsequently expect to recover $(\cP_M , \hat\Omega_M )$ from $(\cS_M , \Omega_M)$ by a suitable gauge reduction at the level of the subregion $M$. This is captured in the following (and tentative) commutative diagram:   
\[\begin{tikzcd}
(\cS , \Omega) \arrow{r}{/\cG} \arrow[swap]{d}{\mathrm{post-selection}} & (\cP , \hat\Omega) \arrow{d}{\mathrm{post-selection}} \\
(\cS_M , \Omega_M) \arrow{r}{/\cG} & (\cP_M , \hat\Omega_M)
\end{tikzcd}
\]
However, the process of gauge reduction being in general highly non-trivial, horizontal arrows are not always readily available in practice. Moreover, even in theories in which the process of moding out by gauge transformations is reasonably well-understood, it can only be implemented at the expense of locality.\footnote{For a lucid and pedagogical illustration of the interplay between locality and gauge redundancy, see \cite{Riello:2021lfl}.} This should not be too surprising, since in gauge theories, working with a redundant description of physical states is precisely the price one pays to preserve the locality of the dynamcis. Given our focus on dynamical questions in the present work (as well as in follow-ups \cite{CH2}), it is highly desirable for our purpose to preserve locality as much as possible. We will therefore implement post-selection at the level of the presymplectic structure. Of course, our construction will take the gauge equivalence relation induced by the gauge group $\cG$ into account, but it will not directly rely on the existence of $(\cP , \hat\Omega )$.

\subsection{Relation to canonical phase space methods*}\label{sec:covariant_to_standard}

*\emph{This section may be skipped by a quick reader. It is only relevant for Sec.~\ref{sec:relational_obs}, where it will be established that frame-dressed observables formulated with covariant methods (generalizing the construction proposed in e.g.\ \cite{Donnelly:2016auv}) can be re-expressed as relational observables in the standard canonical setting (as defined in e.g.\ \cite{Dittrich:2004cb}).}

\medskip

As in the mechanical example of Sec.~\ref{sec:mechanical}, the field space $\cF$ is not a phase space; the presymplectic structure features degenerate directions which must be factored out. Indeed, the presymplectic current density $\omega=\delta\Theta$ is a $(2,d-1)$ form. Hence, in order to  turn it into a genuine $2$-form on $\cF$, we have to choose a Cauchy slice $\Sigma\cup\bar\Sigma$ in the spacetime $M\cup\bar M$ and integrate to obtain the presymplectic structure $\Omega=\int_{\Sigma\cup\bar\Sigma}\omega$. Outside (but not on) the subspace of solutions $\cS\subset\cF$ this $\Omega$ depends on the choice of Cauchy slice (see Sec.~\ref{sec:presympl_global} and \cite{Lee:1990nz}). In interesting field theories, such as Maxwell and Yang-Mills theory, or general relativity, the degenerate directions of $\Omega$ consist of vector fields $X$ on $\cF$ with vanishing support on $\Sigma\cup\bar\Sigma$. Such vector fields only generate changes in field configurations away from the Cauchy slice; thus, they do not encompass gauge transformations $\fX_\alpha$ with $\alpha$ such that it and its first derivatives feature non-trivial support on the Cauchy slice \cite{Lee:1990nz}. This is because, in contrast to the mechanical model of Sec.~\ref{sec:mechanical}, these theories yield only on-shell constraints (see also Sec.~\ref{sec:examples}). For this reason, factoring out the degenerate directions $X$ of the presymplectic structure, which define an equivalence relation $\sim$ for spacetime field configurations, results in a kinematical phase space $\cP_{\rm kin}=\cF/\!\sim$ with non-degenerate symplectic form $\Omega_{\rm kin}$. This is the standard phase space of the field theory of interest, parametrized by the pullback of field configurations to $\Sigma\cup\bar\Sigma$, as well as their normal derivatives on the Cauchy slice. Furthermore, for this reason, the restriction of the equivalence classes to the subspace of solutions $\cS/\!\sim$, in contrast to the mechanical toy model, does \emph{not} coincide with $\cP_{\rm kin}$, but rather resides as a constraint surface $\cC$ in $\cP_{\rm kin}$, on which the pullback of $\Omega_{\rm kin}$ is degenerate (for details on these statements, we refer the reader to \cite{Lee:1990nz}).

Next, we can ask what sort of functionals descend from the field space $\cF$ to the quotient space $\cP_{\rm kin}=\cF/\!\sim$. It is clear that these must be functionals which are constant along the degeneracy orbits in $\cF$. These are functionals $\Psi$ on $\cF$ which can be written in the pullback  form $\tilde\Psi\circ\pi=\Psi$, where $\pi$ denotes the natural projection from $\cF$ to $\cP_{\rm kin}$ and $\tilde\Psi$ is a functional on $\cP_{\rm kin}$. Suppose we are given a local functional $\Phi$, i.e.\ a functional on $\cF\times\left(M\cup\bar M\right)$ as the relational observables in equation~\eqref{fieldrelobs1}. Given the above observations, we can turn it into a functional on $\cF$ that descends to $\cP_{\rm kin}$ if we pull it back and smear it appropriately across the chosen Cauchy slice $\Sigma\cup\bar\Sigma$ so as to remove non-trivial spacetime dependence away from the Cauchy slice. In particular, we can choose a Dirac delta function as a smearing function in order to obtain a local functional on $\cP_{\rm kin}$, which now depends on $x\in\Sigma\cup\bar\Sigma$. 

Similarly, we may ask what sort of vector fields $\cZ$ project down from $\cF$ to $\cP_{\rm kin}$ as $\pi_*\cZ$, where $\pi_*$ is the natural push forward from the tangent space $T_f\cF$ to $T_{\pi(f)}\cP_{\rm kin}$ defined by the projection $\pi:\cF\to\cP_{\rm kin}$, where $f\in\cF$. As explained in \cite{Lee:1990nz}, these vector fields are characterized by the conditition that $\cL_X\cZ\cdot\Omega=0$ for all degeneracy vector fields $X$; i.e.\ the change of $\cZ$ along the degeneracy orbits is a degenerate direction itself. In this sense, the transversal part of $\cZ$ is then constant on the degeneracy orbit and thereby admits a well-defined projection. In order to yield a non-vanishing vector field on $\cP_{\rm kin}$, $\cZ$ has to feature non-vanishing transversal components to the degeneracy orbits. For gauge theories such as Maxwell or Yang-Mills theory, as discussed later in this work, this is the case for any gauge symmetry directions $\fX_\alpha$ with $\alpha$ such that it features non-trivial support on the chosen Cauchy slice $\Sigma\cup\bar\Sigma$ \cite{Lee:1990nz}.\footnote{In general relativity, one needs to restrict to the space of solutions in order to also obtain a well-defined projection of non-spatial diffeomorphisms \cite{Lee:1990nz}.} We will henceforth denote the so projected gauge directions on $\cP_{\rm kin}$ by $\tilde\fX_\alpha:=\pi_*\fX_\alpha$ and restrict to field-independent $\alpha$ for now.

This permits us to consider the constraint functionals $\tilde C[\alpha]$, associated with the Lie-algebra-valued spacetime functions $\alpha$, defining the constraint surface $\cC:=\cS/\!\sim$ inside $\cP_{\rm kin}$. These are (non-uniquely) defined by 
\beq \label{pkincons}
\tilde\fX_\alpha=\Omega^{-1}_{\rm kin}\delta \tilde C[\alpha]\,,
\eeq
where here $\delta$ denotes the exterior derivative on $\cP_{\rm kin}$ \cite{Lee:1990nz}.\footnote{As emphasized in \cite{Lee:1990nz}, non-degeneracy of $\Omega_{\rm kin}$ does not imply its invertibility, given that $\cP_{\rm kin}$ is (assumed to be) an infinite-dimensional manifold. Invertibility of $\Omega_{\rm kin}$ is thus an additional assumption.}
The $\tilde\fX_\alpha$ are degenerate directions of the pullback of $\Omega_{\rm kin}$ to $\cC$. Note also that the $\tilde C[\alpha]$ will be local constraint functionals that are smeared across the Cauchy slice $\Sigma\cup\bar\Sigma$; they are related to the smeared field space constraints $C[\alpha]$ we shall later encounter in Sec.~\ref{sec:examples} by $C[\alpha]=\tilde C[\alpha]\circ\pi$.

\subsection{Towards a regional covariant phase space: main questions to be addressed}

We conclude this section by listing the main questions to be addressed in the remainder of the paper.
\begin{enumerate}
    \item How can we account for gauge-invariant regional degrees of freedom, i.e.\ degrees of freedom in $M$ or $\bar M$, in a minimally non-local way?

\item After post-selecting global solutions with respect to suitable gauge-invariant boundary conditions on $\Gamma$, what type of presymplectic structure descends from the global to the local field space? 

\item Given a variational principle for $M \cup \bar M$ together with a suitable set of boundary conditions on $\Gamma$, is there a systematic way of constructing a variational principle for subregion $M$?
\end{enumerate}
In Sec.~\ref{sec_gaugesplit}, we will explain how to gauge-invariantly dress boundary degrees of freedom on $\Gamma$ by suitable reference frames drawn from $\bar M$, hence providing an answer to the first problem. This will allow us in Sec.~\ref{sec:geometry} to focus on a post-selected subset of solutions, defined by gauge-invariant conditions at the interface $\Gamma$. In this subspace, we will be able to canonically induce a presymplectic structure for the subregion $M$ from the global one, thus providing a resolution to the second problem. Finally, the third question will be answered in a systematic and algorithmic manner in Sec.~\ref{sec:splitting}. In conjunction, this post-selection procedure can be viewed as deriving the covariant phase space formulation for finite regions with boundaries, e.g.\ as established  in \cite{Harlow:2019yfa,Geiller:2019bti,Chandrasekaran:2020wwn}, from the covariant phase space construction for global spacetimes, e.g.\ as established  in \cite{Lee:1990nz}. We illustrate some of the constructions in Maxwell and Yang-Mills theory along the way, and provide an in-depth exposition of example theories in Sec.~\ref{sec:examples}.

\section{Edge modes as ``internalized'' external frames and gauge-invariant observables}\label{sec_gaugesplit}

Gauge symmetries can be understood as a manifestation of the intrinsically relational nature of certain observables \cite{Rovelli:2013fga,Rovelli:2020mpk}. A valuable concept for dealing with such observables is that of a \emph{dynamical reference frame} \cite{Giacomini:2017zju,Vanrietvelde:2018pgb,Vanrietvelde:2018dit,Hohn:2018iwn,Hohn:2018toe,Hohn:2019cfk,Hoehn:2020epv,Krumm:2020fws,Hoehn:2021flk,Hoehn:2021wet,delaHamette:2020dyi,Castro-Ruiz:2019nnl,Giacomini:2020ahk,Ballesteros:2020lgl}: it is constituted by a collection of dynamical degrees of freedom which can be used to keep track of the evolution of the rest of the system in a relational and gauge-invariant way. We will make use of a particular class of such reference frames to correctly account for the dynamics of gauge fields in a finite subregion $M$ relative to its complement $\bar M$. These frames are the edge modes which, as in the mechanical setting of the previous section, we identify as ``internalized'' external frames for $M$. As such, they are \emph{not} additional degrees of freedom one has to add to the theory; they are already ingredients of the \emph{global} theory for $M\cup\bar M$ one starts with. After  discarding any additional dynamical information from  $\bar M$, the reference frame turns into an independent dynamical field living on the time-like boundary of $M$. This observation explains why, in gauge theories, dynamical edge modes can emerge at finite boundaries. As we will see, another advantage of the reference frame formalism is that it clarifies under which operational assumptions such edge modes are to be considered as physically relevant: namely, when we have the operational means to measure gauge-variant  observables in $M$ relative to a reference frame in $\bar M$, which jointly, give rise to a gauge-invariant \emph{relational} or \emph{dressed} observable with support on both $M$ and $\bar M$. These relational observables will in later sections become crucial for formulating the boundary conditions inducing foliations of the global space of solutions for $M\cup\bar M$ into distinct subregion theories for $M$ (compare also with the mechanical case in Secs.~\ref{ssec_mechpost} and~\ref{ssec_3types}).

\subsection{Dynamical reference frames in gauge field theory}\label{sec_fieldrfs}

Let us first specify what we mean by a dynamical reference frame for some local (gauge) symmetry Lie group $G$ acting  on some system $S$. Loosely speaking, it should be a dynamical subsystem $R$ internal to the total system $S$ that is as asymmetric under $G$-transformations as possible. Its purpose in life is to constitute a dynamical coordinate system for parametrizing $G$-orbits in the configuration space of the total system $S$. As such, $R$ constitutes a dynamical reference system relative to which we can describe the remaining degrees of freedom of $S$ as they transform under $G$ in such a way that the relations translate into gauge-invariant observables of the theory. Subsystems that are already invariant therefore constitute the worst possible choices of frames. In a nutshell, this is the philosophy underlying much of the recent efforts on quantum reference frames \cite{Giacomini:2017zju,Vanrietvelde:2018dit,Vanrietvelde:2018pgb,Hohn:2018iwn,Hohn:2018toe,Hoehn:2021wet,Hohn:2019cfk,Hoehn:2020epv,Krumm:2020fws,Hoehn:2021flk,delaHamette:2020dyi,Castro-Ruiz:2019nnl,Giacomini:2020ahk,Ballesteros:2020lgl}. In gauge field theory, we will take $G$ to be the local structure group rather than the (global) gauge group, the latter coinciding in this context with the group of gauge transformations (which can be described locally as $G$-valued functions). Our construction will therefore be spacetime local by nature. In particular, the system $S$ and reference frame $R$ will consist of local degrees of freedom at (or in a neighborhhod of) a spacetime point $x$.

More precisely, in order for $R$ to parametrize $G$-orbits, it must (locally in spacetime) feature a set of at least $\dim G$ configuration degrees of freedom that transform freely under $G$. In other words, $R$ must be such that $G$ acts freely on its configuration space $\cR$ (i.e.\ the isotropy group for each configuration is trivial). If $G$ acts furthermore regularly on $\cR$ (i.e.\ it also acts transitively, connecting any two configurations), then $\cR$ is a principal homogeneous space for $G$\,---\,so $\dim\cR=\dim G$\,---\,and $R$ features as many configurations as there are group elements. In the language of \cite{Bartlett:2007zz}, such a frame $R$ is called a \emph{complete reference frame}. If, by contrast, $G$ acts freely, but not regularly on $\cR$, the latter contains more configurations than there are group elements, in which case we may call $R$ an \emph{overcomplete reference frame}. It is also possible to consider \emph{partial reference frames} characterized by the action of $G$ on $\cR$ not being free, e.g.\ see \cite{periodic}, but we shall not explore them here much further.  Henceforth, we shall refer to the configurations of $R$ as its \emph{frame orientations}. For example, in Sec.~\ref{sec:mechanical}, we considered particles as complete reference frames for the translation group, whose positions were their frame orientations.

\medskip

Let us now consider the general case of a gauge theory with local symmetry governed by a connected Lie group $G$ (which may be non-Abelian), and for definiteness we assume it to be a compact matrix group (which could be weakened). The group of gauge transformations $\cG$ is a space of $G$-valued functions on spacetime, $x \mapsto g(x)$, with suitable fall-off conditions at asymptotic infinity. The group law is the pointwise multiplication in $G$: given two gauge transformations $g_1$ and $g_2$, $g_1 g_2(x)= g_1(x)g_2(x)$. The connection is a $\mathfrak{g}$-valued one-form $A$, transforming under the group action:
\beq
g \acts A = g A g^{-1} - \extd g g^{-1}\,.
\eeq
Indeed, we can check that $(g_1 g_2)\acts A = g_1 \acts g_2 \acts A$. The generators of such gauge transformations are vector fields $\fX_\alpha$ on field configuration space $\cF$, labeled by $\mathfrak{g}$-valued spacetime functions $\alpha$:\footnote{This is true at least locally, and relative to a choice of trivialization. Gauge transformations are more generally understood as $G$-valued sections, but this will not play any significant role in the present paper.}  
\beq
\fX_\alpha (A) = \left[ \alpha , A \right] - \extd \alpha\,,
\eeq
where $[\cdot,\cdot]$ denotes the Lie bracket. 
In particular, $e^{t \fX_\alpha (\cdot)} = e^{t \alpha}\acts (\cdot)$ for any $t\in \mathbb{R}$. 

Let us suppose that the gauge theory contains other fields, which are dynamically independent from $A$, and that we collectively denote by $\Psi$. We can now consider the three types of reference frames $R$ alluded to above for the field theory (the system $S$) parametrized by $A,\Psi$. The following applies both to the bulk and boundary of spacetime regions, so for the moment we do not restrict to edge modes. The simplest possible type of frame $R$ is a  \emph{$G$-valued reference frame}: its configurations are described by a functional $U[\Psi,A]$ with values in the space of gauge transformations, which transforms by left-multiplication under $\acts$. Namely, at least locally in field space $\cF$, $U[\Psi ,A]$ assigns a group element $U[\Psi ,A ] (x) \in G$ to every point $x$ in spacetime and:
\beq\label{eq:group_frame}
g\acts U[\Psi ,A] = g U[\Psi ,A] \,. 
\eeq
The frame orientation space $\cR$ is therefore parametrized by these $G$-valued functions on spacetime.
When such a construction is possible, $U[\Psi ,A]$ provides a field-space-local and complete reference frame for the local structure group $G$ at every point in spacetime, since the action of $G$ on itself by left multiplication is regular. Owing to the Gribov obstruction, this will in general not be possible globally in field space $\cF$; for if it was possible, one could use $U[\Psi,A]$ to define global gauge fixing conditions for the $G$-orbits in $\cF$.

A group-valued reference frame is characterized by a particular choice of free group action of the local structure group $G$ on the configuration space $\cR$: namely, the action of $G$ on itself by left multiplication, which enters into the definition \eqref{eq:group_frame} (so that, in particular, $\cR=G$ and the action is also regular). It is straightforward to generalize this construction to other types of reference frames, with $\cR \neq G$; the free character of the $G$-action on $\cR$ is the only necessary condition we need to preserve in order to obtain a reference frame whose configurations can be used to parametrize $G$-orbits in field space. Indeed, suppose we are given a free group action of the local symmetry group $G$ on $\cR$, that is, a (smooth) map $D: G \times \cR \rightarrow \cR\,,\; (g,R)\to g\cdot R$ such that: for any $(g,R)\in G\times \cR$, $g\cdot R = R$ implies that $g=e$ (where $e$ is the identity element in $G$). A local reference frame for the structure group $G$ can then be obtained by constructing a $\cR$-valued functional $R[\Psi,A]$ transforming as:
\beq\label{eq:faithful_gauge}
(g\acts R[\Psi ,A])(x) = g(x) \cdot R[\Psi ,A](x)\,.
\eeq
If the group action $D$ is regular on $\cR$, then the latter defines a complete reference frame for $G$ at every point $x$ in spacetime; if it is only free (hence not transitive), the resulting local reference frame is overcomplete.  Again, on account of the Gribov obstruction, the frame will generally have these properties only locally in field space $\cF$.\footnote{A mechanical toy model for dynamical reference frames in the presence of the Gribov problem is the $N$-body problem in 3D subject to rotation and translation symmetry, where globally valid gauge-fixing conditions likewise do not exist on the constraint surface \cite{Vanrietvelde:2018dit}. This leads to an absence of globally valid internal frame perspectives, in some analogy to the absence of globally valid coordinate systems on general spacetimes. }

Finally, we could imagine relying on group actions that fail to be free to construct partial reference frames. Such structures may be of interest to describe situations in which the observer has incomplete knowledge of her frame of reference. For simplicity, we will ignore this possibility here, and instead adopt the view that, at least in principle, it is always possible to construct (field-space-local) complete reference frames for the gauge degrees of freedom of interest. Concretely, this may require the (explicit or implicit) introduction of a sufficiently rich set of ancilla systems into our description.\footnote{This is best illustrated with an example based on the natural action of $G=\rm{SO}(3)$ on the unit two-sphere $S^2$ of $\mathbb{R}^3$. This group action is not free since a point on the sphere is left invariant by a ${\rm U}(1)$ subgroup of $G$. For the same reason, the action of $\rm{SO}(3)$ on three copies of $S^2$ fails to be free. However, by restricting the latter to the orbit of a normal basis $(e_1, e_2, e_3)$, we do obtain a free group action of $G$. In other words, while a single unit vector does not constitute a complete local reference frame in $\mathbb{R}^3$, a basis does (for a related discussion in quantum theory, see \cite{delaHamette:2020dyi}). Likewise, in gauge field theory, a matter-dressed Wilson line $R[A](x):=\cP\exp\left(-\int_{\gamma_x} A\right)\psi(\gamma(0))$, where $\gamma:[0,1]\to M\cup\bar M$ is a path in spacetime and $\psi$ denotes a charged spinor field, transforms as $(g\acts R[A])(x) =g(x) \cdot R[A](x)$ under gauge transformations, where $\cdot$ is a group action on spinors. This action is in general not free, so that $R$ will only constitute a partial frame for $G$ at $x$. Nonetheless, it is always possible to construct a complete reference frame from a suitable collection of such matter-dressed Wilson lines, in a conceptually analogous manner to the previous example. Anticipating Sec.~\ref{ssec_edgefields}, this shows that edge reference frames (or edge modes) which are not group-valued are in principle relevant in gauge field theory, even though, for simplicity, we will focus exclusively on group-valued ones from Sec.~\ref{sec:concrete} onwards.} For instance, in pure Maxwell theory, we could imagine adding test charges to make operational sense of holonomies along open paths, which can in turn be used to construct group-valued reference frames at their open ends. In most of the remainder of this manuscript, we shall work with group-valued frames.

\subsection{Frame-dressed observables are relational observables}\label{sec:relational_obs}

Suppose now that we have access to a group-valued\,---\,and hence complete\,---\,reference frame $U[\Psi]$, which for simplicity does not depend on the connection $A$. This permits us to decompose any local functional $\Phi[A]$ of the connection canonically into:
\beq
\Phi = \Phi_{\rm inv} + \Phi_{\rm gauge}\,,
\eeq 
where $\Phi_{\rm inv}$ is the gauge-invariant \emph{frame-dressed} field
\beq
\Phi_{\rm inv} := U[\Psi]^{-1} \acts \Phi\,,
\eeq
and 
\beq
\Phi_{\rm gauge} := \Phi - U[\Psi]^{-1} \acts \Phi 
\eeq
carries the gauge dependence of $\Phi$. Such a frame associated decomposition is similar to the splitting into vertical and horizontal components relative to a field-space connection, as advocated by Gomes and Riello \cite{Gomes:2016mwl, Gomes:2018shn}. While the latter splitting was done for a spatial slice, here we take a covariant picture, applicable to spacetime regions. As an illustration, we find for the connection and field strength:
\begin{align}
    A_{\rm inv} &= U[\Psi]^{-1} A U[\Psi] - \extd U[\Psi]^{-1} U[\Psi] \,,\qquad\qquad
    A_{\rm gauge} = - U[\Psi]^{-1} \extd U[\Psi]  - \sum_{n\geq1} \frac{1}{n!} [A, u]_n\,, \label{genArad}\\
    F_{\rm inv} &= U[\Psi]^{-1} F U[\Psi]\,,\qquad\qquad\quad\,\qquad\qquad\qquad F_{\rm gauge}=-\sum_{n\geq1} \frac{1}{n!} [F, u]_n\,,\label{genFrad}
\end{align}
where we have defined the Lie algebra generator $u$ via $U[\Psi](x):= e^{u(x)}$, used the general formula $\mathrm{Ad}_{e^X} = e^{\mathrm{ad}_X}$, and introduced the short-hand $[A,u]_n:= [[[A,\ldots],u],u]$. Owing to the Gribov obstruction, this gauge-invariant dressing only needs to hold locally in field space $\cF$. When $G= \mathrm{U}(1)$, a special case we will restrict to in some examples, $A_{\rm inv} = A - \extd \vphi$ and $A_{\rm gauge} = \extd \vphi$, where $U[\Psi](x)= e^{i \vphi(x)}$ (and we have absorbed a factor of $i$ in $A$ so that $A_\mu(x) \in \mathbb{R}$). 

While the choice of dynamical reference frame $U$ is not a choice of gauge, standard gauge-fixing conditions can be recovered from this formalism and translate into specific restrictions on the orientation of $U$. For instance, in a $\mathrm{U}(1)$ gauge theory where $U(x)=e^{i \varphi(x)}$, we can view the Lorenz gauge-fixing condition $\nabla_\mu A^\mu = 0$ as an equation for $\varphi$: $\nabla_\mu \nabla^\mu \varphi = - \nabla_\mu A^\mu_{\rm rad}
$, with the gauge-independent contribution on the right-hand side playing the role of source.

This construction extends to the case of a frame $R[\Psi]$ not valued in $G$ that, however, transforms freely under a $G$-action $D$ (as specified in equation \eqref{eq:faithful_gauge}). The only extra ingredient we need in order to make sense of frame-dressings in this context is an equivariant isomorphism between $G$ and $\cR$; that is, a (smooth) invertible map $E: G \to \cR$ such that:
\beq\label{eq:equivariance}
\forall g, g' \in G \,, \qquad E(g g') = g \cdot E(g')\,.
\eeq
Since $E$ is one-to-one, we can use its inverse $E^{-1}: \cR \to G$ to introduce the decomposition:
\beq
\Phi_{\rm inv}(x):= [E^{-1} ( R[\Psi](x))]^{-1} \acts  \Phi(x)\,, \quad \Phi_{\rm gauge}(x) = \Phi(x) - [E^{-1} ( R[\Psi](x))]^{-1} \acts  \Phi(x)\,.
\eeq
Owing to equation~\eqref{eq:faithful_gauge}, together with the fact that $E^{-1}$ is itself equivariant,\footnote{That is, for any $(g,R)\in G\times \cR$:
\beq
g E^{-1}(R) = E^{-1}(g \cdot R)\,,
\eeq
which is a direct consequence of equation \eqref{eq:equivariance}, evaluated for $g'=E^{-1}(R)$.} the frame-dressed field $\Phi_{\rm inv}$ is again gauge-invariant. What we have left to do is to justify the existence of an equivariant and invertible map $E$. But it suffices to define $E(g):= g\cdot R_0$, where $R_0$ is some arbitrary (and fixed) configuration in $\cR$. As can be checked explicitly, this is an equivariant map, and given that $G$ acts freely on $\cR$, it is also injective. To make sure it is surjective, all we need to do is to restrict the configuration space to its image $E(G)$, namely, define $\cR := G\cdot R_0$.

In fact, recalling the mechanical example from Secs.~\ref{ssec_intpart} and~\ref{ssec_extHeis}, we can slightly generalize the above construction of the gauge-invariant frame-dressing. Restricting to group-valued frames for simplicity, we can introduce a dependence on the frame orientation $g\in G$ into the frame-dressed observable via \emph{right multiplication} on the frame (cf.\ equation~\eqref{relobs0}): 
\beq \label{fieldrelobs1}
 O_{\Phi,U}(g;x):=\left(U[\Psi](x)\,g^{-1}(x)\right)^{-1}\acts\Phi[A](x)
 \eeq
for any local functional $\Phi[A]$ of the connection. We have here written the dependence on the spacetime point $x$ to emphasize that these observables are functionals on $\cF\times\left(M\cup\bar M\right)$. This observable is clearly gauge-invariant, $g'\acts O_{\Phi,U}(g)=\left(g'Ug^{-1}\right)^{-1}\acts\left(g'\acts\Phi\right)=\left(Ug^{-1}\right)^{-1}g'^{-1}\acts\left(g'\acts\Phi\right)=O_{\Phi,U}(g)$. It is a \emph{relational observable}, describing in a gauge-invariant manner how the functional $\Phi$ transforms locally under $G$ relative to the frame $U$; it encodes the question ``what is the value of $\Phi$ when the frame $U$ is in orientation $g\in G$?''. Indeed, this can be seen as follows: since $O_{\Phi,U}(g)$ is gauge-invariant, we can evaluate it on a gauge-fixing surface in field space without it changing value. In particular, we can evaluate it on the local gauge-fixing condition $U[\Psi]=g$, in which case the observable simply reduces to $\Phi$. Relational observables are therefore so-called gauge-invariant extensions of gauge-fixed quantities \cite{Henneaux:1992ig,Dittrich:2004cb}. As a special case, we note for later use that ``$\Phi$ when the frame is in the origin $e$'' coincides with the invariant observables above
 \beq 
 O_{\Phi,U}(e)\equiv \Phi_{\rm inv}\,.
 \eeq 

In the remainder of this subsection, we shall demonstrate that the frame-dressed observables in equation~\eqref{fieldrelobs1} indeed constitute the covariant phase space version of the relational observables originally constructed in \cite{Dittrich:2005kc}, using the standard Legendre transformation based approach to phase spaces (and generalizing earlier proposals \cite{Rovelli:1989jn,Rovelli:1990jm,Rovelli:1990ph,Rovelli:1990pi}). To this end, the reader is invited to remember how the standard phase space formulation in field theory is obtained from the covariant one, which is discussed in Sec.~\ref{sec:covariant_to_standard}. One elementary aspect to keep in mind is that, in their canonical form, and in contrast to their covariant incarnation above, the relational observables depend on a point on a Cauchy slice, rather than a spacetime point. To map from the covariant to the canonical picture, we therefore need the canonical projection map $\pi:\cF \to \cP_{\rm kin}$, where $\cP_{\rm kin}$ denotes the standard kinematical phase space, the construction of which is outlined in Sec.~\ref{sec:covariant_to_standard}.

With this key ingredient at hand, we are in a good position to discuss the ``projection'' of the frame-dressed observables in equation~\eqref{fieldrelobs1} from $\cF$ to $\cP_{\rm kin}$. Recalling, $e^{\fX_\alpha(\cdot)}=e^{\alpha}\acts(\cdot)$ and setting $Ug^{-1}=e^{u}e^{-\alpha}=:e^{u'}$, for $u'\in\mathfrak{g}$, we can also write them as
\beq \label{fielddepgauge}
O_{\Phi,U}(g;x)=\restr{e^{\fX_\alpha}\Phi[A](x)}{\alpha(x)=-u'(x)}\,.
\eeq
Provided $x\in\Sigma\cup\bar\Sigma$ and the reference frame functional $U[\Psi](x)$, which may depend non-locally on dynamical fields, only has support in this chosen Cauchy slice, it will ``project'' to a group valued local reference frame functional on $\cP_{\rm kin}$ via $U=\tilde U\circ\pi$. We shall henceforth assume that this is the case (which may require adjusting the Cauchy slice accordingly).\footnote{For example, in the below subregion problems, this will be the case for holonomy constructed frames whose holonomies do not have support outside $\Sigma\cup \bar \Sigma$.} In this case, also the field-dependent Lie-algebra-valued spacetime functions $u'$ descend to the kinematical phase space as $u'=\tilde u'\circ\pi$ since pointwise addition and multiplication of ``projectable'' functions is preserved by the projection $\pi$.  Altogether, if $\Phi[A](x)$ is a local field functional and $x\in\Sigma\cup\bar\Sigma$, $O_{\Phi,U}(g;x)$ constitutes a functional which only depends on the degeneracy orbits in $\cF$ and therefore descends to a functional on $\cP_{\rm kin}$, i.e.
\beq
O_{\Phi,U}(g;x) = \tilde O_{\tilde\Phi,\tilde U}(g;x)\circ\pi\,.
\eeq
Specifically, the assumption on $u'$ restricts us to $\alpha$ with non-trivial support in the Cauchy slice. As noted above, the corresponding $\fX_\alpha$ contain transversal directions to the degeneracy orbits in $\cF$. Since $\Phi[A](x)$ is constant along the degeneracy orbits for $x\in\Sigma\cup\bar\Sigma$, we have that only the transversal directions contribute to the derivative $\fX_\alpha(\Phi)$. This implies $\fX_\alpha(\Phi)=\tilde\fX_\alpha(\tilde\Phi)\circ\pi$ (transversal derivatives are preserved by the projection $\pi$) and thus ultimately
\beq
\restr{e^{\tilde\fX_\alpha}\tilde\Phi}{\alpha=-\tilde u'}\circ\pi=\restr{e^{\fX_\alpha}\Phi}{\alpha=-u'}\,.
\eeq
Recalling equation~\eqref{pkincons}, this yields for $x\in\Sigma\cup\bar\Sigma$
\beq \label{Bianca}
\tilde O_{\tilde\Phi,\tilde U}(g;x)=\restr{e^{\tilde\fX_\alpha}\tilde\Phi}{\alpha=-\tilde u'}=\restr{e^{\{\tilde C_\alpha,\tilde\Phi\}}}{\alpha=-\tilde u'}=\restr{ \sum_{n=0}^\infty\frac{1}{n!}\{\tilde C_\alpha,\tilde\Phi\}_n}{\alpha=-\tilde u'}\,,
\eeq
where $\{f,g\}_n=\{f,\{f,g\}_{n-1}\}$ denotes the $n^{\rm th}$-nested Poisson bracket with convention $\{f,g\}_0=g$. We emphasize that the Poisson bracket is here the one defined in terms of the kinematical symplectic form $\Omega_{\rm kin}$ on $\cP_{\rm kin}$. This is precisely the power-series representation of relational observables in field theory originally constructed in \cite[Eqs.\ (4--6)]{Dittrich:2005kc} using the notion of a kinematical phase space. 

Equation~\eqref{fieldrelobs1} therefore constitutes the covariant phase space formulation of relational observables and we conclude that all frame-dressed observables are in fact relational observables. We also note that it is more convenient to construct relational observables covariantly as done here; this  does not necessitate to choose a Cauchy surface and neither to choose a (possibly non-locally defined) reference frame geared to this Cauchy slice. 

So far we have considered relational observables $O_{\Phi,U}$ on the full field space, but ultimately one is interested in their restriction to the space of solutions $\cS$, which in the kinematical phase space formulation means restricting equation~\eqref{Bianca} to the constraint surface $\cC$. Note that, while the construction of $\cP_{\rm kin}$ depends on the choice of Cauchy slice, the construction of $\cC$ in fact is independent of it since $\Omega$ does not depend on it on solutions. This permits one to also dynamically reconstruct relational observables on spacetime points $x$ not lying on the chosen Cauchy slice by solving the equations of motion. By contrast, off-shell, one would have to construct a new kinematical phase space.

\subsection{Edge modes as dynamical reference frame fields}\label{ssec_edgefields}

\begin{figure}[htb]
    \centering
    \includegraphics[scale= .7]{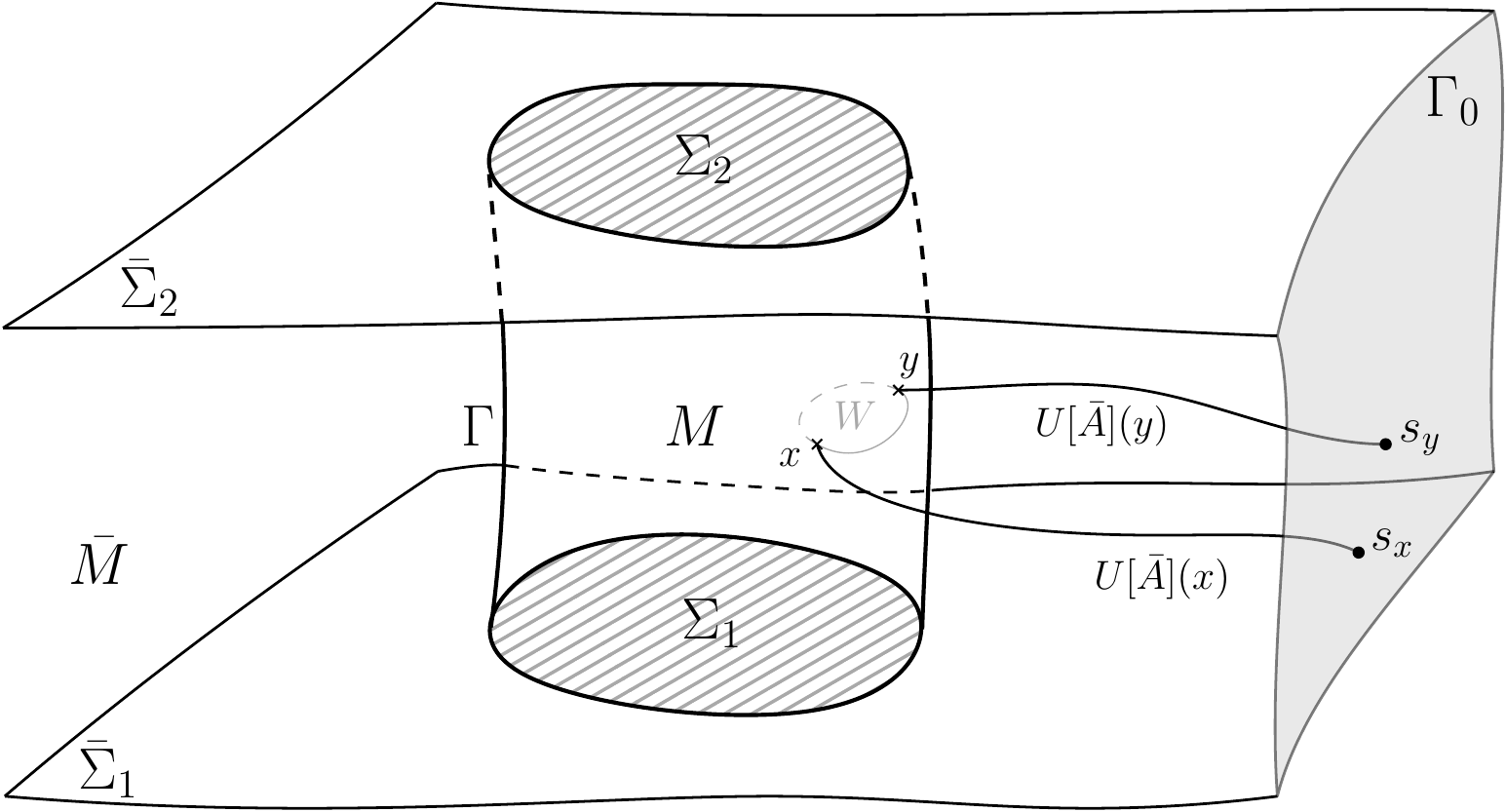}
    \caption{The spacetime region $M$ and its complement. From a system of Wilson lines $\{ \gamma_x \}$ anchored at asymptotic source $\{ s_x \}$, we construct the dynamical edge field $U[\bar A]$, which provides a reference frame for the time-like boundary $\Gamma$. It allows us to decompose gauge-invariant observables with support on both $M$ and $\bar M$, such as the Wilson loop $W$, into composites of regional gauge-invariant observables.}
    \label{fig:regions}
\end{figure}

We now apply this general formalism to a subregion $M$, delimited by initial and final partial Cauchy surfaces $\Sigma_1$ and $\Sigma_2$, and a time-like boundary $\Gamma$. In particular, we shall henceforth only consider dynamical frames and relational observables on $\Gamma$: that is, we should content ourselves with a realization of formula \eqref{eq:faithful_gauge} in which $x$ spans the time-like boundary $\Gamma$ rather that the whole spacetime. For simplicity, we assume that the hypersurface $\Gamma$ has the topology of $S_{d-2}$ times an interval. We can arbitrarily extend $\Sigma_1$ and $\Sigma_2$ into complete Cauchy surfaces $\Sigma_1 \cup \bar\Sigma_1$ and $\Sigma_2 \cup \bar\Sigma_2$. We then denote by $\bar M$ the complement of $M$ in the spacetime slab delimited by $\Sigma_1 \cup \bar\Sigma_1$ and $\Sigma_2 \cup \bar\Sigma_2$. See Fig.~\ref{fig:regions}.

\begin{figure}[htb]
    \centering
    \includegraphics[scale= .7]{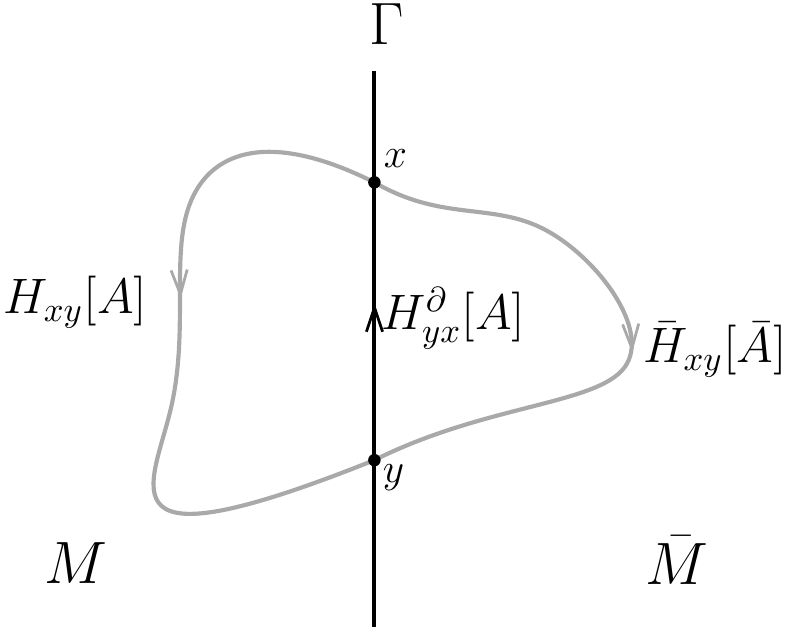}
    \caption{Wilson loop $W=\Tr\left(\bar H_{xy}[\bar A]^{-1}\,H_{xy}[A]\right)$, with support on both $M$ and $\bar M$.}
    \label{fig:split_holonomies}
\end{figure}

We collectively denote local field configurations in $M$ (resp.\ $\bar M$) by $\Phi$ (resp.\ $\bar \Phi$). In the presence of gauge symmetries, there might exist gauge-invariant observables with support in $M \cup \bar M$, which cannot be reconstructed from the knowledge of gauge-invariant observables which are solely supported in $M$ or $\bar M$. Let us consider two examples, we shall revisit below:\footnote{In the mechanical toy model of sec.~\ref{ssec_mechpost}, examples would be intergroup relative distances, such as, say, $Q_{j|\bar{j}}$ between particle $j$ of group $M$ and particle $\bar{j}$ of  group $\bar M$.}  
\begin{itemize}
    \item[(a)] Consider a Wilson loop $W$ with support in both $M$ and $\bar M$, that intersects $\Gamma$ at two points $x$ and $y$, see Fig.~\ref{fig:regions}. Calling $\bar H_{xy}[\bar A]:=\cP\exp\left(-\int_{\gamma_{xy}}\bar A\right)$ (resp.\ $H_{xy}[A]$) the holonomy from $x$ to $y$ along the portion of the loop that lies inside $\bar M$ (resp.\ $M$),\footnote{Note that in the special case $G=\rm{U}(1)$ our convention to have real angles entails that $\bar H_{xy}[\bar A]=\cP\exp\left(-i\int_{\gamma_{xy}}\bar A\right)$.} we have
\beq \label{Wilsonloop}
W:=\Tr\left(\bar H_{xy}[\bar A]^{-1} H_{xy}[A]\right)
\eeq 
and neither $H_{xy}[A]$ nor $\bar H_{xy}[\bar A]$ are gauge-invariant. If $G$ is Abelian this can be decomposed into gauge-invariant contributions from $M$ and $\bar M$, but not otherwise. Indeed, consider the closed Wilson line $H_{x x}[A]=H_{xy}[A]\,H^\p_{{yx}}[A]=\cP\exp\left(-\oint_\gamma A\right)$, where $\gamma$ is the closed path consisting of the previous path from $x$ to $y$ through $M$ and some path from $y$ to $x$ that passes solely through the interface $\Gamma$. Similarly, we can define $\bar H_{xx}[\bar A]=\bar H_{xy}[\bar A]\,H^\p_{yx}[A]$, so that the Wilson loop can be written as
\beq 
W=\Tr\left(\bar H_{xx}[\bar A]^{-1}\,H_{xx}[A]\right)\,,
\eeq 
see Fig.~\ref{fig:split_holonomies}. When $G$ is Abelian, both $H_{xx}[A]$ and $\bar H_{xx}[\bar A]$ are gauge-invariant. In the case that $G=\rm{U}(1)$, the trace is trivial, so that the Wilson loop $W$ becomes the product of the purely regional Wilson loops $H_{xx}[A]$ and $\bar H_{xx}[\bar A]$ \cite{Riello:2021lfl}. In the non-Abelian case, neither $H_{xx}[A]$, nor $\bar H_{xx}[\bar A]$ are gauge-invariant. The following example shows that the decomposition of $W$ into regional gauge-invariant quantities in the Abelian case is  a degenerate feature that, regardless of the group, does not hold anymore as soon as dynamical coupling to matter is included into the picture.

\item[(b)] Consider a Wilson line passing through the point $y$ in the interface $\Gamma$ 
\beq \label{chargedWilson}
\psi^\dag(x)\,H_{xy}[A]\,\bar H_{y\bar{x}}[\bar A]\,\bar\psi(\bar x)\,
\eeq 
and connecting a charged antiparticle $\psi^\dag(x)$ in $M$ to a charged particle $\bar\psi(\bar x)$ in $\bar M$.\footnote{To avoid confusion with our convention to equip fields in $\bar M$ with a $\bar{}\,$, we denote the antiparticle field with a $\dag$.} Regardless of the group, neither $\psi^\dagger(x) H_{xy}$ nor $\bar H_{y\bar{x}} \bar\psi(\bar x)$ are gauge-invariant.
\end{itemize}

More abstractly, in the quantum theory, the failure to decompose gauge-invariant observables with support in both $M\cup\bar M$ into gauge-invariant observables from $M$ and $\bar M$ is rooted in the appearance of a constraint-induced non-trivial common center in the algebras of gauge-invariant observables of $M$ and $\bar M$, e.g.\ see \cite{Casini:2013rba}; given any Cauchy surface, this center does not commute with arbitrary gauge-invariant cross-boundary observables.\footnote{For instance, the algebra generated by $\cA_\Sigma \cup \cA_{\bar\Sigma}$, where $\cA_\Sigma$ (resp.~$\cA_{\bar\Sigma}$) is the algebra of gauge-invariant observables supported on the partial Cauchy slice $\Sigma$ (resp.~$\bar \Sigma$) is obtained by taking the bicommutant $(\cA_\Sigma \cup \cA_{\bar\Sigma})''$. Since both the commutant and bicommutant contain the center, the result cannot include arbitrary gauge-invariant cross-boundary observables; that is, $(\cA_\Sigma \cup \cA_{\bar\Sigma})'' \subsetneq \cA_{\Sigma \cup \bar \Sigma}$. } 

In such a situation, two natural options can be considered when projecting the global dynamics down to the subregion $M$:
\begin{itemize}
\item if one is only interested in gauge-invariant observables which are fully supported in $M$, it is sufficient to construct a variational principle for the regional configuration variables $\Phi$, and completely forget about the configuration variables $\bar\Phi$;\footnote{Sec.~\ref{ssec_intpart} considers the analogous situation in mechanics.}
\item by contrast, in order to be able to reconstruct all gauge-invariant observables with support in $M \cup \bar M$ (including those which cannot be reconstructed from regional gauge-invariant observables), it is necessary to keep track of a sufficient amount of relational information between $M$ and $\bar M$.  
\end{itemize}
In the present work, we are interested in the second option because it is more general, but we emphasize that there is nothing wrong with following the first route: it is more restrictive, but completely appropriate in a physical situation where one has no operational access to the region $\bar M$.

The main idea is to construct a reference frame $R[\bar\Phi, \Phi]$ on the time-like boundary $\Gamma$, with a non-trivial dependence on degrees of freedom $\bar \Phi$ (in a sense we will make more precise below), and possibly on $\Phi$ as well. Namely, we assume we can construct a functional:
\beq\label{eq:faithful_frame}
R[\bar \Phi , \Phi]: \Gamma \to \cR\,, \qquad \forall x \in \Gamma, \; (g\acts R[\bar\Phi , \Phi])(x) = g(x) \cdot R[\bar\Phi , \Phi](x)\,,
\eeq 
that is kinematically independent (i.e.\ can be varied independently) from the pull-backs $\phi := \restr{\Phi}{\Gamma}$, as we will elucidate in more detail in the following subsection.\footnote{We emphasize that $\restr{\Phi}{\Gamma}$ is our notation for the \emph{pull-back} of $\Phi$ to $\Gamma$. To avoid any potential confusion, we will never invoke the distinct notion of restriction of a field (for instance, the restriction of a normal vector field to $\Gamma$).} This functional $R[\bar\Phi , \Phi]$ constitutes an edge mode field on $\Gamma$, which we can use as a reference frame to decompose any $\phi$ on $\Gamma$ as:
\beq
\phi = \phi_{\rm rad} + \phi_{\rm gauge}
\eeq
where
\beq\label{eq:phi_rad}
\phi_{\rm rad}(x) := [E^{-1} (R[\bar\Phi , \Phi](x))]^{-1} \acts  \phi(x)\,,  \quad \phi_{\rm gauge}(x) := \phi(x) - [E^{-1} (R[\bar\Phi , \Phi](x))]^{-1} \acts  \phi(x)\,,
\eeq
and $E$ is an equivariant map as defined in \eqref{eq:equivariance}.
For a group-valued edge mode frame field, which for distinction we will continue to denote by the letter $U$, we also consider more generally the edge relational observables (cf.\ equation~\eqref{fieldrelobs1})
\beq \label{edgerelobs}
O_{\phi,U}(g):=\left(U[\bar\Phi,\Phi]\,g^{-1}\right)^{-1}\acts\phi\,.
\eeq
Because this decomposition operates solely on the time-like boundary, we call the gauge-invariant dressed field $\phi_{\rm rad}$ the \emph{radiative component} of $\phi$ relative to ${R}[\bar\Phi , \Phi]$.\footnote{We are here adopting the suggestive nomenclature introduced in the related work \cite{Gomes:2019xto, Riello:2020zbk, Riello:2021lfl}, even though our set-up is not identical.} We see that the joint data $(\phi , {R}[\bar\Phi , \Phi])$ allows us to define the gauge-invariant observable $\phi_{\rm rad}$, which cannot in general be reconstructed from $\phi$ alone (unless gauge symmetries act trivially on $\phi$). It is therefore crucial to keep track of $R[\bar\Phi , \Phi]$ when restricting to the subregion $M$. A natural way of doing so is to extend the configuration space for $M$ spanned by $\Phi$ by an $\cR$-valued field on $\Gamma$, that we will simply denote by $R$. We will therefore construct a regional variational principle for the pair of bulk and edge fields $(\Phi, R)$. Such an extension (without variational principle) was originally introduced in \cite{Donnelly:2016auv}, but we emphasize that from the point of view of the  post-selection procedure discussed in the next sections (see also Sec.~\ref{ssec_mechpost} for the mechanical analog), $R$ can be directly interpreted as a projection of $R[\bar\Phi , \Phi]$ from the global variational problem for $M\cup\bar M$ to the regional one for $M$. This clarifies why $R$ must be dynamical, since $R[\bar\Phi , \Phi]$ is kinematically independent  from $\Phi$ in the global field space associated with $M\cup\bar M$. Likewise, this point of view naturally explains why, when gluing back the two regions $M$ and $\bar M$, $R$ will serve as a placeholder for $R [\bar\Phi , \Phi]$ \cite{CH2}.

Hence, $R[\bar\Phi,\Phi]$ does not constitute a set of degrees of freedom that has to be added to the theory: it is part of the global theory to start with, however, only becomes relevant when we restrict our attention to the subregion $M$ under the premise that we would like to know how it relates to its complement $\bar M$. We can therefore view the edge mode field $R$, in the sense of Sec.~\ref{ssec_extHeis}, as an ``internalized'' external reference frame for $M$. In particular, the relational observables \eqref{edgerelobs} encode how $M$ relates to its complement $\bar M$. As we will illustrate in the next subsection, once we explicitly include the edge frame into the picture, we \emph{can} recover the cross-boundary observables in examples (a) and (b) above from regional relational observables.

\subsection{Concrete realization of group-valued edge reference frames}\label{sec:concrete}

As an illustration of the general framework outlined in the previous subsection, let us discuss how group-valued edge reference frames can be constructed from a gauge connection $A$.

A natural idea is to construct a system of paths $\{ \gamma_x \vert x \in \Gamma \}$ such that the path $\gamma_x$ ends at $x$ and originates from the interior of the complementary region $\bar M$.\footnote{$\gamma_x$ may or may not intersect $M$, but it must have some support in $\bar M$.} One can then define $U[\bar A , A](x)$ as the holonomy of $(\bar A , A)$ along that path. In order for $U[\bar A, A ](x)$ to transform as in \eqref{eq:group_frame}, the Wilson line must originate from a hypersurface where gauge transformations vanish.  We can for instance imagine that $\bar M$ has an asymptotic boundary where gauge transformations are constrained to fall-off sufficiently rapidly, so that we only consider bulk-supported gauge transformations in the global field space.\footnote{In doing so, we are ignoring potentially interesting effects associated to asymptotic symmetries and soft modes. Understanding the interplay between our general construction and asymptotic limits is an important objective that we leave for future work.} Such a boundary provides a non-dynamical anchor for the Wilson lines, which can be understood as a proxy for background matter degrees of freedom which are not explicitly included in the description (such as heavy degrees of freedom constituting measurement apparata). We can for example assume that the path $\gamma_x$ originates from a reference point $s_x$ lying on an asymptotic boundary, and that the latter is equipped with a background reference frame $U_0(s_x)$ that no longer transforms under $G$, as illustrated in Fig.~\ref{fig:regions}. 

To be more precise, we will assume that we have a continuous map $\gamma: \left[0 , 1 \right] \times \Gamma \rightarrow M\cup\bar M$, $(t,x) \mapsto \gamma_x(t)$, such that $\gamma (\cdot , x) = \gamma_x$ for any $x\in \Gamma$. In particular, the reference set $\Gamma_0 := \gamma(0, \Gamma) = \{ s_x \,, x \in \Gamma \}$ is homotopic to $\Gamma = \gamma( 1, \Gamma)$, which will prove important later on. We also assume sufficient regularity of $\gamma$ (such as, for instance, differentiablity) to ensure that $U[\bar A, A]: \Gamma \to G$ is well-defined and differentiable. Lastly, in order for $U[\bar A, A]$ to qualify as a maximally rich and interesting external reference frame for the region $M$, we will require it to be a non-trivial functional of the fields in $\bar M$, namely:\footnote{Even though these conditions make the formalism physically more transparent, they are not strictly necessary for our general construction to go through. One could for instance consider a situation in which, because of a non-generic choice of system of paths, $U[\bar A, A]$ happens to be determined uniquely across $\Gamma$ once we know its value at one point. In such a situation, $U[\bar A, A]$ would effectively generate a reference frame valued in a single copy of the structure group $G$, rather than in a space of group-valued functions. But this would not preclude its use as a dressing field.}
\begin{itemize}
\item for any $x \in \Gamma$, $U[\bar A, A](x)$ can be varied independently from $A$ in $M$; 
\item for any $x, y \in \Gamma$, $U[\bar A, A](x)$ and $U[\bar A, A](y)$ can be varied independenly from each other whenever $x \neq y$.
\end{itemize}
We will assume that those conditions hold at the dynamical level, that is to say in $\cS$, in which case we will say that $U[\bar A, A](x)$ is \emph{dynamically independent} from $\restr{A}{M}$. This may be hard to check in full generality,\footnote{For example, this assumption holds in the mechanical toy model of Sec.~\ref{sec:mechanical}.} but this is a reasonable assumption that we can expect to hold true for many systems of paths. Furthermore, we can easily justify that the conditions do hold in $\cF$, a weaker statement we may call \emph{kinematical independence}. Indeed, at the kinematical level, the first condition is fullfilled by our assumption that $\Gamma_0 \subset \mathring{\bar M}$, where $\mathring{}$ denotes the interior of a region: it is then always possible to change the value of $U[\bar A, A](x)$ by a variation of $A$ with support in $\gamma_x (\left[ 0,1 \right]) \cap \mathring{\bar M}$. Similarly, one can enforce the second condition by requiring, for any $x\neq y$, the existence of $t_x$ and $t_y$ such that: $\gamma_x(t_x) \in \mathring{\bar M}$, $\gamma_y(t_y) \in \mathring{\bar M}$, and $\gamma_x(t_x) \neq \gamma_y(t_y)$. In words, any two distinct paths are assumed to be non-overlapping somewhere inside $\bar M$. For conceptual clarity, and without significant loss of generality, we will from now on assume that the paths $\gamma_x$ are entirely supported in $\bar M$, as illustrated in Fig.~\ref{fig:regions}, so that the reference frame 
is a functional of $\bar A$ alone and reads explicitly
\beq \label{edgemodeexplicit}
U[\bar A]:=\cP\exp\left(-\int_{\gamma_x}\bar A\right)\,.
\eeq

\medskip

Once we have selected a dynamical reference frame for the boundary $\Gamma$, it is possible to decompose any gauge-invariant observable as a composite of regional gauge-invariant observables, relative to that particular frame. For instance, 
we can equivalently write the cross-boundary Wilson loop in equation~\eqref{Wilsonloop} as:
\beq
W= \Tr\left({\bar H}_{s}[\bar A]  H_{xy}[A]_{\rm rad} \right)\,,
\eeq
where
\begin{align}
    H_{xy}[A]_{\rm rad} &:= U[\bar A]^{-1} \acts H_{xy}[A] = U[\bar A](y)^{-1}  H_{xy}[A] U[\bar A](x) \,, \\ 
    {\bar H}_{s}[\bar A] &:= U[\bar A](x)^{-1} H_{xy}[\bar A]^{-1} U[\bar A](y)\,,
\end{align}
and both of these quantities are gauge-invariant. 
Similarly, the dressed Wilson line in equation~\eqref{chargedWilson} can be split into
\beq
\psi^\dag(x)\,H_{xy}[A]\,H_{y\bar{x}}[\bar A]\,\bar\psi(\bar x)= (\psi^\dag(x)\,H_{xy}[A])_{\rm rad} \, \bar\psi_{s}[\bar A](\bar x)\,,
\eeq
where now
\beq 
(\psi^\dag(x)\,H_{xy}[A])_{\rm rad} := \psi^\dag(x)\,H_{xy}[A]\,U[\bar A](y)\,,\qquad\qquad \bar\psi_{s}[\bar A](\bar x):= U[\bar A](y)^{-1}\,H_{y\bar{x}}[\bar A]\, \bar\psi(\bar x)\,.
\eeq

\subsection{Symmetries as edge frame reorientations}\label{ssec_edgeframereorient}

Now that we have set up an edge frame field, we are in the position to consider two types of frame transformations as in the mechanical toy setting in Secs.~\ref{ssec_extHeis} and~\ref{ssec_rfc}: frame reorientations and frame changes. Both types of transformations act on the right of the frame $U[\bar \Phi]$, in contrast to gauge transformations \eqref{eq:faithful_frame} which act on the left. The essential difference between the two is that frame reorientations are \emph{field-independent}, while frame changes depend on the gauge-invariant relation between the old and new frames. We will begin with the former and discuss the latter in the next subsection.

These observations are particularly clear in the example of the group-valued reference frame $U[\bar A]$ we have just introduced. A reorientation of the edge mode $U[\bar A]$ can be interpreted as a \emph{change of background reference frame}, parametrized by a group-valued function $g_0 (s_x)$ on the asymptotic boundary $\Gamma_0$ where the Wilson lines are anchored.\footnote{By contrast, in the case when we use an open ended Wilson line dressed by a charge in the bulk or on the asymptotic boundary of $\bar M$ and running solely through $\bar M$ to $\Gamma$, i.e.\ $\tilde U [\bar A](x)=\cP\exp(-\int_{\gamma_x}\bar A)\bar\psi(s_x)$, the reorientation of the edge frame corresponds to a physical transformation of the charge $\bar\psi(s_x)\mapsto g_0(s_x)^{-1}\bar\psi(s_x)$.} The effect of this change of asymptotic background frame on the holonomy along $\gamma_x$ can be written as 
\beq 
\cP\exp\left(-\int_{\gamma_x}\bar A\right)\rightarrow\cP\exp\left(-\int_{\gamma_x}\bar A\right)\,g_0^{-1}(s_x)\,
\eeq
and constitutes an asymptotic change of trivialization. Since $\gamma_x(0)=s_x$, this induces a group-valued map $g: x \mapsto g_0 (s_x)$ on $\Gamma$, which acts on $U[\bar A]$ in equation \eqref{edgemodeexplicit} from the right as: 
\beq
U[\bar A](x) \to U[\bar A](x) g(x)^{-1}\,,
\eeq
where we emphasized that $g$, corresponding to a change of background frame, is \emph{independent} from dynamical fields. In particular, it is generated by a vector field $\fY_{\rho}$ on field space $\cF$ where $\rho$ is some non-dynamical Lie-algebra-valued field on $\Gamma$, corresponding to the group element $g$. 

Transformations of this kind that act ``on the other side'' of the edge mode from gauge transformations have been identified in \cite{Donnelly:2016auv} as \emph{symmetries}.\footnote{In the convention of \cite{Donnelly:2016auv}, gauge transformations act on the right of the edge mode, while symmetries act on the left, in contrast to here. We would obtain the convention of \cite{Donnelly:2016auv} if we chose the opposite orientation for the holonomies defining our edge mode frame.}
Our construction elucidates what they mean physically: they are edge frame reorientations. In \cite{Donnelly:2016auv,Geiller:2019bti}, these symmetries have been associated with charges $Q[\rho]$ obtained via $\delta Q[\rho]=\fY_\rho\cdot\Omega_M$, where $\Omega_M$ is the regional presymplectic structure for $M$. In the mechanical toy model, we have shown the analog in Sec.~\ref{ssec_extHeis} and demonstrated in Sec.~\ref{ssec_3types} that, in the presence of boundary conditions, one has to distinguish symmetries into three further types. In order to expand on this using our construction, we will need more structure. In particular, we will explain in Sec.~\ref{sec:geometry} how to obtain $\Omega_M$ from the global presymplectic structure on $M\cup\bar M$ via a post-selection procedure involving the edge modes. This will then permit us to explore the charges in Secs.~\ref{sec:geometry} and~\ref{sec:examples} in detail in Maxwell, Chern-Simons and Yang-Mills theories, including the further distinction of symmetries into genuine symmetries, meta-symmetries and boundary gauge transformations.

For now, we observe that frame reorientations change the physical situation, as they change the gauge-invariant relation between the system of interest (here the fields $\phi$ on $\Gamma$) and the frame. This is clear from the associated transformations of the relational observables in \eqref{edgerelobs} (see also \eqref{mechrelabel} for the mechanical analog):
\beq \label{ickweessochnich}
O_{\phi,Ug^{-1}}(g')=O_{\phi,U}(g'g)\,.
\eeq
This constitutes a change of frame orientation label $g(x)$, $x\in\Gamma$, in the family of gauge-invariant observables $O_{\phi,U}(g)$. Hence, this amounts to a change of observable within this family and therefore a change of relation since the family encodes the value of $\phi$ when the frame $U$ is in orientation $g$.

In analogy to the left group action $\acts$ corresponding to field-independent gauge transformations $e^{\fX_\alpha(\cdot)}=e^\alpha\acts(\cdot)$, we can define the right group action $\odot$ corresponding to symmetries on functionals $f$ depending on boundary fields $\phi,U$, as well as their derivatives by (cf.\ equation~\eqref{symactsmech} for the mechanical analog)
\beq \label{symactfield1}
g(\rho)\odot f[\phi,U]:=e^{\fY_\rho}f=f[\phi,Ug(\rho)]\,.
\eeq 
In particular, frame reorientations take the form $g^{-1}\odot U=Ug^{-1}$, leaving other degrees of freedom invariant, and we can rewrite the relational observable transformation in equation~\eqref{ickweessochnich} as (cf.\ equation~\eqref{obssymtransmech} for the mechanical analog)
\beq \label{symactfield2}
g^{-1}\odot O_{\phi,U}(g')=O_{\phi,g^{-1}\odot U}(g')=O_{\phi,U}(g'g)\,.
\eeq

\subsection{Edge frame changes and transformations of relational observables}\label{ssec_edgerfc}

\begin{figure}[htb]
    \centering
    \includegraphics[scale= .7]{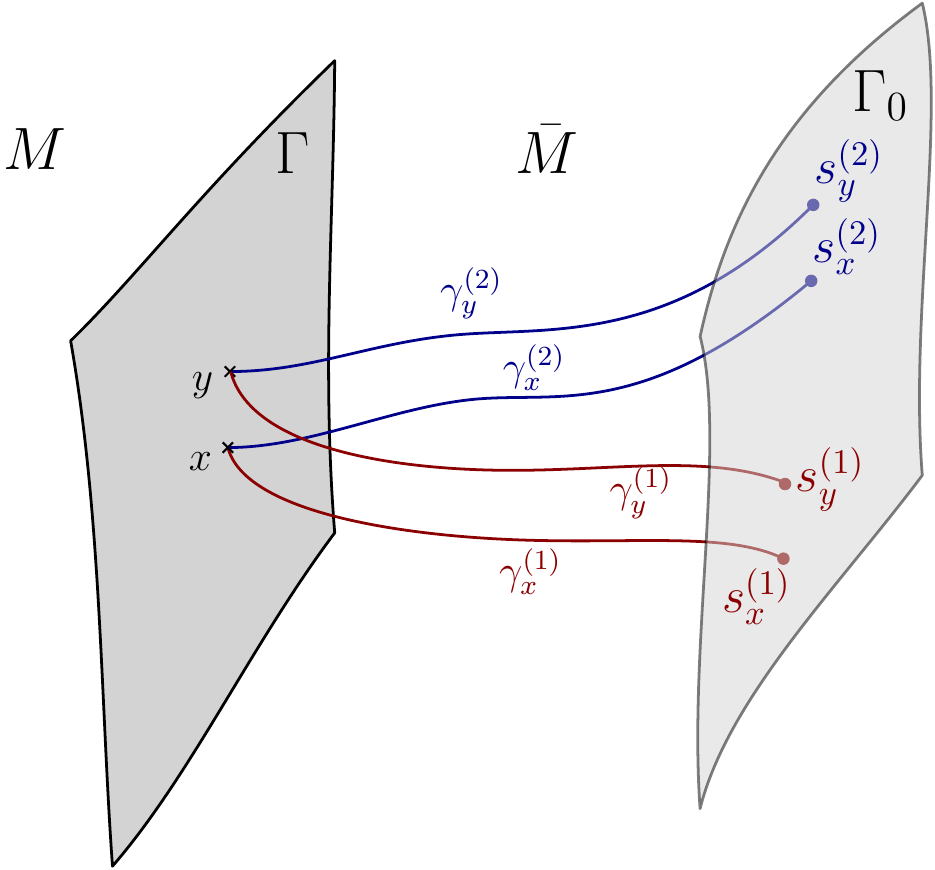}
    \caption{Two systems of paths $\gamma^{(1)}$ and $\gamma^{(2)}$, supporting distinct families of Wilson lines. They generate two distinct group-valued reference frames $U_1[\bar A]$ and $U_2[\bar A]$ for the boundary $\Gamma$.}
    \label{fig:different_paths}
\end{figure}

We have provided a prescription for how to construct an edge mode frame field on $\Gamma$ using a system of paths $\{\gamma_x\vert x\in\Gamma\}$. However, this construction is by no means unique, as illustrated in Fig.~\ref{fig:different_paths}. The global theory in $M\cup\bar M$ supports a whole plethora of different edge mode fields on the interface $\Gamma$, each corresponding to a different system of paths, e.g.\ from the asymptotic boundary $\Gamma_0$ to $\Gamma$. Each ensuing such edge mode field constitutes a \emph{different} external frame field for the subregion $M$ of interest, which we could then ``internalize'' by extending the field space associated with $M$ correspondingly, as sketched at the end of Sec.~\ref{ssec_edgefields} (and expanded on in Sec.~\ref{sec:geometry} below).\footnote{As mentioned in Sec.~\ref{ssec_extHeis}, this is related to an extension of the Heisenberg cut in the quantum theory. In fact, the analogy is stronger in the field theory case here because of the local nature of the gauge constraints: extending the field space with additional edge mode fields does not change the form of the gauge constraints. As such, the gauge-invariant physics on $\Gamma$ \emph{intrinsic} to $M$ (i.e.\ not involving the edge modes) is contained in the same form in any of such field space extensions. By contrast, in the mechanical setting of Sec.~\ref{sec:mechanical}, the total momentum ``constraint'' as well as the action contribution involving the particles of group $M$ change with every addition of a new particle to $\bar M$ (see also Sec.~\ref{ssec_mechpost}).} For each such edge mode field, we can construct relational observables on $\Gamma$ according to equation \eqref{edgerelobs} and the relational observables relative to different edge frame choices will encompass the different ways in which $M$ relates to its complement $\bar M$.

Since there is therefore no distinguished choice of edge frame, we better explain how to relate different frame choices, especially how to translate the relational observables relative to one frame into relational observables relative to another. 
The following is essentially a translation of the reference frame changes discussed in the mechanical toy model in Sec.~\ref{ssec_rfc} into the field theory context. Consider two edge frame fields $U_1,U_2$ and the corresponding relational observables (cf.\ equation~\eqref{edgerelobs}) $O_{\phi,U_k}(g_k)=\left(U_k[\bar A]\,g_k^{-1}\right)^{-1}\acts\phi$, $k=1,2$.
The relation between the two sets of observables depends on the relation between the two frames ``when the second frame $U_2$ is in the origin $e\in G$'', encoded in the gauge-invariant group-valued function
\bes
x\mapsto g_{21}[\bar A](x)&:=&O_{U_1,U_2}(e,x)=U_2[\bar A](x)^{-1}\acts U_1[\bar A](x)\\
&=&U_2[\bar A]^{-1}(x)U_1[\bar A](x)\,,
\ees
which is now dynamical. We can then write the transformation from frame $U_1$ to frame $U_2$ as the \emph{relation-conditional} right action $\odot$ on the frame $U_1$ (cf.\ the mechanical counterpart in equation~\eqref{mechobsrfc}): 
\begin{align}
U_1[\bar A]\quad&\longmapsto\quad U_2[\bar A]=g_{21}[\bar A]^{-1}\odot U_1[\bar A]:= U_1[\bar A]\,g_{21}[\bar A]^{-1}\\
O_{\phi,U_1}(g_1)\quad&\longmapsto\quad O_{\phi,U_2}(g_2)=\left(g_1^{-1}\,g_2\,g_{21}\right)^{-1}\odot O_{\phi,U_1}(g_1):=O_{\phi,(g_1^{-1}\,g_2\,g_{21})^{-1}\odot U_1}(g_1)\,.\label{fieldredef}
\end{align}
Note that this amounts to an extension of the right action $\odot$ defined in equation~\eqref{symactfield1} to field-dependent frame reorientations, in analogy to how equation~\eqref{fielddepgauge} extends field-independent gauge-transformations to field-dependent ones. Similarly to how equation~\eqref{fielddepgauge} generates gauge-invariant observables, equation~\eqref{fieldredef} produces observables invariant under reorientations of frame $R_1$.
Indeed, we have: 
\beq 
O_{\phi,U_2}(g_2)=\Big(U_1[\bar A]\left(U_1[\bar A]^{-1}\,U_2[\bar A]g_2^{-1}g_1\right)g_1^{-1}\Big)^{-1}\acts\phi=\left(U_1[\bar A]^{-1}\,U_2[\bar A]g_2^{-1}g_1\right)\odot\Big(\left(U_1[\bar A]g_1^{-1}\right)^{-1}\acts\phi\Big)\,.\nn
\eeq 
We note that this is equivalent to (cf.\ the field-independent frame reorientation in equation~\eqref{symactfield2})
\beq 
O_{\phi,U_2}(g_2)=O_{\phi,(g_1^{-1}\,g_2\,g_{21})^{-1}\odot U_1}(g_1) \equiv O_{\phi,U_1}\big(g_1(g_1^{-1}g_2g_{21}[\bar A])\big)=O_{\phi,U_1}\big(g_2g_{21}[\bar A]\big)
\eeq 
and hence a relation-conditional right action on the frame orientation label of the relational observable. This encompasses the tautological observables describing the edge frames relative to themselves
\bes
O_{U_1,U_1}(g_1)=g_1\quad&\longmapsto&\quad O_{U_1,U_2}(g_2)=g_2g_{21}[\bar A]\,,\\\
O_{U_2,U_1}(g_1)=g_1g_{21}[\bar A]^{-1}\quad&\longmapsto&\quad O_{U_2,U_2}(g_2)=g_2\,.
\ees

In view of the previous subsection, we can thus interpret frame changes as \emph{relation-conditional frame reorientations}. However, in contrast to the field-independent frame reorientations of the previous subsection, frame changes are \emph{not} symmetries (and not generated by charges). Symmetries change the physical situation, while frame changes only change the description of it. In particular, equation~\eqref{fieldredef} constitutes a field redefinition for the gauge-invariant data on $\Gamma$. As such, a change of edge frame simply amounts to a change of coordinates on the field space $\cF_M$ (which includes the edge modes). Recall also that field-independent frame reorientations map observables from one family $O_{\phi,U}(g)$ of relational observables into others within the \emph{same} family (only changing the orientation label $g$ of the frame in a field-independent manner). By contrast, the frame changes here map the observables from one family $O_{\phi,U_1}(g_1)$ of relational observables (associated with frame $U_1$) to a \emph{distinct} family $O_{\phi,U_2}(g_2)$ (associated with frame $U_2$).

In order to say that this coordinate change is equivalent to a symplectomorphism, as in the mechanical case of Sec.~\ref{ssec_rfc}, we first need control of the presymplectic structure. As argued in Sec.~\ref{ssec_mechrfcfol} for the mechanical setting, different edge frame choices induce distinct families of foliations of the solution space $\cS$ which are crucial for the post-selection procedure; a frame change relates a foliation within the family associated with the first frame to a foliation in the family associated with the second frame. Accordingly, frame changes relate different implementations of the post-selection procedure. This will also be true in the field theory case, as we will argue in Sec.~\ref{sec:geometry}.

\subsection{Frame-reorientation-invariant field-space form}\label{subsec:Maurer-Cartan}

We finally introduce the \emph{field-space right-invariant Maurer-Cartan form} $\delta U U^{-1}$ which, being invariant under frame reorientations, will play an important role in our construction. For notational convenience, we will from now on keep the dependence of the frame $U$ on $\bar A$ implicit. Its variation is given by
\beq
\delta( \delta U U^{-1}) = (\delta U U^{-1}) (\delta U U^{-1}) = \frac{1}{2} \left[ \delta U U^{-1} , \delta U U^{-1} \right]\,,
\eeq
where $[\cdot , \cdot]$ is to be understood as the Lie-bracket on field-space forms.\footnote{That is, $[\omega_1 , \omega_2 ] := \omega_1 \omega_2 - (-1)^{pq} \omega_2 \omega_1$ if $\omega_1$ is a $p$-form and $\omega_2$ a $q$-form (as before, the field-space wedge product is kept implicit).} As in finite dimensions, one can understand this equation as a flatness condition for a field-space gauge potential $- \delta U U^{-1} = U \delta U^{-1}$.\footnote{Following e.g.~\cite{Gomes:2016mwl, Gomes:2018shn,  Gomes:2018dxs}, one could thus define a flat covariant differential $\cD := \delta - \delta U U^{-1}$ for fields on $\Gamma$. We will not rely on such a formalism in the present paper, but it would be interesting to explore it further.} The Maurer-Cartan form also transforms nicely under field-dependent gauge transformations, namely:
\beq\label{eq:gauge_transfo_maurer}
g \acts \delta U U^{-1} = g \delta U U^{-1} g^{-1} + \delta g g^{-1}\,,
\eeq
where we have kept the field-dependence of $g$ implicit. At the infinitesimal level, this yields:
\beq
\cL_{\fX_\alpha} \delta U U^{-1} = \fX_\alpha \cdot \delta (\delta U U^{-1}) + \delta \fX_\alpha \cdot \delta U U^{-1} = [\alpha , \delta U U^{-1}] + \delta \alpha\,,
\eeq
Finally, the differential of the field-space Maurer-Cartan form $\delta U U^{-1}$ can be related to the variation of the spacetime Maurer-Cartan form $\extd U U^{-1}$, as follows:
\beq
\extd( \delta U U^{-1}) = \delta (\extd U U^{-1}) + [\extd U U^{-1}, \delta U U^{-1}]\,.
\eeq

\section{Symplectic geometry of splitting post-selection}\label{sec:geometry}

We shall now explain how to extend the splitting post-selection procedure, illustrated in the mechanical example in Sec.~\ref{sec:mechanical}, to field theory. The underlying idea of our construction is to take the \emph{total} space of solutions $\cS$ to the global equations of motion in $M\cup\bar M$ and post-select the subset of solutions on it, which satisfy certain boundary conditions $X = X_0$ on $\Gamma$. $X$ will be some differential form, locally constructed out of the pullbacks $\phi^a:=\Phi^a|_\Gamma$ of the fields to $\Gamma$ and their derivatives, as well as of $U$ and its derivatives, and $X_0$ is a non-dynamical background field. We require the boundary conditions\,---\,and thus $X$\,---\,to be gauge-invariant, so as to cleanly distinguish between physical boundary conditions and gauge fixing conditions.

First implementing splitting post-selection on-shell, rather than at the level of the action,  has several advantages. From the point of view of $M$, this allows us to cleanly disentangle bulk dynamical equations from boundary conditions (which, as we will explain in Sec.~\ref{sec:splitting}, will eventually be enforced dynamically). More generally, it is useful to investigate the relation between the covariant phase spaces of $M$ and $\bar M$ independently from a choice of variational principle. By focusing on the (pre-)symplectic geometry underlying post-selection, we will be readily able to connect our work to previous literature on edge modes.

\subsection{Radiative presymplectic structure from a dressing flow in field space}\label{subsec:radiative_sympl}

Let us first proceed to show that the group-valued dressing of edge fields introduced in Sec.~\ref{sec:concrete} induces a natural and unambiguous splitting of the presymplectic structure on $\Gamma$ into radiative and gauge contributions. This is important, as only the radiative component will be subject to boundary conditions, while, on-shell, the gauge component will essentially constitute the corner term.

Dressing being implemented by a \emph{field-dependent} gauge transformation $\phi \mapsto U^{-1} \acts \phi$, this operation does not commute with the field-space exterior derivative $\delta$. As a result, it is not completely obvious how to extend a prescription like \eqref{eq:phi_rad} to field forms such as $\theta:= \restr{\Theta}{\Gamma}$ and $\restr{\omega}{\Gamma}$.\footnote{In the following, we will simply denote $\restr{\omega}{\Gamma}$ by $\omega$, keeping the pull-back to $\Gamma$ implicit.}  To circumvent this difficulty, we will implement dressing by means of a particular flow in field space. The advantage of this point of view is that it will readily generalize to field forms by means of the concept of Lie derivative. 

Given our working hypotheses for the system of paths $\{ \gamma_x , x \in \Gamma \}$, we can find a homotopy between $U(x)$ and the identity frame field $\one: \, \Gamma \to G$, $x \mapsto e$ (where $e$ is the identity element in $G$). Indeed, for any $t \in [0,1]$ and $x \in \Gamma$, we can define $\tilde{U}_t (x)$ as the holonomy between the points $s_x$ and $\gamma_x (t)$, along the path $\gamma_x$. We then have $\tilde{U}_1 = U$, $\tilde{U}_0 = \one$, and given that $(t,x)\mapsto \tilde{U}_t(x)$ is continuous, we conclude that $U$ is homotopic to the identity frame field $\one$. As a result, $U$ is in the image of the exponential on the Lie algebra of the gauge group. In particular, there exists a \emph{continuous} $\mathfrak{g}$-valued function $u$ on $\Gamma$, such that: for any $x \in \Gamma$, $U(x)= e^{u(x)}$.\footnote{$u$ is not unique, but at least locally in field-space, one can choose a smooth map $U \mapsto u$ from group- to algebra-valued fields, defining local logarithmic coordinates.} As anticipated, it is therefore legitimate to view the dressing $\phi \mapsto \phi_{\rm rad} = U^{- 1} \acts \phi$ as a field-dependent gauge transformation (in this work, we do not consider large gauge transformations as part of the gauge group). 

In order to extend this dressing to field-space forms, it is convenient to realize it as a flow. To this effect, and given a fixed field-configuration $A$, we consider a path $\{ A_t := g_t \acts A\vert t \in [0,T]\}$ in field-space, where $\{ g_t \vert t \in [0,T]\}$ is a one-parameter family of (global) field-dependent gauge transformations. All we demand from $g_t$ is that $g_0 = \mathrm{Id}$, and $g_T (x) = U^{-1}[A](x)$ for any $x \in \Gamma$.\footnote{Such a path necessarily exists since $U$, seen as a field on $\Gamma$, is connected to the identity. $A_t$ is actually highly non-unique, but we will see that the resulting dressing of edge modes is independent from such ambiguity.} Given those definitions, $\phi_{\rm rad}$ (where we again collectively denote pull-backs of fields on $\Gamma$ by $\phi$) can be recovered as the edge configuration induced by $A_T$: 
\beq
\phi_{\rm rad} = \phi_T \,.
\eeq
Likewise, the frame field flows from its initial value $U$ to the identity frame field $U^{-1}\acts U = \one$.\footnote{It is crucial at this level that $U$ transforms nicely under \emph{field-dependent} gauge transformations.} In other words, the two defining properties of the path $A_t$ are that: 1) it is confined to the gauge orbit of $A_0 = A$, and 2) $U[\bar A_T] = \one$. In particular, we can find a vector field $\fX_\alpha$ such that:
\beq\label{eq:flow_A}
\frac{\extd}{\extd t} A_t = \fX_\alpha (A_t ) = \alpha \acts A_t\,,
\eeq
where the infinitesimal gauge parameter $\alpha$ is \emph{a priori} field-dependent.

The advantage of this somewhat contrived definition of the dressing is that it immediately generalizes to field forms, upon replacement of the vector field $\fX_\alpha$ by the Lie derivative $\cL_{\fX_{\alpha}}$. In more detail, given a form $p$ on the edge field-space, we introduce $p_t$ as the solution to the flow 
\beq
\frac{\extd }{\extd t} p_t = \cL_{\fX_{\alpha}} p_t = \fX_{\alpha} \cdot \delta p_{t} + \delta \fX_{\alpha} \cdot p_{t}\,,
\eeq
with initial condition $p_0=p$. We can then define the \emph{radiative component of the form $p$} by 
\beq
p_{\rm rad} := p_T 
\,.
\eeq
Owing to $[\cL_{\fX_{\alpha}},\delta]=0$, which directly follows from Cartan's magic formula, an immediate consequence of this definition is that:
\beq\label{eq:rad_commutes}
(\delta p)_{\rm rad} = \delta (p_{\rm rad})\,.
\eeq
In particular, $\theta_{\rm rad}$ is a potential for $\omega_{\rm rad}$: 
\beq\label{omegaradconstruct}
\omega_{\rm rad} = \delta \theta_{\rm rad}\,.
\eeq
Furthermore, using the Leibniz property of $\cL_{\fX_\alpha}$, we infer that 
\beq\label{eq:rad_wedge}
(p q)_{\rm rad} = p_{\rm rad} q_{\rm rad}\,,
\eeq 
for any two forms $p$ and $q$. Crucially, this observation is sufficient to show that our dressing prescription is independent from the choice of vector field representative $\fX_\alpha$, and in particular, of the choice of path $A_t$. Indeed, $\phi_{\rm rad} = U^{-1} \acts \phi$ is itself independent from $\fX_\alpha$, and therefore the statement holds for $0$-forms. Invoking \eqref{eq:rad_commutes} and \eqref{eq:rad_wedge}, we can then extend this unicity to forms of arbitrary degree.\footnote{Alternatively, we could have directly defined the dressing of forms by equations \eqref{eq:phi_rad}, \eqref{eq:rad_commutes} and \eqref{eq:rad_wedge}.} Considering that $\phi_{\rm rad}$ is invariant under arbitrary \emph{field-dependent} gauge transformations, a similar reasoning invoking \eqref{eq:rad_commutes} and \eqref{eq:rad_wedge}, allows us to extend this invariance to any dressed form $p_{\rm rad}$. In particular, we have
\beq
\cL_{\fX_{\alpha}} \theta_{\rm rad} = 0 \qquad \mathrm{and} \qquad \cL_{\fX_{\alpha}} \omega_{\rm rad}  =0 \,,
\eeq
for any gauge parameter $\alpha$ (including field-dependent ones). 

Finally, we define the gauge contributions of the presymplectic potential and the presymplectic current as:
\beq\label{defomrad}
\theta_{\rm gauge} = \theta - \theta_{\rm rad} \quad \mathrm{and} \quad \omega_{\rm gauge} = \omega - \omega_{\rm rad}\,.
\eeq

\medskip

For completeness, let us illustrate our construction with a concrete choice of vector field $\fX_\alpha$. Since $U$ is in the image of the exponential, it is natural to choose suitable logarithmic coordinates $U(x)= e^{u(x)}$, for instance by demanding $u$ to be minimal relative to some well-suited norm.\footnote{One could for instance minimize the $L^2$ norm. That is, given a group-valued $U$, we look for an algebra-valued $u$ such that $U= e^u$ and $\| u \|:=\left( \int_\Gamma (u(x)\cdot u(x))  \extd x \right)^{1/2}$ is minimal (where $\cdot$ denotes here the Killing form on $\mathfrak{g}$). The solution will be unique, except on a set of field configurations of measure zero. For those special configurations, we choose one solution in an arbitrary way.} We can then require the field-dependent gauge parameter $\alpha$ to verify: $\alpha[A](x) = - u[A](x)$, for any $x\in \Gamma$. Given such a generator, it follows that:
\beq
\fX_{\alpha} \phi = - u \acts \phi\,,
\eeq
for any local field on $\Gamma$. We can then observe that the following flow, which implements a homotopy between the edge mode and identity frame field, is well-defined and unique:\footnote{The norm $\| u \|$ is strictly decreasing along the flow, so that one never encounters field configurations where  $u$ is not uniquely defined, except possibly at $t=0$.}
\beq
\frac{\extd }{\extd t} (\phi_t , U_t ) = \fX_{\alpha} (\phi_t , U_t ) \,, \quad \phi_0 = \phi \,, \quad U_0 = U\,.
\eeq
Its solution is:
\beq
(e^{(-1 + \tau(t)) u} \acts \phi , e^{\tau(t) u})\,, \quad \tau(t) = e^{- t} \,, t \in [ 0 , + \infty [\,,
\eeq
which one can check by direct computation. Using that $\tau'(t)= -\tau(t)$ in conjuction with the transformation rule $\alpha \acts U = uU$ for the reference frame, we indeed find that
\begin{align}
\frac{\extd }{\extd t}(e^{(-1 + \tau(t)) u} \acts \phi , e^{\tau(t) u}) &= ( (\tau'(t) u e^{(-1 + \tau(t)) u})\acts \phi , \tau'(t) u e^{\tau(t) u}) = (- \tau(t) u) \acts (e^{(-1 + \tau(t)) u} \acts \phi , e^{\tau(t) u}) \\
& = \fX_{\alpha} (e^{(-1 + \tau(t)) u} \acts \phi , e^{\tau(t) u})\,.
\end{align}
Consequently, $\phi_{\rm rad} = U^{-1} \acts \phi$ can be recovered as the limit:
\beq
\phi_{\rm rad} = \underset{t \to +\infty}{\lim} \phi_{t}\,,
\eeq
and, as result, any field-space form can be dressed by following the induced Lie flow.

\subsection{Foliation with respect to field configurations on the time-like boundary}\label{subsec:foliation}

Splitting post-selection proceeds in three steps (see Sec.~\ref{ssec_mechpost} for this procedure in mechanics): one
\begin{enumerate}
    \item foliates the global space of solutions $\cS$ with respect to suitable field configurations on $\Gamma$; then
    \item  restricts to a particular leave in this foliation; and finally
    \item  discards the complementary region $\bar M$ to obtain the looked for presymplectic form $\Omega_M$. 
\end{enumerate}
We call this process \emph{splitting post-selection} as the second step ``post-selects'' those solutions in $\cS$, which satisfy a specific choice of boundary conditions on the interface $\Gamma$, thereby effectively splitting the global space of solutions $\cS$ for $M\cup\bar M$ into two, one for each  spacetime subregion. Upon the third step, one then focuses on the dynamical problem for the subregion $M$ of interest, defined by the selected boundary conditions. Since the latter define the dynamical theory for $M$ (and, likewise, for $\bar M$), we can think of each leaf in the foliation of $\cS$ as a particular subregion theory. As such,  the global solution space $\cS$ assumes the role of a space of subregion theories, in this sense constituting a meta-theory for the local subregions. This will lead us to distinguish symmetries (frame reorientations) on $\Gamma$ into (i) genuine symmetries that leave the subregion theory invariant, (ii) meta-symmetries, which change leaf, i.e.\ subregion theory within $\cS$, and (iii) symmetries that become gauge transformations on particular boundary conditions. We stress that case (ii) only constitutes meta-symmetries for the subregion theories, not, however, for the global dynamics in $M\cup\bar M$, since it  maps one solution into another one of the same global theory.\footnote{This global theory $\cS$ too is to be defined by some asymptotic boundary conditions that are unaffected by the meta-symmetries (ii).} 

In more detail, we look for a decomposition of $\cS$ of the form 
\beq\label{eq:foliation}
\cS = \underset{X_0}{\bigsqcup}\,  \cS_{X_0} \,, \qquad  \cS_{X_0} := \left\lbrace \Phi \in \cS \vert X\left[\phi_{\rm rad}\right] = X_0 \right\rbrace 
\eeq 
for some suitable choice of $X$ and $X_0$, where $X_0$ is a (set of) background field(s) that spans a function space on $\Gamma$ appropriate for a well-posed boundary value problem, and $X$ is a map from field configurations on $\Gamma$ to field configurations on $\Gamma$ (including the reference frame $U$, which enters the definition of $\phi_{\rm rad}$). The post-selected leaf $\cS_{X_0}$ can be thought of as (the space of solutions to) a specific subregion theory. The boundary condition $X\left[\phi_{\rm rad} \right] = X_0$ must fix sufficiently many degrees of freedom to give rise to a well-posed dynamical problem in $M$, and not too many so that partial Cauchy data in the interior of $\Sigma$ remains freely specifiable. In particular, we will require that there is no physical symplectic flux through $\Gamma$ after pulling-back $\Omega$ to $\cS_{X_0}$. A natural and sufficient requirement seems at first to be that $\restr{\omega}{\Gamma, \cS_{X_0}} = 0$. However, in the presence of gauge symmetries such a condition may be overly restrictive, in the sense that it is not always gauge-invariant. In such a situation, $\restr{\omega}{\Gamma, \cS_{X_0}} = 0$ would amount to imposing a vanishing flux, as well as a non-trivial gauge-fixing condition. To make sure that we only constrain physical degrees of freedom, we instead impose the weaker condition:
\beq\label{eq:omega_gamma_eq}
\restr{\omega_{\rm rad}}{\cS_{X_0}} = 0\,.
\eeq
In the following, we will denote equalities that only hold on the leaf $\cS_{X_0}$, that is on-shell of both the bulk equations of motion and the boundary conditions, by the symbol $\heq$. In particular, $\omega_{\rm rad} \heq 0$. This ensures that the presymplectic current $\omega$ is gauge-equivalent to zero when restricted to $\Gamma$ and a specific leaf $\cS_{X_0}$, but not necessarily identically zero.

To make sure that we do not constrain $\omega_{\rm rad}$ more than necessary, we will impose boundary conditions on a Lagrangian submanifold of $\cS$. In practice, we will choose local and gauge-invariant Darboux coordinates $X$ and $Y$ (which are fields on $\Gamma$) such that $\omega_{\rm rad} = \delta X  \delta Y$. To implement \eqref{eq:omega_gamma_eq}, it is then sufficient to impose e.g.\ $\delta X \heq 0$, while leaving $Y$ dynamical. 

\medskip

Let us look at a simple but illuminating example in Maxwell theory (which we will discuss again in more depth in Sec.~\ref{subsec:maxwell}). The presymplectic current can be expressed in the canonical Darboux coordinates as $\omega = \delta A \wedge \delta \star F$. Given a $\U(1)$ reference frame with phase $\vphi(x)$ one can define the radiative data (see equations~\eqref{genArad} and~\eqref{genFrad}):
\beq\label{1stArad}
A_{\rm rad} = A - \extd \vphi \,, \quad (\star F)_{\rm rad} = \star F\,, \quad \omega_{\rm rad} = \delta A_{\rm rad} \wedge \delta (\star F)_{\rm rad} \approx \delta A \wedge \delta \star F - \extd (\delta \vphi \delta \star F)\,.
\eeq
These coordinates induce two natural foliations. We can foliate the solution space $\cS$ with respect to the pull-back of $\star F$, which amounts to defining $X[A_{\rm rad}, (\star F)_{\rm rad}]:= (\star F)_{\rm rad}$. Since $\star F$ is gauge-invariant to start with, it follows that $\restr{\omega}{\Gamma} \heq 0$. Relative to this particular choice of foliation, the dressing by $U$ can be safely ignored. By contrast, it becomes relevant again if one decides to impose a boundary condition on $A_{\rm rad}$, that is, if we foliate the solution space with respect to $X[A_{\rm rad}, (\star F)_{\rm rad}]:= A_{\rm rad}$. In that case, $\restr{\omega_{\rm rad}}{\cP_f} \heq 0$ while $\restr{\omega}{\Gamma} \heq \extd (\delta\varphi \delta\star F)$ does not vanish in general. Note that the boundary condition on $A_{\rm rad}$ fixes the pull-back of $F$ to $\Gamma$ completely, but is stronger in that it also contains relational information about $\restr{A}{\Gamma}$ and $U$. Finally, one can impose boundary conditions on linear combinations of $A_{\rm rad}$ and $(\star F)_{\rm rad}$, which we will investigate thoroughly in Sec.~\ref{sec:examples}.

\medskip

Let us now briefly discuss the two types of transformations one can perform on the canonical splitting of $A$ relative to the reference frame $U= e^{i \vphi}$:\footnote{We focus exclusively on the connection because its conjugate $\star F$ is gauge-invariant in the present Abelian context.}
\beq\label{eq:splitting_A}
\restr{A}{\Gamma} = A_{\rm rad} + \extd \vphi\,.
\eeq
The gauge contribution $A_{\rm gauge} = \extd \vphi$ is unconstrained by purely local gauge-invariant observables, such as $\restr{F}{\Gamma}$, but is fixed unambiguously by the reference frame data. Gauge transformations act on both $A$ and $\vphi$ in a way that leaves $A_{\rm rad}$ invariant: 
\beq\label{eq:gauge_transfo_maxwell}
\delta_\alpha A = \extd \alpha  \,, \qquad \delta_\alpha \vphi = \alpha \qquad \Rightarrow \qquad \delta_\alpha A_{\rm rad} = 0\,.
\eeq 
By contrast, frame reorientations take the form of a symmetry transformation $\Delta_\rho$ which acts solely on $\vphi$, and thereby changes the radiative part of the connection: 
\beq\label{eq:symmetry_Maxwell1}
\Delta_\rho A = 0\,,  \qquad  \Delta_\rho \vphi = - \rho \qquad \Rightarrow \qquad \Delta_\rho A_{\rm rad} =   \extd \rho \,.
\eeq
Depending on the choice of boundary condition, the solutions might not be stable under this transformation, in which case $\Delta_\rho$ generates a \emph{meta-symmetry}, i.e.\ a change of post-selected theory $\cS_{X_0}$ within the global solution space $\cS$. For instance, when foliating the solution space with respect to $A_{\rm rad}$, $\Delta_\rho$ acts transversally to the leaves and thereby changes the boundary condition (that is, the value of $X_0$ in \eqref{eq:foliation}), unless $\rho$ is constant on $\Gamma$ (in which case $\Delta_\rho$ does generate a symmetry). As we will see below, this meta-symmetry is a symplectomorphism within $\cS$. The definition \eqref{eq:symmetry_Maxwell1} is an Abelian incarnation of the symmetry transformation originally introduced by \cite{Donnelly:2016auv} in Yang-Mills theory. Our construction clarifies its relation to reference frames, and its interplay with boundary conditions. 

Our construction has so far been implicit about the choice of edge frame $U[\bar A]$ that goes into the construction of the radiative data $\phi_{\rm rad}$ (cf.\ equation~\eqref{eq:phi_rad}) defining the boundary conditions $X_0$. For a fixed frame choice, we can choose different maps $X$, each corresponding to a distinct foliation of $\cS$. We can consider them as part of one family of foliations, associated with the frame $U$. They will be related by the transformations relating the different choices of $X$ (e.g., in the mechanical setting of Sec.~\ref{ssec_mechrfcfol}, these were canonical transformations). As argued in Sec.~\ref{ssec_edgerfc}, there is a plethora of distinct choices of edge frame fields. It is clear that each of them will lead to a distinct family of foliations \eqref{eq:foliation} of the global solution space $\cS$. Any choice of boundary conditions $X_0$ fixes a set of edge relational observables associated with some frame. Since the edge frame transformations of Sec.~\ref{ssec_edgerfc} map the fixed set of relational observables into relational observables associated with a different frame, they also relate the different frame associated families of foliations of $\cS$.\footnote{Since the frame changes are field-dependent, the new set of relational observables will generally not be fixed via the boundary conditions $X=X_0$. } The frame changes, as specific coordinate changes on $\cS$, can thereby be invoked to link physically distinct implementations of post-selection.

\subsection{Presymplectic structure of the subregion field space}\label{subsec:presympl}

Let us now introduce a presymplectic structure $\Omega_M$ on the space of dynamical field configurations in $M$, which includes both bulk fields and the boundary reference frame and which is consistent with the global presymplectic structure $\Omega$ for fields in $M\cup\bar M$. In the pure gauge-theory examples we are interested in, this is defined as:\footnote{Under the assumptions we made in Sec.~\ref{sec:concrete}\,---\,namely, that $U[\bar A]$ can be varied independently from $\restr{A}{M}$ at the \emph{dynamical} level \,---, we can conclude that $\cS_M^{X_0}$ factorizes as a solution space for $\restr{A}{M}$ times a space of $G$-valued functions on $\Gamma$. As a result, $\cS_M^{X_0}$ can be understood as a boundary extended functional space for region $M$, in the spirit of \cite{Donnelly:2016auv}. If \emph{dynamical independence} were not to hold, $\cS_M^{X_0}$ could still be embedded in the same extended functional space, hence our construction could straightforwardly be generalized to include this case. To avoid introducing further notation that would only cloud the main conclusions of the paper, we refrain from demonstrating this explicitly.}
\beq
\cS_{M}^{X_0} := \left\lbrace (\restr{A}{M}, U[\bar A])\vert (A, \bar A) \in \cS_{X_0}\, \right\rbrace\,.
\eeq
We first assume that $\Omega_M$ is locally identical to $\Omega$ in the bulk of $M$ (resp.\ $\bar M$), and that the regional presymplectic forms are additive, in line with the idea that the global degrees of freedom are constituted by the sum of the local degrees of freedom: 
\beq
\Omega_M + \Omega_{\bar M} \heq \Omega\,.
\eeq
We are then naturally led to the following Ansatz:
\beq\label{eq:ansatz_sympl}
\Omega_M = \int_\Sigma \omega + \int_{\partial \Sigma }
 \omega_\partial \,, \qquad \mathrm{and} \qquad \Omega_{\bar M} = \int_{\bar \Sigma} \omega - \int_{\partial \Sigma}
 \omega_\partial\,.
 \eeq
In this equation, $\omega_\partial$ is a $(2, d-2)$-form on $\Gamma$ that may depend on both bulk and boundary fields and constitutes a corner term for $\Omega_M$. To ensure that $\delta \Omega_M = 0$, we will furthermore assume that $\omega_\partial = \delta \theta_\partial$ for some $(1,d-2)$-form $\theta_\partial$, so that for every Cauchy surface $\Sigma$,
\beq
\Theta_\Sigma := \int_\Sigma \Theta + \int_{\partial \Sigma}
 \theta_\partial
\eeq
defines a potential (which may depend on $\Sigma$) for $\Omega_M$. We call $\omega_\partial$ (resp.\ $\theta_\partial$) the \emph{boundary presymplectic current} (resp.\ \emph{boundary presymplectic potential}). We can take advantage of this splitting ambiguity to enforce two additional desirable properties for $\Omega_M$, namely: its independence from the choice of partial Cauchy surface, and its invariance under field-dependent gauge transformations. As we shall see, the latter property follows from the former.

\begin{figure}[htb]
    \centering
    \includegraphics[scale= .7]{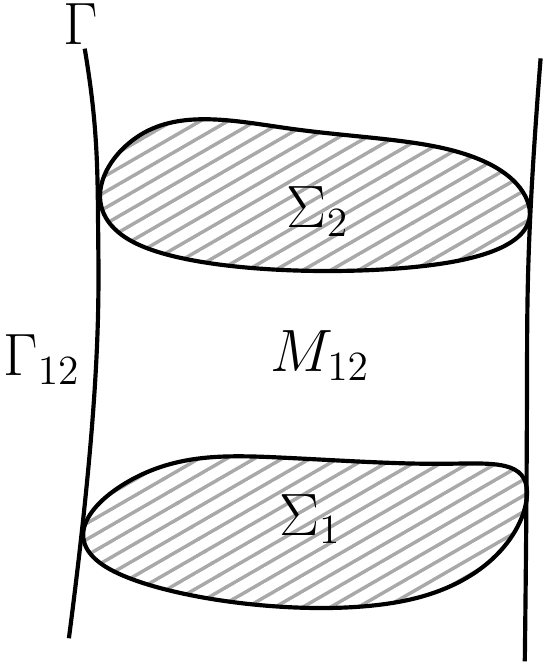}
    \caption{We now allow $\Sigma_1$ and $\Sigma_2$ to be arbitrary surfaces anchored on $\Gamma$; we call $\Gamma_{12}$ the portion of $\Gamma$ delimited by $\partial\Sigma_1$ and $\partial\Sigma_2$, while $M_{12}$ designates the portion of $M$ delimited by $\Sigma_1$ and $\Sigma_2$.}
    \label{fig:regions_varied}
\end{figure}

Asking for \eqref{eq:ansatz_sympl} to be conserved across distinct Cauchy surfaces is quite constraining. Let us consider an initial and final surface $\Sigma_1$ and $\Sigma_2$, which we now permit to vary; see Fig.~\ref{fig:regions_varied}. We call $\Gamma_{12}$ the portion of $\Gamma$ whose boundary is $\partial \Sigma_1 \cup \partial \Sigma_2$, and $M_{12}$ the subregion of $M$ delimited by $\Sigma_1$ and $\Sigma_2$.  
From $\extd \omega \approx 0$, it follows by Stokes' theorem that:
\begin{equation}
    \int_{\Sigma_2} \omega - \int_{\Sigma_1} \omega + \int_{\Gamma_{12}} \omega = \int_{M_{12}} \extd \omega   \approx 0 \,.
\end{equation}
Inserting this relation into the consistency condition
\beq\label{eq:consistency}
\int_{\Sigma_1} \omega + \int_{\partial \Sigma_1 }
 \omega_\partial \heq \int_{\Sigma_2} \omega + \int_{\partial \Sigma_2 }
 \omega_\partial  \,,
\eeq
we find
\beq
\int_{\partial \Sigma_2 }
 \omega_\partial - \int_{\partial \Sigma_1 }
 \omega_\partial \heq \int_{\Gamma_{12}} \omega\,.
\eeq
For this relation to hold for any choice of $\Sigma_1$ and $\Sigma_2$, we must have: 
\beq
\restr{\omega}{\Gamma} \heq - \extd \omega_\partial \,.
\eeq
Irrespective of the choice of gauge-invariant boundary condition, equations~\eqref{defomrad} and~\eqref{eq:omega_gamma_eq} entail that $\restr{\omega}{\Gamma} \heq \omega_{\rm gauge}$ on $\cS_{X_0}$, and therefore:
\beq\label{eq:omega_bdy}
\omega_{\rm gauge} \heq - \extd \omega_\partial\,.
\eeq
This fixes $\omega_\partial$ up to a closed form, provided that $\omega_{\rm gauge}$ is itself exact (at least when pulled-back to the leaf $\cS_{X_0}$). We will find out that this consistency requirement is fulfilled in Maxwell, Chern-Simons and Yang-Mills theory. 
Finally, equation \eqref{eq:omega_bdy} implies that 
\beq\label{eq:theta_bdy}
\theta_{\rm gauge} \heq - \extd \theta_\partial - \delta \ell_\partial\,.
\eeq
for some $(0,d-1)$-form $\ell_\partial$. 

\medskip

In Maxwell theory, this leads to:
\beq\label{eq:boundary_sympl_Maxwell}
\extd \omega_\partial = - \delta \extd \vphi \wedge \delta \star F = - \extd ( \delta \vphi \delta \star F) + \delta \vphi  \delta \extd \star F \approx - \extd ( \delta \vphi \delta \star F) \,,
\eeq
where we have used the equation of motion $\restr{\extd \star F}{\Gamma} = 0$ in the last equality.\footnote{Note that it is also possible to consistently include boundary matter sources: it is indeed sufficient to assume that $\restr{\extd \star F}{\Gamma} \heq j$ where $j$ is some background field for the consistency condition $\extd \omega_\partial \heq \restr{\omega}{\Gamma}$ to hold.} Henceforth, we can define $\omega_\partial$ as
\beq
\omega_\partial = - \delta \vphi \delta \star F + c\,,\label{edgesympform}
\eeq
where $c$ is some closed form on $\Gamma$. If $\partial \Sigma$ is simply-connected and has no boundary, $c$ is also exact and does not contribute to the presymplectic form. We will assume we are in this situation and choose $c=0$.\footnote{Given our working hypotheses, $\partial \Sigma$ is homeomorphic to $S^1$ in $2+1$ dimensions, and therefore fails to be simply-connected. In that case, including a closed form $c$ with $\int_{\partial \Sigma} c \neq 0$ might be relevant to consistently account for non-trivial winding modes of $U(x)$. However, in our set-up, $U(x)$ is always homotopic to the identity ($\int_{\partial \Sigma} \extd \vphi =0$), and $c=0$ remains a consistent choice.}
If we decide to impose a boundary condition on $\star F$, we find out that $\extd \omega_\partial \heq 0$; $\omega_\partial = 0$ is then a consistent choice, as expected. By contrast, as soon as the boundary condition constrains $A_{\rm rad}$,\footnote{That is, for Dirichlet and Robin boundary conditions, as discussed in Sec.~\ref{subsec:maxwell}.} thereby leaving at least some components of $\star F$ dynamical on the boundary, the boundary presymplectic current contributes non-trivially.  

\medskip

Let us now consider the question of gauge invariance. In Yang-Mills theories, the presymplectic current $\omega$ and presymplectic potential $\Theta$ are gauge-invariant to start with, but are not invariant under field-dependent gauge transformations. In \cite{Donnelly:2016auv}, the authors postulated such an extended invariance, which in turn motivated the introduction of abstract edge modes and led to the extended presymplectic structure $\Omega_M$. In the present work, we are somewhat reversing the logic. Edge modes are physical from the start and already present in the global phase space. The way they must contribute to the regional presymplectic structure $\Omega_M$ has then been determined by the physical requirement of a vanishing symplectic flux through $\Gamma$. As we will see now, invariance under field-dependent gauge transformations is a necessary by-product of this construction, but conceptually secondary.  

Field-dependent gauge transformations $\fX_{\alpha[A]}$ verify two interesting properties:
\begin{itemize}
\item[(a)] Irrespective of the choice of gauge-invariant boundary condition $X[\phi_{\rm rad}] = X_0$, $\cS_{M}^{X_0}$ is stable under the flow of $\fX_{\alpha[A]}$, since the latter maps solutions to solutions and leaves $\phi_{\rm rad}$ invariant.
\item[(b)] $\fX_{\alpha[A]}$ generate local gauge symmetries, which can be tuned independently in disjoint neighborhoods of $M$. 
\end{itemize}
Given a Cauchy surface $\Sigma$, we can always pick up a second Cauchy surface $\Sigma'$ such that $\Sigma$ and $\Sigma'$ are disjoint. By \eqref{eq:consistency} and property (a), we know that:
\beq\label{eq:gauge_two_sigma}
\int_\Sigma \fX_{\alpha[A]}\cdot \omega + \int_{\partial \Sigma} \fX_{\alpha[A]}\cdot \omega_{\partial} \approx \fX_{\alpha[A]}\cdot \Omega_M \approx \int_{\Sigma'} \fX_{\alpha[A]}\cdot \omega + \int_{\partial \Sigma'} \fX_{\alpha[A]}\cdot \omega_{\partial}\,.
\eeq
Furthermore, thanks to property (b), we can always find a gauge transformation $\tilde{\alpha}[A]$ that agrees with $\alpha[A]$ on $\Sigma$ but vanishes on $\Sigma'$:
\beq
\forall x \in \Sigma \,, \quad \alpha[A](x)=\tilde{\alpha}[A](x)\,, \qquad \mathrm{and} \qquad \forall y \in \Sigma' \,, \quad \tilde{\alpha}[A](y)=0\,.
\eeq 
It follows from \eqref{eq:gauge_two_sigma} that
\beq\label{eq:generalized_gauge}
\fX_{\alpha[A]}\cdot \Omega_M \approx 0\,.
\eeq
Using Cartan's magic formula, this implies gauge-invariance of the regional presymplectic structure:
\beq \label{sympinvgauge}
\cL_{\fX_{\alpha[A]}}\Omega_M \approx 0\,,
\eeq 
which can be equivalently written as 
\beq\label{eq:gauge_inv_Theta_sigma}
\cL_{\fX_{\alpha[A]}} \Theta_\Sigma \approx \delta(\fX_{\alpha[A]}\cdot \Theta_\Sigma )\,.
\eeq
In our main examples of Sec.~\ref{sec:examples}, we will see that $\fX_{\alpha[A]}\cdot \Theta_\Sigma$ is a constraint (a fact we will discuss explicitly for Maxwell theory in Sec.~\ref{subsec:maxwell}), so that $\cL_{\fX_{\alpha[A]}} \Theta_\Sigma \approx 0$. Such a formula played the role of a postulate in the construction of the extended phase space for Yang-Mills theory first outlined in \cite{Donnelly:2016auv}. Our starting point was different, but we recover it in a second step. One can directly check that \eqref{eq:generalized_gauge} holds in Maxwell's theory:
\beq
\fX_{\alpha[A]}\cdot \Omega_M = \int_{\Sigma} \extd \alpha[A] \wedge \delta \star F - \int_{\partial \Sigma} \alpha[A] \delta \star F \approx \int_{\Sigma} \extd ( \alpha[A] \delta \star F) - \int_{\partial \Sigma} \alpha[A] \delta \star F = 0\,.  
\eeq

It is interesting to contrast the invariance of $\Omega_M$ under field-dependent gauge transformations with its behaviour under the reference frame symmetries $U \mapsto U g^{-1}$. For definiteness and simplicity, let us focus on Maxwell theory (Chern-Simons and Yang-Mills theory will be discussed in Sec.~\ref{sec:examples}, accompanied with a more in-depth discussion of Maxwell theory). The reference frame symmetries are then generated by the transformations $\Delta_\rho$ defined in \eqref{eq:symmetry_Maxwell1}; let us denote the associated field-dependent vector fields by $\fY_{\rho[A]}$. The status of $\fY_{\rho[A]}$ depends on the type of boundary condition one is imposing. If $X[A_{\rm rad}, (\star F)_{\rm rad}]= (\star F)_{\rm rad}$, any $\fY_{\rho[A]}$ leaves $\cS_{M}^{X_0}$ invariant, so that condition (a) holds. Since condition (b) also holds, we are led to the conclusion that the vector fields $\fY_{\rho[A]}$ generate \emph{gauge} transformations rather than symmetries. Indeed, we find that the charge vanishes by virtue of the boundary condition $\restr{\star F}{\Gamma} = X_0$
\beq
\fY_{\rho[A]}\cdot \Omega_M = \int_{\partial \Sigma} \rho[A] \delta \star F \heq 0\,.
\eeq
For \emph{field-independent} $\rho$, we can thus define, \emph{in addition} to the standard $\rm{U}(1)$ bulk constraints $C[\alpha]=\int_\Sigma\alpha\extd\star F$ (see Sec.~\ref{subsec:maxwell} for further details), the edge constraints
\beq\label{edgeconstraint}
C_\partial[\rho]:=\int_{\partial\Sigma}\rho(\star F- X_0 )\,,
\eeq 
which, by virtue of $\Delta_\rho \star F=0$, are first-class
\beq 
\{C_\partial[\sigma],C_\partial[\rho]\}=\fY_\rho\cdot \fY_\sigma \cdot\Omega_M=0\,.
\eeq
Note that, by \eqref{edgesympform}, the edge constraints are conjugate to the edge mode $\varphi$ of Maxwell theory. Hence, for Neumann boundary conditions in Maxwell theory the edge modes assume a somewhat redundant role. The phase space has been extended, but also subjected to another set of first-class momentum-type constraints so that the phase space extension does not amount to an increase in the number of \emph{independent} gauge-invariant degrees of freedom on $\Gamma$, despite the presence of edge modes as reference frames.

By contrast, when $X[A_{\rm rad}, (\star F)_{\rm rad}]= A_{\rm rad}$, only the vector fields $\fY_{\rho[A]}$ with $\extd \rho[A] = 0$ leave the solution space $\cS_M^{X_0}$ invariant. That is, $\rho[A]$ must be a constant on $\Gamma$. For such transformations, condition (a) still holds but (b) does not. As a consequence, they escape our general argument. Indeed, we can explicitly verify that such transformations do generate non-trivial symmetries, recorded by integrable charges $Q[\rho]$ when $\rho$ is furthermore assumed to be \emph{field-independent}:\footnote{In particular, note that the Dirichlet boundary condition $A_{\rm rad}=X_0$, in contrast to the case of Neumann boundary conditions, is now a holonomic constraint (devoid of momentum degrees of freedom) and such a constraint does not constitute a generator of gauge transformations. Accordingly, the extension of the phase space through edge modes does in this case lead to additional independent gauge-invariant degrees of freedom.}
\beq
\fY_\rho\cdot \Omega_M = \delta Q[\rho] \quad \mathrm{with} \quad Q[\rho]  :=  \int_{\partial \Sigma} \rho \star F \,.
\eeq
Finally, transformations $\fY_\rho$ with $\rho$ field-independent but not constant on $\Gamma$ are neither symmetries nor gauge: since they affect the boundary conditions themselves, they can be understood as non-trivial transformations of background fields, which map one solution space $\cS_M^{X_0}$ to another $\cS_M^{X_0'}$, both descending from the same foliation of the global solution space $\cS$ in \eqref{eq:foliation}. They are symplectomorphisms with non-vanishing charges, but between \emph{different} phase spaces. Indeed, Cartan's magic formula implies $\cL_{\fY_\rho}\Omega_M=\delta\left(\fY_\rho\cdot\Omega_M\right)=\delta^2Q[\rho]=0$; they therefore constitute meta-symmetries.

\section{Boundary actions from splitting post-selection}\label{sec:splitting}

After exploring splitting post-selection directly at the level of solution spaces, let us extend this discussion by constructing a suitable variational principle for it. Post-selecting the \emph{total} space of solutions $\cS$ of $M\cup\bar M$ on certain boundary conditions $X = X_0$ on $\Gamma$ will have the effect of turning the global variational problem for $M\cup\bar M$ into two separate variational problems for $M$ and $\bar M$. We will begin by  considering  a \emph{fictitious} boundary $\Gamma$ between the two subregions $M$ and $\bar M$, i.e.\ a boundary which  is not physically distinguished through sources or currents residing on it. In other words, the regular bulk equations of motion of the theory have to hold on the interface $\Gamma$. We shall address the question of \emph{physical} boundaries in a follow-up work \cite{CH2}.

In order to have well-defined equations of motion everywhere on the space of field configurations $\cF$, we shall assume the latter to only contain field configurations $\Phi^a$ which are $C^q$ across $M\cup\bar M$ (incl.\ $\Gamma$) if the equations of motion are of order $q$ in derivatives. 

A variational problem for the subregion $M$ of interest can be formulated in two equivalent ways. (1) If the field configuration space $\cF_M$ associated with $M$ only contains configurations which satisfy the desired boundary conditions, the role of the boundary action will be limited to canceling the presymplectic potential on the interface $\Gamma$ on-shell. This is necessary in order to make the boundary value problem well-defined and consistent with the global variational problem. (2) If instead $\cF_M$ contains all $C^q$ field configurations in $M$, it accordingly requires a second type of contribution to the  boundary action, with the purpose of imposing the desired boundary conditions as boundary equations of motion. 
Post-selecting the global solution space on $X=X_0$ will first yield the former formulation, which we however then use to construct the equivalent one in terms of boundary actions.

Once translated into the covariant phase space language, post-selection will result in a simple but general algorithm which, given a set of admissible (Dirichlet, Neumann, Robin or mixed) boundary conditions,  produces a family of appropriate boundary actions implementing these boundary conditions.

\subsection{Splitting a global variational problem into local ones via post-selection}\label{ssec_globsplit}

Suppose we are given some field theory and would like to formulate its variational principle for the subregion $M$ subject to the boundary conditions $X^a=X^a_0$ on the interface $\Gamma$, where the $X^a$ are some functionals locally constructed out of the pullbacks of the radiative  fields $\phi_{\rm rad}$  and the $X_0^a$ denote a corresponding choice of background fields. 

A natural way to proceed is to start with the space of solutions $\cS\subset\cF$ associated with the global action 
\beq
S_{M\cup\bar M}=\int_{M\cup\bar M} L=S_M+S_{\bar M}\label{Stot0}
\eeq 
and post-select (i.e.\ restrict to) the subset $\cS_{X_0}\subset\cS$ of solutions satisfying the desired conditions on $\Gamma$. Noting that the boundary conditions are integrable, $\delta X=0$,\footnote{Since $X_0$ is a background field, the condition $X-X_0=0$ entails the differential constraint $\delta X=0$, which thus is integrable and, in turn, holonomic.} this can be achieved by imposing them as a holonomic constraint, yielding the total action
\beq
S_{\rm tot}:=S_{M\cup\bar M}+\int_\Gamma \lambda_a(X^a-X^a_0)\,,\label{Stot}
\eeq
where $\lambda_a$ denote a set of Lagrange multiplier densities defined on $\Gamma$.\footnote{We emphasize that $X^a=X^a_0$ does \emph{not} constitute a gluing condition as will be discussed in our follow-up work \cite{CH2} (see also \cite{Blommaert:2018oue,Geiller:2019bti}) since $X^a_0$ is a non-dynamical background field and we are here not identifying \emph{a priori} independent degrees of freedom from $M$ and $\bar M$ on $\Gamma$.}  In order to maintain gauge-invariance of the total action when there is a gauge symmetry, we shall require the Lagrange multipliers $\lambda_a$ to transform trivially under the relevant gauge transformations. Furthermore, since we are dealing with an interface between two regions, we will have to choose a convention as to which orientation of  $\Gamma$ we tacitly invoke in the integral of \eqref{Stot}; henceforth, we shall always choose the `perspective' of our subregion of interest $M$, i.e.\ orient $\Gamma$ relative to the normal pointing outward from $M$. 

The action \eqref{Stot} constitutes a well-defined variational problem since (assuming asymptotic fall-off conditions of $\Phi^a$ in $\bar M$) 
\bes
\delta S_{\rm tot}&=&\int_{M\cup\bar M}(E_a\delta\Phi^a+\extd\Theta)+\int_\Gamma \left(\delta\lambda_a(X^a-X^a_0)+\lambda_a\delta X^a\right)\nn\\
&=&\int_{M\cup\bar M}E_a\delta\Phi^a+\int_{\Sigma_2\cup\bar\Sigma_2-\Sigma_1\cup\bar\Sigma_1}\Theta+\int_\Gamma \left(\delta\lambda_a(X^a-X^a_0)+\lambda_a\delta X^a\right)\,,\label{globvar}
\ees
where we used the shorthand notation $\int_{A-B} := \int_A - \int_B$.
Note that there is no contribution of the presymplectic potential $\Theta$ on $\Gamma$; the contribution $\int_\Gamma \theta$  coming from $M$ cancels the corresponding $\int_{\bar \Gamma} \theta = - \int_{\Gamma} \theta$  coming from its complement $\bar M$ because their integrals along $\Gamma$ feature opposite orientation.\footnote{$\bar \Gamma$ here denotes the interface equipped with the canonical orientation induced by $\bar M$.} 
In other words, around field configurations satisfying the bulk equations of motion $E_a=0$ and the interface equations of motion $X^a=X^a_0$ and $\lambda_a=0$, the total variation is stationary up to terms on the future and past Cauchy surface 
\beq 
\delta S_{\rm tot}\heq \int_{\Sigma_2\cup\bar\Sigma_2-\Sigma_1\cup\bar\Sigma_1}\Theta\,.
\eeq 

We would like to turn this global variational problem into two \emph{independent} variational problems, one for $M$ and one for its complement $\bar M$, which together are equivalent to the global one. This can be achieved by noting that the two action contributions for $M$ and $\bar M$ in \eqref{Stot0} and \eqref{Stot} are only uniquely defined up to the addition and subtraction of a local integral over their interface $\Gamma$. More precisely, we are free to replace $S_M$ and $S_{\bar M}$ in \eqref{Stot} with
\beq
\tilde S_M:=S_M+\int_\Gamma\ell_{\rm corr}\,,\qquad\tilde S_{\bar M}:=S_{\bar M}-\int_\Gamma\ell_{\rm corr}
\eeq
without changing $S_{\rm tot}$, where $\ell_{\rm corr}$ is any density locally constructed out of $\phi^a$, the pullback of the derivatives of $\Phi^a$ to $\Gamma$, as well as the reference frame $U$.\footnote{In both modified actions we oriented $\Gamma$ according to our convention above and, accordingly, the volume form $\epsilon_\Gamma$ and any normal derivatives in $\ell_{\rm corr}$ are defined relative to $M$.} This argument is completely analogous to the one used in Sec.~\ref{sec:geometry} to motivate the Ansatz \eqref{eq:ansatz_sympl}, now transposed onto the level of the action. In fact, we can understand what follows as an off-shell extension of our previous construction, with respect to both equations of motion and boundary conditions.

In particular, we can exploit this ambiguity and choose $\ell_{\rm corr}$ such that 
\beq
\delta\tilde S_M \heq \int_{\Sigma_2-\Sigma_1}\Theta + \int_{\partial\Sigma_2-\partial\Sigma_1} \ C\,,\qquad \delta\tilde S_{\bar M} \heq \int_{\bar\Sigma_2-\bar\Sigma_1}\Theta + \int_{\partial\bar\Sigma_2-\partial\bar\Sigma_1} C\,.\label{locvar}
\eeq
 We allow for a possible corner contribution $C$, which we will view below as arising from a $(1,d-2)$-form intrinsic to $\Gamma$ and locally constructed out of the fields, their variations and derivatives. 
 Note that we then have 
\beq
\delta S_{\rm tot}\heq\delta\tilde S_M +\delta\tilde S_{\bar M}
\eeq
since the two corner contributions will cancel one another, seeing as they come with opposite orientation on $\partial \Sigma_i$ and $\partial\bar\Sigma_i$, $i=1,2$. Hence, upon invoking the boundary conditions through post-selection, the global variational problem in \eqref{globvar} splits into two independent subregion variational problems in \eqref{locvar}, which together are equivalent to the global one.

We can render the condition on $\ell_{\rm corr}$ in \eqref{locvar} more explicit by noting that the first equality, which is associated to the subregion of interest $M$, is equivalent to 
\beq 
\int_\Gamma \left( \theta +\delta\ell_{\rm corr} \right) \heq -\int_{\Gamma} \extd C \,. 
\label{intform}
\eeq
Using Stokes' theorem, the corner terms have been reexpressed as an integral over $\Gamma$, where the extra minus sign is explained by the fact that the orientations of the integrals over $\partial\Sigma_2$ and $\partial\Sigma_1$ induced by $\Sigma_2$ and $\Sigma_1$ are opposite to the orientation of the integral over $\partial\Gamma$ which is induced by $\Gamma$. If we require the condition in \eqref{intform} to hold for any choice of initial and final time slice $\Sigma_1$ and $\Sigma_2$ (which is tantamount to varying $\Gamma$), it reduces to a local one: 
\beq 
\theta_{\rm corr} := \theta +\delta\ell_{\rm corr} \heq - \extd C\,.\label{shifttheta}
\eeq 
That is to say, we have a shifted presymplectic potential $\theta_{\rm corr}$ on $\Gamma$ which is such that it vanishes up to a corner term on field configurations satisfying the boundary conditions. Since, however, $\omega_{\rm corr}:=\delta\theta_{\rm corr}=\delta\theta=\restr{\omega}{\Gamma}$, we can already see that this will not affect the symplectic flux through the interface and we shall discuss this further below. A sufficient condition for equation~\eqref{shifttheta} to hold is that
\beq 
\theta_{\rm corr}\approx -\extd C+Y_a\delta X^a\label{shifttheta1}
\eeq 
for some suitable set of densities $Y_a$ constructed locally out of the $\phi^a$ and the pullbacks of the derivatives of $\Phi^a$ to $\Gamma$, as well as of $U$ and its derivatives.

The conditions \eqref{shifttheta} and \eqref{shifttheta1} are similar to the sufficient condition for a well-defined subregion variational principle put forth in \cite{Harlow:2019yfa}, however, are also somewhat different. Here we arrive at them from a global variational principle using post-selection, and $\ell_{\rm corr}$ is a boundary Lagrangian whose role is \emph{not} to impose the desired boundary conditions, which are already satisfied in \eqref{shifttheta}. $\ell_{\rm corr}$ rather assumes two different roles for the subregion $M$ that are otherwise played by the complement $\bar M$ and which are necessary in order to render the subregion variational problem well-defined and consistent with the global one: 
\begin{itemize}
    \item[(i)] Varying the fields in $M$ generates a presymplectic potential $\theta$  on $\Gamma$ which is cancelled by the corresponding contribution $-\theta$  from varying the fields in $\bar M$. This is important for the global variational problem as $\theta$  does not in general vanish (or reduce to a corner term) individually when the boundary conditions are fulfilled. From the perspective of $M$ the interface Lagrangian $\ell_{\rm corr}$ mimics the effect of the complementary region $\bar M$  in such a way that $\theta$ gets cancelled up to a corner term once the boundary conditions are satisfied. This enables us to split the global variational problem into two independent ones. As we shall see below in theories admitting Darboux coordinates, $\ell_{\rm corr}$ constitutes a generating function of a canonical transformation that implements a change of polarization on field space. This in turn ensures that the boundary conditions $X^a=X^a_0$ become truly holonomic constraints that only depend on field configurations, but not their derivatives.\footnote{Note that, since we are considering canonical transformations of $\delta \theta$, the relevant field space here is the space of fields pulled-back to $\Gamma$. As a result, constraints involving normal derivatives of the fields but no intrinsic derivative on $\Gamma$ are holomomic.} 
    
    \item[(ii)] In some topological theories, such as Chern-Simons theory (see Sec.~\ref{subsec:CS}), the action $S_M$ associated with a finite subregion may not be gauge-invariant, with its non-invariance coming from a boundary piece. This boundary piece has to be cancelled by the complement $S_{\bar M}$ in order to ensure invariance of the global action in the bulk of $M\cup\bar M$. For such theories, $\ell_{\rm corr}$ will also mimic the role of the complement $\bar M$ by cancelling any non-invariance of $S_M$ and thereby establishing gauge-invariance of the subregion variational principle. In other words, $\ell_{\rm corr}$ is only gauge-invariant if $S_M$ is.
\end{itemize}

So far we have split the global variational problem into independent subregion problems \emph{after} globally imposing the boundary conditions. The ensuing variational problem for subregion $M$ is defined by the action $\tilde S_M$ on the post-selected field configuration space $\restr{\cF_M}{X=X_0}$, containing all $C^q$ field configurations in $M$ which satisfy the boundary conditions $X=X_0$ on $\Gamma$. As in \cite{Carlip:1995cd,Margalef-Bentabol:2020teu} the boundary conditions are thus directly built into the field configuration space. Clearly, we can equivalently formulate this subregion variational problem on the larger space $\cF_M$ of all $C^q$ field configurations in $M$ if we dynamically impose the boundary conditions through Lagrange multipliers. On $\cF_M$ the variational principle for the subregion is therefore defined by 
\beq
\tilde S_{M\cup\Gamma}:=\tilde S_M+\int_\Gamma \lambda_a(X^a-X^a_0)=S_M+S_{\rm ps}\,,
\eeq
where
\beq 
S_{\rm ps}:=\int_\Gamma\left(\ell_{\rm corr}+\lambda_a(X^a-X_0^a)\right)\label{Ssp}
\eeq 
is the post-selection boundary action. Condition \eqref{shifttheta1} implements a polarization of field configuration space in which the $X^a$ are configuration degrees of freedom. This guarantees that we can consistently impose the boundary conditions $X^a=X^a_0$ as  genuine holonomic constraints. In problems with holonomic constraints  the Lagrange multipliers have physical significance: they can be solved for as the forces imposing the constraints and this solution can be reinserted into the action, defining the variational principle for the remaining degrees of freedom. In our case, \eqref{shifttheta1} entails that the constraint forces read $\lambda_a=-Y_a$ which yields the final subregion action on $\cF_M$
\beq 
S_{M\cup\Gamma}:=S_M+S_\Gamma\,,
\eeq
with boundary action
\beq 
S_\Gamma :=\restr{S_{\rm ps}}{\lambda=-Y}=\int_\Gamma\left(\ell_{\rm corr}-Y_a(X^a-X_0^a)\right)\,.\label{bdryact}
\eeq 
As we will illustrate below, this procedure will provide a systematic algorithm for generating boundary actions, encompassing various well-known ones in the literature as well as producing new ones. This will also elucidate the appearance of edge modes in boundary actions.

\subsection{Local Darboux coordinates, appearance of edge modes and non-uniqueness of the boundary action}

In this subsection, we shall render the above general construction more concrete, using the edge mode dressed observables and field space forms of Secs.~\ref{sec_gaugesplit} and~\ref{sec:geometry}. We shall also discuss the non-uniqueness of the boundary action implementing a specific boundary value problem.

\subsubsection{General algorithm}\label{ssec_genalg}

In order to render this construction more explicit, let us assume we are given a theory admitting the use of Darboux coordinates on $\cF$, so that the presymplectic potential reads
\beq\label{Thetaassump}
\Theta = \Pi_a\delta\Phi^a\,,
\eeq 
for some density $\Pi_a$ and where $\Phi^a$ denote the fields in the Lagrangian. Its pullback to $\Gamma$ will be denoted by
\beq
\theta=\pi_a\delta\phi^a.\label{darbouxtheta}
\eeq
For example, this assumption is true in all theories whose Lagrangian depends on the fields $\Phi^a$ and their first, but not higher order derivatives \cite{Lee:1990nz};  this is the case e.g.\ in Maxwell, Chern-Simons and Yang-Mills theory as discussed later. In this case, $\Pi_a$ typically (but not always) depends on only the first derivatives of $\Phi_a$. This assumption can, however, also be realized in theories with Lagrangians featuring higher orders of derivatives, which might require either a relabeling of field variables or taking care of boundary terms.\footnote{For example, also the presymplectic potential of general relativity can be put into the form of equation~\eqref{Thetaassump} by taking care of boundary terms appearing in the Gauss-Codazzi equations \cite{Lee:1990nz,Freidel:2020xyx}. }

When dealing with a gauge theory, it will be convenient to split the field degrees of freedom $\phi^a$ and the momenta $\pi_a$ on $\Gamma$ into pure gauge and \emph{radiative} parts, as explained in Sec.~\ref{sec_gaugesplit}:
\beq 
\phi^a= \phi^a_{\rm rad} + \phi^a_{\rm gauge} \qquad \mathrm{and} \qquad \pi_a= \pi_a^{\rm rad} + \pi_a^{\rm gauge} 
\,.\label{split1}
\eeq 
Owing to the symmetry induced redundancy, this can always be done in many different ways in the global covariant phase space, but it is uniquely defined once we have specified a reference frame.

Next, we shall make one further assumption, namely that we can use the split into pure gauge and radiative degrees of freedom to decompose the presymplectic potential on solutions to the bulk equations of motion on $\Gamma$ as
\beq
\theta = \theta_{\rm rad} + \theta_{\rm gauge} \,, \qquad  \theta_{\rm rad} = \pi_a^{\rm rad} \delta \phi^a_{\rm rad}\qquad \theta_{\rm gauge} \approx - \extd\theta_\partial - \delta \ell_\partial\,,\label{assump1}
\eeq
where $\theta_\partial$ and $\ell_\partial$ have already been introduced in equation \eqref{eq:theta_bdy}. The extra assumption we are making is that equation \eqref{eq:omega_bdy} (from which \eqref{eq:theta_bdy} follows) can be extended off-shell of the boundary conditions.  In other words, the non-invariant part of $\theta$ is in the kernel of $\delta \extd$ on-shell of the equations of motion. This assumption holds true in many gauge theories and we shall illustrate it in Maxwell, Chern-Simons and Yang-Mills theory below. The assumption is essentially that we can find local Darboux coordinates for the degrees of freedom on $\Gamma$ that split into gauge-invariant and -variant pairs, where the `momenta' of the latter pair are simply the constraints of the theory which vanish on-shell. Such a split is related to a so-called \emph{constraint abelianization} \cite{Henneaux:1992ig}. 

The gauge-invariant pairs $(\phi^a_{\rm rad},\pi_a^{\rm rad})$ constitute the set of degrees of freedom on which boundary conditions will be imposed. Typically, one imposes boundary conditions either on the fields (Dirichlet), their normal derivatives (Neumann), or linear combinations (Robin). Since the latter contain the former two as special cases, we shall directly formulate the variational problem for Robin boundary conditions. In fact, we shall be even slightly more general and permit the nature of the boundary conditions to change as $\partial\Sigma$ evolves through $\Gamma$, so that we encompass \emph{mixed} boundary conditions, i.e.\ time-dependent Robin boundary conditions.  To this end, it will suffice to consider linear combinations of `position' and `momentum' variables, which for simplicity we write in matrix notation 
\beq
\left(\begin{array}{c}\mathbf{X} \\\mathbf{y}\end{array}\right)=\mathbf{M}\left(\begin{array}{c}\mathbf{\phi}_{\rm rad} \\\mathbf{p}^{\rm rad}\end{array}\right)\,,\qquad\qquad\mathbf{M}=\left(\begin{array}{cc}\mathbf{a} & \mathbf{b} \\\mathbf{c} & \mathbf{d}\end{array}\right)\,,\label{lincomb}
\eeq
where $\mathbf{a},\mathbf{b},\mathbf{c},\mathbf{d}$ are real $n \times n$-matrix background fields and $\mathbf{p}^{\rm rad}$ is the $n$-vector with scalar components $p^{\rm rad}_a$, defined by $\pi_a^{\rm rad}=p_a^{\rm rad}\epsilon_\Gamma$, where $\epsilon_\Gamma$ is the volume form on $\Gamma$, and similarly $\mathbf{y}$ is the $n$-vector whose components are defined by a new `momentum' density $Y_a=y_a \epsilon_\Gamma$. In order for this to define a symplectic transformation, we shall require
\beq
\omega_{\rm rad}=\,\delta\pi^{\rm rad} .\delta \phi_{\rm rad} = \delta\mathbf{Y} . \delta\mathbf{X}\,,\label{symptr}
\eeq
where $.$ denotes the dot product between $n$-vectors, which is equivalent to 
\bes
\mathbf{d}^\top\mathbf{a}-\mathbf{b}^\top\mathbf{c}&=\mathbf{1}\,, \qquad\qquad \mathbf{c}^\top\mathbf{a} &= \mathbf{a}^\top\mathbf{c} \,,  \nn\\
\mathbf{a}^\top\mathbf{d}-\mathbf{c}^\top\mathbf{b}&=\mathbf{1}\,, \qquad\qquad \mathbf{d}^\top\mathbf{b} &= \mathbf{b}^\top\mathbf{d}  \,, \label{Sp}
\ees
so that $\mathbf{M}\in\rm{Sp}(2n,\mathbb{R})$. 

We are now in a position to determine the shape of the interface action $\ell_{\rm corr}$ ensuring consistency of the subregion variational problem from the condition in \eqref{shifttheta1}  
\bes
\delta\ell_{\rm corr} &\approx& \mathbf{Y}.\delta\mathbf{X}-\theta-\extd C\nn\\
&\approx& \mathbf{Y}.\delta\mathbf{X}-\pi^{\rm rad} .\delta\phi_{\rm rad}+\extd(\theta_\partial -C)\label{ellcorr0} + \delta \ell_\partial \\
&=&\frac{1}{2}\delta\left(\mathbf{Y}.\mathbf{X}-\pi^{\rm rad}. \phi_{\rm rad} + 2 \ell_\partial \right) +\extd(\theta_\partial -C)\,,\nn
\ees
where we have invoked \eqref{Sp} in arriving at the last line. To solve this consistency relation, we henceforth define $C$ and $\ell_{\rm corr}$ on $\cF_M$ by:\footnote{It is implicitly assumed here that quantities such as $\theta_\partial$ and $\ell_\partial$ can be naturally extended from forms on $\cS_M$ to forms on $\cF_M$.}
\beq
C:= \theta_\partial \,, \qquad \ell_{\rm corr}:= \ell_\partial + \frac{1}{2}\left(\mathbf{Y}.\mathbf{X}-\pi^{\rm rad} .\phi_{\rm rad}  \right)\,.\label{ellcorr1}
\eeq

The above observations permit us to shed new light on the so-called \emph{extended presympletic potential}, as considered, for instance, in \cite{Donnelly:2016auv,Geiller:2019bti}:\footnote{The field-space exact term $\delta \ell_\partial$ was not included in those works, as it happens to be irrelevant in many situations of interest; we keep it for full generality.}
\beq
\theta_e:=\theta+\extd \theta_\partial  + \delta \ell_\partial \,.
\eeq
The corner term $\theta_\partial$ has been added to $\theta$ in \cite{Donnelly:2016auv,Geiller:2019bti} to render the potential gauge-invariant even for gauge transformations with support on $\Gamma$, and independently of any choice of boundary conditions. 
Noting the decomposition \eqref{assump1}, on-shell the extended potential becomes the pure radiative contribution
\beq
\theta_e\approx \pi^{\rm rad}. \delta\phi_{\rm rad} =\theta_{\rm rad}\,.\label{extpot}
\eeq
This is a consequence of our variable split \eqref{split1} which renders the meaning of the extended potential especially transparent. Similarly, the extended symplectic current $\omega_e:=\delta\theta_e$ of \cite{Donnelly:2016auv,Geiller:2019bti} then coincides on-shell with the radiative contribution constructed in equation~\eqref{omegaradconstruct},
\beq 
\omega_e\approx\omega_{\rm rad}\,.
\eeq
It is thus the extended symplectic structure which ultimately is post-selected to vanish boundary on-shell in equation~\eqref{eq:omega_gamma_eq}, $\omega_{\rm rad}\heq0$, so as to eliminate gauge-invariant symplectic flux through the interface $\Gamma$ (see Sec.~\ref{sec:geometry}).

The boundary condition $\mathbf{X}=\mathbf{X}_0$ in \eqref{Ssp} is clearly the integrated version of the differential constraint $\delta\mathbf{X}=0$, which thus is holonomic. Holonomic constraints can be written as functions of only the configuration degrees of freedom. This is not the case in the old variables $(\phi_{\rm rad},\pi^{\rm rad})$, unless $\mathbf{b}=\mathbf{c}=0$, but clearly in the new boundary variables $(\mathbf{X},\mathbf{Y})$. The interface action $\ell_{\rm corr}$ thus implements a change of polarization of the field configuration space, as can already be inferred from \eqref{shifttheta1}. Indeed, $\ell_{\rm corr} - \ell_{\rm \partial}$ can be interpreted as a generating function of a canonical transformation in multiple ways. For example, if $\mathbf{b}$ is invertible and $\ell_\partial = 0$, one can use \eqref{lincomb} to rewrite $\ell_{\rm corr}$ as a type I generating function of a canonical transformation 
\beq
\ell_{\rm corr} =\frac{1}{2}\left( \mathbf{X}.(\mathbf{d}\mathbf{b}^{-1}\mathbf{X})-2(\mathbf{b}^{-1}\mathbf{X}).\phi_{\rm rad}+(\mathbf{b}^{-1}\mathbf{a}\phi_{\rm rad}).\phi_{\rm rad} \right) \epsilon_\Gamma \,, \nn
\eeq
which can be checked to satisfy the functional derivative relations 
\beq
\frac{\delta\ell_{\rm corr} (x)}{\delta\phi_{\rm rad}^a (y)}=-\pi_a^{\rm rad}(x) \delta^{(d-1)}(x,y)\,,\qquad\qquad\frac{\delta\ell_{\rm corr} (x)}{\delta X^a (y)}=Y_a(x)  \delta^{(d-1)}(x,y) \,.\nn
\eeq
Similarly, when $\mathbf{d}$ is invertible, one can write $\ell_{\rm corr}$ in terms of a type II generating function, etc. 

Finally, we can now also write the boundary action for the subregion variational principle on $\cF_M$ in \eqref{bdryact} explicitly 
\beq
S_\Gamma = \int_\Gamma \ell \,, \qquad \ell :=  \ell_{\rm corr}-\mathbf{Y}.(\mathbf{X}-\mathbf{X}_0) = \mathbf{Y}.\mathbf{X}_0-\frac{1}{2}\left(\mathbf{X}.\mathbf{Y}+\pi^{\rm rad}.\phi_{\rm rad} \right) + \ell_\partial\,. \label{bdryaction}
\eeq
In more detail, we find that
\beq
\ell = \ell_\partial + (\mathbf{a}\phi_{\rm rad} + \mathbf{b}\pi^{\rm rad}).X_0 - \frac{1}{2} \phi_{\rm rad}.(\mathbf{a}^\top \mathbf{c} \phi_{\rm rad}) - \frac{1}{2} \pi^{\rm rad}.(\mathbf{b}^\top \mathbf{d} \pi^{\rm rad}) - \pi^{\rm rad}.(\mathbf{d}^\top \mathbf{a} \phi_{\rm rad})  \,.
\eeq
When $\ell_\partial$ vanishes, this action is gauge-invariant by construction. This is for instance the case in Yang-Mills theories. By contrast, in the case that $S_M$ is not gauge-invariant, as for example in Chern-Simons theory, one still has the freedom to add the corresponding non-invariant piece on $\Gamma$ to $\ell_{\rm corr}$ (via $\ell_\partial$) as mentioned under (ii) in Sec.~\ref{ssec_globsplit}. Such a requirement may be imposed \emph{in addition to}  our main on-shell consistency condition \eqref{ellcorr0}. 

This concludes the general procedure for post-selecting on the gauge-invariant variables $\mathbf{X}$.

\subsubsection{Non-unicity of the boundary action}

The boundary action \eqref{bdryaction} was derived from a particular choice of polarization $(\mathbf{X}, \mathbf{Y})$ for the symplectic current $\omega_{\rm rad} = \delta \pi^{\rm rad} . \delta \phi_{\rm rad} = \delta \mathbf{Y} . \delta \mathbf{X}$. Since we are imposing boundary conditions on $\mathbf{X}$, any other polarization $(\mathbf{X}', \mathbf{Y}')$ such that $\mathbf{X}'$ is related to $\mathbf{X}$ by an invertible transformation is also admissible. More precisely, any two such polarizations are related by an $\mathrm{Sp}(2 n, \mathbb{R})$ transformation of the form
\beq
\left(\begin{array}{c}\mathbf{X}' \\\mathbf{y}'\end{array}\right)=\mathbf{N}\left(\begin{array}{c}\mathbf{X} \\\mathbf{y}\end{array}\right)\,,\qquad\qquad\mathbf{N}=\left(\begin{array}{cc}\mathbf{a} & 0 \\\mathbf{c} & \mathbf{d}\end{array}\right)\,,\label{ambiguity}
\eeq
where $\mathbf{d}^\top\mathbf{a}=\mathbf{1}$ and $\mathbf{c}^\top\mathbf{a} = \mathbf{a}^\top\mathbf{c}$. It is convenient to parametrize such transformations by the invertible matrix $\mathbf{a}$ and the symmetric matrix $\mathbf{s}= \mathbf{a}^\top\mathbf{c}$. 
They generate the subgroup of $\mathrm{Sp}(2 n , \mathbb{R})$
\beq
\cN = \left\lbrace \left(\begin{array}{cc}\mathbf{a} & 0 \\ (\mathbf{a}^\top)^{-1} \mathbf{s} & (\mathbf{a}^\top)^{-1}\end{array}\right) \vert \mathbf{a} \in \mathrm{GL}(n , \mathbb{R})\,, \mathbf{s} \in \mathrm{Sym}(n , \mathbb{R}) \right\rbrace\,.
\eeq
We can view this set as the matrix product $\cN = \cD \cT$, where 
\beq
\cD = \left\lbrace \left(\begin{array}{cc}\mathbf{a} & 0 \\ 0 & (\mathbf{a}^\top)^{-1}\end{array}\right) \vert \mathbf{a} \in \mathrm{GL}(n , \mathbb{R})\right\rbrace \simeq \mathrm{GL}(n , \mathbb{R})
\eeq
and
\beq
\cT = \left\lbrace \left(\begin{array}{cc}\mathbf{1} & 0\\ \mathbf{s}  & \mathbf{1}\end{array}\right) \vert \mathbf{s} \in \mathrm{Sym}(n, \mathbb{R})\right\rbrace \simeq \mathrm{Sym}(n, \mathbb{R})\,.
\eeq
As a group, $\cN$ is in fact a semidirect product of $\cD$ and $\cT$. 
Indeed, introducing the notation $\mathbf{N}(\mathbf{a}, \mathbf{s}):=\left(\begin{array}{cc}\mathbf{a} & 0 \\ (\mathbf{a}^\top)^{-1} \mathbf{s} & (\mathbf{a}^\top)^{-1}\end{array}\right)$, we are led to the semidirect group multiplication law:
\beq
\mathbf{N}(\mathbf{a}_1, \mathbf{s}_1)\mathbf{N}(\mathbf{a}_2, \mathbf{s}_2) = \mathbf{N}(\mathbf{a}_1 \mathbf{a}_2, \mathbf{a}_2^\top\mathbf{s}_1\mathbf{a}_2 + \mathbf{s}_2)\,.
\eeq
We therefore conclude that
\beq
\cN \simeq \mathrm{GL}(n , \mathbb{R}) \ltimes \mathrm{Sym}(n , \mathbb{R})\,.
\eeq
 
Among the admissible changes of polarization encoded in $\cN$, which transformations affect the functional form of the boundary action $\ell$? From the construction of the previous subsection, and in particular equation \eqref{ellcorr0}, it is clear that any canonical transformation that preserves the flux $\cE := \mathbf{Y} \delta \mathbf{X}$ will also preserve $\delta\ell$ (and thereby $\ell$, up to an irrelevant additive ambiguity). Such transformations generate the (normal) stabilizer subgroup $\cD \subset \cN$.
Ambiguities in the definition of $\ell$ are therefore parametrized by the Abelian subgroup $\cT$ (which is not normal). Acting with such a transformation on the polarization $(\mathbf{X},\mathbf{y})$ yields new coordinates
\beq
\mathbf{X}' = \mathbf{X} \,, \qquad \mathbf{y}' = \mathbf{s}\mathbf{X}+ \mathbf{y}\,,
\eeq
where $\mathbf{s}$ is a real symmetric matrix. By application of \eqref{bdryaction}, the boundary Lagrangian produced by our algorithm in the new coordinates is
\begin{align}
\ell' &= \mathbf{Y}'.\mathbf{X}_0-\frac{1}{2}\left(\mathbf{X}.\mathbf{Y}'+\pi^{\rm rad}.\phi_{\rm rad}\right) + \ell_\partial = \ell + (\mathbf{s} \mathbf{X}).\mathbf{X}_0 - \frac{1}{2} \mathbf{X}.(\mathbf{s} \mathbf{X})  \\
&= \ell - \frac{1}{2} (\mathbf{X} - \mathbf{X}_0). \left(\mathbf{s}(\mathbf{X} - \mathbf{X}_0)\right) + \frac{1}{2} \mathbf{X}_0 . \left(\mathbf{s} \mathbf{X}_0 \right)\,,
\end{align}
where we have used the fact that $\mathbf{s}$ is symmetric in the last line.

We see that $\ell'$ only differs from $\ell$ by:
\begin{itemize}
\item a non-trivial quadratic term $- \frac{1}{2} (\mathbf{X} - \mathbf{X}_0). \left(\mathbf{s}(\mathbf{X} - \mathbf{X}_0)\right)$;
\item and a field-independent additive contribution $\frac{1}{2} \mathbf{X}_0 . \left(\mathbf{s} \mathbf{X}_0\right)$.
\end{itemize}
The second term can be ignored since, by construction, $\ell$ is only defined up to such additive contributions. By contrast, the first term affects the functional form of $\delta \ell$ and is \emph{a priori} non-trivial. However, because it is quadratic in $(\mathbf{X} - \mathbf{X}_0)$, we have:
\beq
\delta \ell' = \delta \ell - (\mathbf{X} - \mathbf{X}_0). \left(\mathbf{s}\delta \mathbf{X}\right)\,,
\eeq
which implies that $\delta \ell$ and $\delta \ell'$ coincide on-shell of the boundary conditions. As a result, this ambiguity only affects the functional form of the boundary action, not the presymplectic structure that results from it.\footnote{In fact, this ambiguity in $\delta \ell$ is of the general form $(X^i - X^i_0) \delta Z_i$, and such contributions have been implicitly discarded in the definition \eqref{shifttheta1}.} It is then clear that we are not changing anything of substance when we change polarization. Likewise, we will show in an upcoming paper that work functionals can be consistently associated to the boundary Lagrangian $\ell$, and are not affected by such quadratic (or higher order) ambiguities. 

\subsection{Summary of the algorithm}\label{sec:summary_algo}

In summary, given a set of boundary conditions ${\bf X}= {\bf X}_0$ to be imposed dynamically, a suitable boundary Lagrangian $(0, d-1)$-form $\ell$ can be obtained by implementation of the following steps:
\begin{enumerate}
\item Decompose the presymplectic potential on $\Gamma$ into radiative and gauge contributions, relative to a frame $U$: $\theta = \theta_{\rm rad} + \theta_{\rm gauge}$.

\item Find local Darboux coordinates for the radiative degrees of freedom, so that $\theta_{\rm rad} = \pi_a^{\rm rad} \delta \phi^a_{\rm rad}$, where $\pi_a^{\rm rad} = p_a^{\rm rad} \epsilon_\Gamma$ is the momentum density dual to the frame-dressed field $\phi^a_{\rm rad}= (U^{-1}\acts \phi)^a$.

\item Find a $(1,d-2)$-form $\theta_\partial$ and a $(0, d-1)$-form $\ell_\partial$ such that, on-shell: $\theta_{\rm gauge} + \extd \theta_{\partial} + \delta \ell_{\partial} \approx 0$.

\item Find a linear transformation to new Darboux coordinates $({\bf Y},{\bf X})$ such that:\footnote{In particular, $X^a$ must be linear in $\phi^a_{\rm rad}$ and $p^{\rm rad}_a$, as we have assumed in our construction.} $\delta \theta_{\rm rad}=\delta {\bf Y}. \delta{\bf X}$.

\item Define the boundary Lagrangian as:
\beq
\ell = \mathbf{Y}.\mathbf{X}_0-\frac{1}{2}\left(\mathbf{X}.\mathbf{Y}+\pi^{\rm rad}.\phi_{\rm rad} \right) + \ell_\partial\,.
\eeq
\end{enumerate}
Then, as we have explained in the main text, the stationary points of the action $S_{M\cup \Gamma} = \int_M L + \int_\Gamma \ell$ in the configuration space $\cF_M$, span the solution space $\cS_M^{X_0}$.

\section{Illustration of splitting post-selection in example field theories}\label{sec:examples}

We showcase the splitting post-selection algorithm for generating boundary actions that impose Robin, Dirichlet and Neumann boundary conditions in the examples of scalar field, Maxwell, Abelian Chern-Simons, and Yang-Mills theories. The latter three examples will more specifically illustrate the appearance and role of edge modes on the boundary $\Gamma$. We also emphasize that, given that the matrix $\mathbf{M}$ implementing the change of polarization in \eqref{lincomb} is a background field, it allows us in principle to continuously interpolate between different types of boundary conditions along $\Gamma$. For clarity of the exposition, we will not consider such mixed boundary conditions in the following examples.

We emphasize that throughout this section, we consider general Lorentzian spacetimes, which in particular can be curved and, except in the Chern-Simons theory case, of arbitrary dimension $d\geq2$.

\subsection{Scalar field theory}

Varying the Lagrangian 
\beq
L = -\left( \frac{1}{2} \nabla_\mu \Phi \nabla^\mu \Phi + V(\Phi) \right) \epsilon
\eeq 
of the bulk scalar field theory action $S_M=\int_ML[\Phi]$ yields
\beq
\delta L = \left( \nabla_\mu \nabla^\mu \Phi - V'(\Phi) \right) \epsilon \delta \Phi + \extd \Theta\,,
\eeq 
where $\epsilon$ is the spacetime volume form and the presymplectic potential reads
\beq
\Theta = \theta \cdot \epsilon\,, \quad \mathrm{with} \quad \theta^\mu = - \nabla^\mu \Phi \delta\Phi \,.
\eeq
The pullback of $\Theta$ to $\Gamma$ takes the form
\beq
\theta=\restr{\Theta}{\Gamma} = -\partial_n \phi\delta\phi\epsilon_\Gamma\,,\label{phitheta}
\eeq
where $\phi:=\restr{\Phi}{\Gamma}$, $\partial_n\phi:=\restr{n_\mu\nabla^\mu\Phi}{\Gamma}$ and $n^\mu$ is the outward pointing normal to $\Gamma$. Since there is no gauge symmetry in this model, we directly have
\beq
\phi_{\rm rad}= \phi\,,\qquad\qquad p^{\rm rad}= -\partial_n \phi\,, \qquad\qquad  \theta_{\rm rad}= \theta \,.
\eeq
The coefficient matrices defining $X$ and $y$ in \eqref{lincomb} are here real functions ${\bf a},{\bf b},{\bf c},{\bf d}\in\mathbb{R}$ satisfying ${\bf a}{\bf d}-{\bf b}{\bf c}=1$, so that $\mathbf{M}$ is an $\rm{SL}(2,\mathbb{R})$ matrix field. The generating function of this canonical transformation reads $\ell_{\rm corr}=\frac{1}{2}\left(YX-\pi^{\rm rad}\phi_{\rm rad} \right)$, where $\pi^{\rm rad}=p^{\rm rad} \epsilon_\Gamma$. 

Equation \eqref{bdryaction} implies that the variational principle for subregion $M$ subject to \textbf{Robin boundary conditions} $X=X_0$ is thus given by 
$
S_{M\cup\Gamma}=S_M+S_\Gamma=\int_M L[\Phi]+\int_\Gamma\ell [\phi]
$
with boundary Lagrangian
\beq
\ell =\left(yX_0-\frac{1}{2}\left(yX-\phi \partial_n \phi\right)\right)\epsilon_\Gamma\,.\label{scalarRobin}
\eeq
Indeed, varying this action yields the shifted symplectic potential $\theta_{\rm corr}=\theta+\delta\ell_{\rm corr}=Y\delta X$ and 
\beq
\delta S_{M\cup\Gamma}=\int_M \left( \nabla_\mu \nabla^\mu \Phi - V'(\Phi) \right) \epsilon \delta \Phi+\int_{\Sigma_2-\Sigma_1}\Theta-\int_\Gamma\left(X-X_0\right)\epsilon_\Gamma\delta y\,,
\eeq
thus dynamically imposing the desired boundary conditions as boundary equations of motion.

In the special case that ${\bf a}={\bf d}=1$ and ${\bf b}=0$, we post-select on $X=\phi$, giving \textbf{Dirichlet boundary conditions} $\phi=\phi_0$, therefore a vanishing generating function $\ell_{\rm corr}=0$, and a boundary  Lagrangian (setting for simplicity also ${\bf c}=0$)
\beq
\ell = \partial_n \phi\left(\phi-\phi_0\right)\epsilon_\Gamma\,.
\eeq
The dual case ${\bf a}=0$ and ${\bf b}=-1=-{\bf c}$ amounts to post-selection on the normal derivative $X= \partial_n \phi$ and thus generates \textbf{Neumann boundary conditions} $\partial_n \phi=N_0$ which are imposed by the simple boundary action (setting ${\bf d}=0$)
\beq
\ell =\phi N_0 \epsilon_\Gamma\,.
\eeq
This leads to a change of polarization compared to \eqref{phitheta}
\beq
\theta_{\rm corr}=\phi \delta \partial_n \phi\epsilon_\Gamma\,.
\eeq

\subsection{Maxwell theory}\label{subsec:maxwell}

While Maxwell theory is the restriction of the Yang-Mills theory construction in Sec.~\ref{ssec_YM} to the special $\rm{U}(1)$ case, we illustrate it here for clarity and its standalone physical interest.
We start from the bulk Lagrangian in $d$ spacetime dimensions:
\beq
L[A] = - \frac{1}{2} F \wedge \star F\,,
\eeq
where $F := \extd A$ is the curvature of a $\U(1)$ connection. The variation of $L$ gives
\beq
\delta L = - \extd \delta A \wedge \star F = - \extd (\delta A \wedge \star F ) - \delta A \wedge \extd \star F \,,
\eeq
from which we recover the bulk equation of motion, $\extd \star F = 0$, and the presymplectic potential
\beq
\Theta = - \delta A \wedge \star F \,,
\eeq
so $A$ and $\star F$ are conjugated.  

On the boundary $\Gamma$, we invoke the split of $A$ into a radiative and a pure gauge part
\beq
\restr{A}{\Gamma}=A_{\rm rad}+\extd\varphi\,,\label{Maxsplit}
\eeq
where $A_{\rm rad}$ is defined relative to the edge mode frame in equations~\eqref{genArad} and~\eqref{1stArad}, which splits the presymplectic potential on $\Gamma$ into $\theta = \theta_{\rm rad} +\theta_{\rm gauge}$, where
\beq
 \theta_{\rm rad} = -\delta A_{\rm rad}\wedge \star F\,, \qquad \theta_{\rm gauge} = \left(\extd\star F\right)\delta\varphi-\extd\left(\star F\delta\varphi\right) \approx -\extd\left(\star F\delta\varphi\right) \,.\label{Maxtheta}
\eeq
In order to formulate the decomposition \eqref{assump1} at the form level, we have to invoke the Hodge star operator $*$ on $\Gamma$. Using that $**\mathrm{v} = (-1)^{(d-1-q)q+1}\mathrm{v}$ for a $q$-form $\mathrm v$ on a $(d-1)$-dimensional Lorentzian manifold, we can write the radiative piece of the presymplectic potential as 
\bes
\theta_{\rm rad}=-\delta A_{\rm rad}\wedge\star F &=& (-1)^d\delta A_{\rm rad}\wedge **\star F = (-1)^d\langle \delta A_{\rm rad},*\star F\rangle\epsilon_\Gamma\nn\\
&=&(-1)^d\langle*\star F,\delta A_{\rm rad}\rangle\epsilon_\Gamma = (-1)^d*\star F\wedge \delta *A_{\rm rad}\,,\label{boundaryHodge}
\ees 
where $\langle\cdot,\cdot\rangle$ denotes the inner product of $1$-forms on $\Gamma$. Accordingly, the assumption in \eqref{assump1} is realized with  $\theta_\partial = \star F \delta \vphi$, $\ell_\partial = 0$, and gauge-invariant momentum one-form $\pi^{\rm rad} = (-1)^{d} \ast \star F$.\footnote{We emphasize that in all expressions defined on $\Gamma$, $\star F$ denotes the pullback $\restr{\star F}{\Gamma}$ which for simplicity of notation we shall not write explicitly. Moreover, we slightly depart from the notations introduced in \eqref{assump1}: we distribute the volume form $\epsilon_\Gamma$ differently, so that $\pi^{\rm rad}$ is a one-form rather than a density. This is for mere convenience due to our reliance on the form language, which otherwise does not affect the applicability of the formalism laid out in Secs.~\ref{sec:geometry} and \ref{sec:splitting}.} Here, $\varphi$ is the phase of a ${\rm U}(1)$ edge reference frame $U(x)= e^{{\rm i} \vphi(x)}$, see Sec.~\ref{sec_gaugesplit}. Clearly, it transforms as $\delta_\alpha\varphi := \fX_\alpha \cdot \delta \vphi = \alpha$ under the infinitesimal gauge transformation $\fX_\alpha$ acting on $\cF$ (see equation~\eqref{eq:gauge_transfo_maxwell}). The momentum density conjugated to the edge mode is a constraint: the bulk equation of motion pulled back to $\Gamma$. Likewise, we find that the contraction of $\Theta_\Sigma = \int_\Sigma \Theta + \int_{\partial \Sigma} \theta_\partial$ with $\fX_\alpha$ is an on-shell constraint, as was already anticipated after equation~\eqref{eq:gauge_inv_Theta_sigma}:
\beq\label{eq:constraint_from_theta}
C[\alpha]:= \fX_\alpha \cdot \Theta_\Sigma = - \int_\Sigma \alpha \extd \star F \approx 0 \,.
\eeq

 The splitting of the presymplectic current into $\omega = \omega_{\rm rad}+ \omega_{\rm gauge}$ can be deduced from \eqref{Maxtheta} by acting with the exterior derivative $\delta$, leading to:
\beq
\omega_{\rm rad} = \delta A_{\rm rad} \wedge \delta \star F \,, \qquad \omega_{\rm gauge} = \delta (\extd \star F) \delta \vphi - \extd (\delta \star F \delta \vphi) \approx - \extd (\delta \star F \delta \vphi) \,. \label{eq:Maxwell_omega-gauge}
\eeq
Following our construction of Sec.~\ref{sec:geometry}, we obtain a presymplectic structure which is conserved on-shell, and gauge-invariant independently of any choice of boundary conditions (see equation~\eqref{sympinvgauge} which here can be directly checked): 
\beq
\Omega_M =\int_\Sigma \omega + \int_{\partial \Sigma} \delta \theta_\partial =  \int_\Sigma\delta A\wedge\delta\star F+\int_{\partial\Sigma}\delta\star F\delta\varphi\,.\label{MaxOmega}
\eeq
This expression coincides with the extended presymplectic structure used in \cite{Geiller:2019bti}, but is here a result of post-selection. As an aside, we note that considering the extended potential as in \cite{Geiller:2019bti}, but using our variable split \eqref{Maxsplit}, yields the purely radiative piece
\beq
\theta_e=\theta+\extd C= -\delta A_{\rm rad}\wedge\star F+(\extd\star F)\delta\varphi\approx -\delta A_{\rm rad}\wedge\star F = \theta_{\rm rad}\,,
\eeq
since the gauge part $\theta_{\rm gauge}$ of $\theta$ is cancelled by $C$ on shell, exemplifying  \eqref{extpot}. 

The split in \eqref{Maxsplit} permits a transparent understanding of the distinction between gauge transformations and symmetries, as originally proposed in \cite{Donnelly:2016auv} (see also \cite{Geiller:2019bti}). Gauge transformations act as 
\beq
\delta_\alpha A=\extd\alpha\,,\qquad\qquad \delta_\alpha A_{\rm rad}=0\,,\qquad\qquad \delta\varphi=\alpha\,,\label{Maxgauge}
\eeq
while symmetry transformations can be understood as leaving the vector potential on $\Gamma$ invariant, but changing its split into radiative and pure gauge parts: 
\beq
\Delta_\rho A =0\,,\qquad\qquad\Delta_\rho A_{\rm rad}=\extd\rho\,,\qquad\qquad\Delta_\rho\varphi=-\rho\,.\label{Maxsymmetry}
\eeq
As frame reorientations, symmetries transform the edge mode and thereby indirectly transform the frame-dressed gauge-invariant observables, such as $A_{\rm rad}=A-\extd\varphi$ (see Sec.~\ref{ssec_edgeframereorient}), while gauge transformations by construction leave all gauge-invariant data invariant. In the present case, a symmetry transformation exploits the fact that the split into radiative and gauge parts is only unique up to gauge transformations and changes the physical part $A_{\rm rad}$ by a total derivative. This has no impact on local gauge-invariant quantities such as $F$ or $\star F$ in $M\cup\Gamma$, but it does affect the relation between the subregion $M$ and the (non-locally defined) reference frame $U$, originating in the complement $\bar M$.

Let us now illustrate post-selection on Robin boundary conditions, from which we obtain  Dirichlet and Neumann conditions as special cases. If we wanted to proceed in the general manner of Sec.~\ref{ssec_genalg} we would have to choose a basis field in the cotangent space of $\Gamma$ to decompose $\restr{A}{\Gamma}$ into  scalar components $\phi^a=A^a$ with $a=0,\ldots,d-2$. This can clearly be done, however, here we choose to work with form language instead, which provides a more elegant split \eqref{Maxsplit} into radiative and gauge parts and a more concise formulation of the boundary value problem. Since we will now build linear combinations directly at the form (rather than component) level in order to define Robin boundary conditions, the matrix $\mathbf{M}$ in \eqref{lincomb} will be a $2\times2$ (rather than $2(d-1)\times 2(d-1)$) matrix. The Robin boundary conditions thereby encompassed will thus only be a subset of all the possible Robin boundary conditions one can formulate in component language. Besides offering a more concise formulation, the advantage is that our boundary value problem will be independent of the choice of spacetime frame field on $\Gamma$. More precisely, while the $1$-form $X=X_a\extd x^a\heq X_0$, which we will use below for imposing boundary conditions, is invariant under local frame (coordinate) changes, the components $X_a$ transform covariantly under the latter. The types of boundary conditions we are excluding by focusing on an ${\rm SL}(2, \mathbb{R})$ subgroup of all canonical transformations are those that, in component form, are not even covariant under local frame transformations.

In order to formulate the linear combinations \eqref{lincomb} at the form level, we recall equation~\eqref{boundaryHodge}. This suggests to define the fields for Robin post-selection in the form 
\beq
X:=\mathbf{a}A_{\rm rad} + (-1)^{d-1} \,\mathbf{b} \ast \star F\,,\qquad\qquad Y:=\mathbf{c} A_{\rm rad} + (-1)^{d-1} \mathbf{d} \ast \star F\,,\label{Maxdecomp}
\eeq
for $\mathbf{a},\mathbf{b},\mathbf{c},\mathbf{d}$ real functions ($\mathbf{d}\star F$ should not be confused with the equation of motion $\extd\star F$), so that $X$ and $Y$ are both $(0,1)$-forms. 
Imposing the ${\rm SL}(2, \mathbb{R})$ constraint $\mathbf{a} \mathbf{d} - \mathbf{b} \mathbf{c}=1$ ensures that \eqref{Maxdecomp} is a canonical transformation of the radiative presymplectic current:\footnote{We use the above identity for applying the Hodge operator twice, and note that $\langle\delta\mathrm{v},\delta\mathrm{u}\rangle=-\langle\delta\mathrm{u},\delta\mathrm{v}\rangle$ since $\delta\mathrm{v},\delta\mathrm{u}$ are $1$-forms on field space.}
\beq
\omega_{\rm rad} = \delta A_{\rm rad} \wedge \delta \star F = \delta X \wedge \ast \delta Y \,.
\eeq

Let us now determine the generating function $\ell_{\rm corr}$. Condition \eqref{ellcorr0} translates here into
\bes 
\delta\ell_{\rm corr}&\approx& - \delta X\wedge\ast Y-\theta-\extd C\nn\\
&\approx& - \delta X\wedge\ast Y +\delta A_{\rm rad} \wedge \star F -\extd(C-\star F\delta\varphi)\nn\\
&=&\frac{1}{2}\delta\left(A_{\rm rad}\wedge \star F - X\wedge \ast Y \right)-\extd(C-\star F\delta\varphi)\,.\label{Maxellcor}
\ees 
In arriving at the last line, we have made use of the various identities above. Henceforth, we define 
\beq 
\ell_{\rm corr}:=\frac{1}{2}\left(A_{\rm rad}\wedge \star F - X\wedge \ast Y\right)\,,\qquad\qquad C:=\star F\delta\varphi\,,
\eeq 
in line with \eqref{ellcorr1}. In turn, we find that
\beq 
\theta_{\rm corr}:= \theta + \delta \ell_{\rm corr} = - \delta X\wedge\ast Y+(\extd\star F)\delta\varphi-\extd(\star F\delta\varphi)\,,
\eeq
to be contrasted with \eqref{Maxtheta}. The post-selection action \eqref{Ssp} on $\cF_M$ thus reads
\beq 
S_{\rm ps}=\int_\Gamma \left(\frac{1}{2}\left(A_{\rm rad}\wedge \star F - X\wedge \ast Y\right)+\lambda\wedge(X-X_0)\right)\,,
\eeq
with the gauge-invariant $(0,d-2)$-form $\lambda$ playing the role of Lagrange multiplier. 

In conjunction, the subregion variational problem defined by post-selection on \textbf{Robin boundary conditions} $X=X_0$ is given by the action $S_{M\cup\Gamma}=\int_M L[A]+\int_\Gamma\ell$ with gauge-invariant boundary Lagrangian (see equation~\eqref{bdryaction})
\beq\label{eq:Maxwell_l}
\ell = - X_0 \wedge \ast Y +\frac{1}{2}\left( A_{\rm rad} \wedge \star F + X \wedge \ast Y \right)\,,
\eeq 
in analogous form to \eqref{scalarRobin} of the scalar field case. Variation of this action yields 
\beq 
\delta S_{M\cup\Gamma} = \int_M \extd\star F\wedge\delta A-\int_{\Sigma_2-\Sigma_1}\delta A\wedge\star F+\int_\Gamma\Big((\extd\star F)\delta\varphi+(X-X_0) \wedge \delta \ast Y-\extd\left(\star F\delta\varphi\right)\Big)\,,
\eeq 
hence implementing the Robin boundary conditions as boundary equations of motion and, in line with our assumption that we are dealing with a fictitious boundary $\Gamma$, also imposing the bulk equation of motion $\extd\star F=0$ on $\Gamma$. 

\textbf{Dirichlet boundary conditions} are obtained in the special case $\mathbf{a}=\mathbf{d}=1$ and $\mathbf{b}=0$, which amounts to post-selection on the radiative part of the connection $X=A_{\rm rad}=A_{\rm rad}^0$. 
The generating function $\ell_{\rm corr}$ vanishes and the boundary Lagrangian becomes (when setting also $\mathbf{c}=0$ for simplicity)
\beq 
\ell = (A_{\rm rad}-A^0_{\rm rad})\wedge\star F=\left(A-\extd\varphi-A_{\rm rad}^0\right)\wedge\star F\,,\label{MaxDirichlet}
\eeq 
where we have restored the standard connection $A$ on $\Gamma$ and the associated edge mode via \eqref{Maxsplit}. This is a generalization of the boundary action for Maxwell theory proposed in \cite{Geiller:2019bti,Mathieu:2019lgi}, which we thus recover through our post-selection algorithm. In particular, our procedure demonstrates that the edge mode in the action does not have to be postulated, but actually is part of the global theory space from the start, and appears thanks to our variable split into radiative and gauge parts relative to the reference frame $U$.\footnote{More precisely, the action in (4.28) of \cite{Geiller:2019bti} is given for $d=4$ in the form $\ell_{\Gamma}=j\wedge (\extd a+A)$, where $a$ is a postulated edge mode transforming as $\delta_\alpha a=-\alpha$ and $j$ is a gauge-invariant dynamical $2$-form that can be viewed as a Lagrange multiplier. One can solve for the Lagrange multiplier, giving $j=\star F$, and the equations of motion also yield $\restr{F}{\Gamma}=0$. In our action, we get the edge mode $\varphi\equiv-a$ with correct transformation property directly from  the reference frame construction of Sec.~\ref{sec_gaugesplit}, which allows us to split the boundary degrees of freedom into radiative and gauge parts as in \eqref{Maxsplit}. Furthermore, the action we obtained generates the more general boundary value $\restr{F}{\Gamma}=\extd A_{\rm rad}^0$ through the background field introduced via post-selection. Hence, \eqref{MaxDirichlet} is a generalization of the action in \cite{Geiller:2019bti},  evaluated on the solution for the Lagrange multiplier $j=\star F$.}

By contrast, post-selection on $\star F$ can be implemented by setting $\mathbf{b}=1$ and $\mathbf{a}=0$, so that $X=(-1)^{d-1} *\star F=(-1)^{d-1}*f_0$, for some background $(d-2)$-form $f_0$. This yields \textbf{Neumann boundary conditions}. If we also set $\mathbf{c}=-1$ and $\mathbf{d}= 0$ for simplicity, we have the `momentum' $Y=-A_{\rm rad}$ and the boundary action reads
\beq 
\ell = A_{\rm rad} \wedge f_0 =\left(A-\extd\varphi\right) \wedge f_0\,.\label{MaxNeumannac}
\eeq 
Accordingly, the shifted presymplectic potential on $\Gamma$ reads on-shell
\beq 
\theta_{\rm corr}\approx\delta\star F\wedge A_{\rm rad}-\extd(\star F\delta\varphi)\,,
\eeq 
which amounts to the dual polarization of equation~\eqref{Maxtheta}, i.e.\ with the roles of radiative configuration and momentum degrees of freedom interchanged.

\medskip

Finally, let us revisit the distinction between gauge transformations and symmetries. The former yield constraints, while the latter give rise to non-vanishing charges. Indeed, let $\fX_\alpha$ denote the vector field on $\cF_M$ inducing the gauge transformation in \eqref{Maxgauge}. On the space of solution $\cS_M$, this defines a degenerate direction of $\Omega_M$ given in \eqref{MaxOmega} 
\beq
\delta C[\alpha]:=\fX_\alpha\cdot\Omega_M = - \int_\Sigma\alpha\delta\extd\star F \approx 0\,,
\eeq
so that the constraint takes the expected form \eqref{eq:constraint_from_theta}.
Since we are dealing with a $\rm{U}(1)$ gauge theory, the constraints yield an Abelian Poisson algebra 
\beq
\{C[\alpha],C[\beta]\}=\fX_{\beta}\cdot(\fX_\alpha\cdot\Omega_M )=0\,.
\eeq

On the other hand, let $\fY_\rho$ with field-independent $\rho$ denote the vector field on $\cF_M$ generating the symmetry transformation in \eqref{Maxsymmetry}. In $\cS_M$ (that is, before boundary conditions are imposed), this does not define a degenerate direction of the presymplectic form, since 
\beq
\delta Q[\rho]:=\fY_\rho\cdot\Omega_M =\int_{\partial\Sigma}\rho\delta\star F
\eeq
does not vanish in general. Instead, this yields the charge
\beq
Q[\rho] = \int_{\partial\Sigma}\rho\star F\,.
\eeq
This expression is in agreement with the one in \cite{Geiller:2019bti}. Constituting the generators of a $\rm{U}(1)$ frame transformation, the charges likewise form an Abelian Poisson algebra 
\beq
\{Q[\rho],Q[\sigma]\}=\fY_\sigma\cdot(\fY_\rho\cdot\Omega_M)=0\,
\eeq
 and they are clearly gauge-invariant boundary observables, $\{C[\alpha],Q[\rho]\}=\fY_\rho\cdot(\fX_\alpha\cdot\Omega)=0$. Finally, the interpretation of those generators in the regional solution space $\cS_M^{X_0}$ is contingent on the type of boundary conditions being considered. As it has been discussed in Sec.~\ref{subsec:foliation} for gauge field theories, as well as in Sec.~\ref{ssec_symvsgaugemech} for a mechanical toy-model, we have three possibilities.
\begin{description}
\item[Symmetries.] These are transformations which leave the boundary conditions invariant, but allow for non-trivial variations of the charge, that is:
\begin{align}
    \Delta_\rho X &= \fY_\rho \cdot \delta X = \mathbf{a} \, \extd \rho \heq 0\,, \\
    \delta Q[\rho] &= \int_{\partial \Sigma} \rho \delta \star F \not\heq 0\,.
\end{align}
Since $\delta\star F$\,---\,and thereby $\delta Q[\rho]$\,---\,vanishes for Neumann boundary conditions, this is only possible for \textbf{Dirichlet} or \textbf{Robin boundary conditions} ($\mathbf{a}\neq0$), and requires $\rho$ to be constant on $\Gamma$. Up to multiplication, there is therefore a unique charge, which records variations of the total flux $\int_{\partial \Sigma} \star F$.

\item[Meta-symmetries.] This corresponds to the situation in which $\fY_\rho$ fails to leave the boundary condition invariant (i.e.\ when $\mathbf{a} \extd \rho \not\heq 0$). In this case, the transformation changes leaf (or subregion theory) $\cS_{X_0}^M$ in the foliation \eqref{eq:foliation} of the global space of solutions $\cS$. This can only happen for $\mathbf{a} \neq 0$, that is for \textbf{Dirichlet} or \textbf{Robin boundary conditions}. In that case, meta-symmetries include any vector field $\fY_\rho$ with $\rho$ not constant on $\Gamma$.  
\item[Boundary gauge symmetries.] Finally, one can have both $\Delta_\rho X \heq 0$ and $\delta Q[\rho]\approx 0$, in which case the generator $\fY_\rho$ is transmuted into a gauge direction due to the boundary condition. This requires $\delta \star F \heq 0$, hence a \textbf{Neumann boundary condition}, and in particular $\mathbf{a}=0$. This is also a sufficient condition, since $\mathbf{a} =0$ automatically implies $\Delta_\rho X \heq 0$, so that any generator $\fY_\rho$ constitutes a boundary gauge symmetry. This case leads to the additional edge constraints $C_\partial[\rho]\heq0$ discussed in equation~\eqref{edgeconstraint}.
\end{description}

Lastly, let us comment on the status of edge modes and the edge-mode-dressed observables in the presence of these various types of symmetries and boundary conditions, especially as far as the reduced phase space is concerned. Let us firstly focus on the case of Dirichlet and Robin boundary conditions. Note that in this case, for any choice of post-selected subsystem theory $\cS_M^{X_0}$ defined by $X=X_0$, there will exist both symmetries and meta-symmetries (but not boundary gauge symmetries), the former acting tangentially to the leaf, while the latter act transversally. In order to construct the regional reduced phase space $\cP^{X_0}_M$ from $\cS_M^{X_0}$, one has to factor out the degenerate directions of the pullback of $\Omega_M$ to $\cS_M^{X_0}$. The symmetry transformation\,---\,unique up to multiplication by a constant (see above)\,---\,does not constitute a degenerate direction and meta-symmetries are not even tangential. The symmetry transformation, which ultimately originates in the edge mode, thus corresponds to a physical direction and thereby a non-trivial transformation on $\cP_M^{X_0}$; specifically, since $Q[\rho]$ is conjugate to the edge mode variable $\varphi$ by equation~\eqref{MaxOmega}, the symmetry will transform any non-trivially frame-dressed observable. At least some such frame-dressed observables must therefore survive on $\cP_M^{X_0}$. By contrast, the standard gauge generators $\fX_\alpha$ will constitute degenerate directions of $\Omega_M$. This entails that a physical solution in $M\cup\Gamma$ is specified by the gauge-invariant boundary condition $X=X_0$, as well as suitable gauge-invariant initial data on some Cauchy slice $\Sigma$ (cf.\ the related discussion in Sec.~\ref{sec:covariant_to_standard}) which must have some non-trivial dependence on the edge mode frame $U$; the reduced phase space $\cP_M^{X_0}$ is thus labeled by the boundary conditions and parametrized by the invariant Cauchy data. 

The invariant Cauchy data will thus not only contain regional gauge-invariant data (i.e.\ only depending on fields in $M$), but also non-local observables involving the edge mode frame (which originates in the complement $\bar M$). For example, it may contain observables relating the connection $A$ at some bulk point $\sigma$ of $\Sigma$ to the edge frame at $x\in\partial\Sigma$, obtained by shooting a Wilson line in from $x$ to $\sigma$ and contracting it with the frame $U$, similarly to the constructions in Sec.~\ref{sec_gaugesplit}. Note that such non-local observables, which ultimately relate bulk properties in $M$ to the edge frame $U$ (and thus $\bar M$), cannot all be reconstructed from purely local gauge-invariant quantities (that is, from the radiative data $X,Y$ on $\partial\Sigma$ together with gauge-invariant bulk variables); they thus amount to independent initial data. In other words, in the case of Dirichlet and Robin boundary conditions, the regional reduced phase space $\cP_M^{X_0}$ contains frame-dressed observables and therefore more than the purely regional gauge-invariant data one would find when ignoring edge modes in the first place. In this case, edge modes are therefore of direct physical significance, besides also being a prerequisite for deriving the boundary value problem from the global variational problem through post-selection.

The situation is, however, quite different for Neumann boundary conditions, in which case \emph{all} symmetry generators $\fY_\rho$ constitute degenerate directions of the pullback of $\Omega_M$ to $\cS_M^{X_0}$, and correspond to additional boundary gauge symmetries encoded in the additional edge constraints $C_\partial[\rho]$. Since the edge constraints are essentially what would be the charges in the case of symmetries, they too are conjugate to the edge mode variable $\varphi$ by  equation~\eqref{MaxOmega}. However, in this case they correspond to gauge directions in $\cS_M^{X_0}$ that are factored out when constructing the regional reduced phase space $\cP_M^{X_0}$. The latter will thus not contain any frame-dressed observables, but exclusively the regional gauge-invariant data one would find when ignoring edge modes. Intuitively, this can be understood from the fact that Neumann boundary conditions correspond to reflective boundary conditions, so that the subregions $M$ and $\bar M$ evolve essentially independently. Note, however, that even in this case, the edge mode is crucial in order to construct the foliation of the space of solutions in equation~\eqref{eq:foliation} and thereby to consistently formulate the boundary value problem through post-selection. Through its appearance in the boundary action $\ell$, it also affects the reduced dynamics on $\cP_M^{X_0}$.

\subsection{Abelian Chern-Simons theory}\label{subsec:CS}

The edge mode construction of \cite{Donnelly:2016auv} was first extended to Chern-Simons theories in \cite{Geiller:2017xad} (see also~\cite{Geiller:2019bti}). As we will now illustrate in our post-selection formalism, an interesting feature of Abelian Chern-Simons theory is that it can support an infinite-dimensional set of boundary symmetries, even after appropriate boundary conditions have been imposed on $\Gamma$. This is a crucial structural difference between Maxwell (or more generally Yang-Mills) and Chern-Simons edge modes, but it is important to note that our general construction and its interpretation apply equally well to both types of theories. In this regard, our formalism differs from that advocated in \cite{Gomes:2018dxs, Gomes:2019xto, Riello:2020zbk, Riello:2021lfl} (see in particular \cite{Riello:2021lfl} for a nice summary of this series of works): by relying on specific features of Yang-Mills theories (for instance, the functional form of the Gauss constraint), this work is not directly exportable to Chern-Simons theories. We also note that Chern-Simons symmetry algebras were previously studied in the context of boundaries, e.g.\ in \cite{Carlip:1994gy,Park:1998yw}, though in a different formalism and without an explicit edge mode construction.

The bulk action $S_{M}=\int_M L[A]$ of Abelian Chern-Simons theory, a topological field theory in $d=3$ spacetime dimensions, features the Lagrangian
\beq
L=A\wedge F\,,
\eeq
where $F=\extd A$ is the field strength of the $\rm{U}(1)$ connection $1$-form $A$, which under gauge transformations transforms as usual: $\delta_\alpha A=\extd\alpha$. The variation of the bulk Lagrangian gives
\beq
\delta L=2\delta A\wedge F-\extd(A\wedge\delta A)\,,
\eeq
yielding the bulk equation of motion $F=0$ and the presymplectic potential
\beq
\Theta = \delta A \wedge A  \,.
\eeq
Its pullback to $\Gamma$ can be written as
\beq 
\theta= \langle \delta A , \ast A \rangle \epsilon_\Gamma\,,
\eeq
where $*$ is the Hodge dual on $\Gamma$.
The presymplectic current
\beq
\omega = - \delta A \wedge \delta A 
\eeq
pulls back to $\Gamma$ in the form
\beq
\omega = \langle \delta A , - \delta \ast A \rangle \epsilon_\Gamma\,.
\eeq 
In local coordinates, this means that the conjugate of a component $A_\mu$ is $(\ast A)^\mu$. $A$ has two components once pulled-back to $\Gamma$; imposing boundary conditions on a Lagrangian submanifold amounts to fixing just one of them.\footnote{While we will ultimately be imposing boundary conditions on the components of the edge-frame-dressed connection $A_{\rm rad}$, we will here briefly explore the frame-(in)dependence of boundary conditions for the gauge-dependent $A$, as the discussion applies in the same manner to $A_{\rm rad}$, but is notation wise simpler.} In local inertial coordinates $(t,x)$, where $x$ is space-like and $t$ time-like, we could for instance impose a boundary condition $A_x = A^{(0)}_x$ for some background field $A^{(0)}_x$. However, such a condition is neither invariant nor covariant under local Lorentz transformations. A better option is to impose a boundary condition on light-cone coordinates $A_{+} = A_t + A_x$ or $A_- = A_t - A_x$. The reason is that $A_\pm$ only changes by a multiplicative constant under a Lorentz transformation, so that the type of boundary condition is invariant: two local inertial observers can agree on the fact that one is imposing a boundary condition on $A_\pm$, even if they might disagree about the value of the background field $A_{\pm}^{(0)}$. Furthermore, the boundary condition $A_{\pm} = 0$ is fully invariant. Therefore, boundary conditions of the form $A_{\pm}= A_{\pm}^{(0)}$ fall in the same class as those already investigated for Maxwell theory. We shall focus exclusively on such boundary conditions.

In coordinate-independent language, we can simply define:
\beq
A^{\pm} := A \pm \ast A \qquad (\ast A^\pm = \pm A^\pm)
\eeq
as the (anti)self-dual part of $A$. Indeed, in the coordinates $(t,x)$ we used above, one can check that $\ast \extd t = \extd x$, $\ast \extd x = \extd t$, and therefore:
\beq
A^{\pm} = (A_t \pm A_x) (\extd t \pm \extd x ) = A_{\pm} (\extd t \pm \extd x )\,.
\eeq
The frame-independent boundary condition $A^\pm=A^\pm_0$ at the form level is thus equivalent to fixing the component $A_\pm$ in a local inertial frame.
We can decompose $\theta$ and $\omega$ as:\footnote{We make use of the following useful fact: 
\beq
\delta A^\pm \wedge A^\pm = \langle \delta A^\pm , \pm A^\pm \rangle \epsilon_\Gamma = \pm \frac{1}{2} \delta \langle A^\pm , A^\pm \rangle \epsilon_\Gamma = \frac{1}{2} \delta (A^\pm \wedge A^\pm) = \frac{1}{2} \left( \delta A^\pm \wedge A^\pm + A^\pm \wedge \delta A^\pm\right) = 0\,.
\eeq}
\begin{align}
\theta &= \frac{1}{4} \left(  \delta A^+ \wedge A^- + \delta A^- \wedge A^+ \right) = \frac{1}{4} \left(  - \langle \delta A^+ , A^- \rangle + \langle \delta A^- , A^+ \rangle \right) \epsilon_\Gamma \,, \\
\omega &= - \frac{1}{2}  \delta A^+ \wedge \delta A^- = \frac{1}{2} \langle \delta A^+ , \delta A^- \rangle \epsilon_\Gamma \,.
\end{align}

On $\Gamma$ we shall once more invoke the split of the connection into a gauge-invariant radiative part and a pure gauge part (see equation~\eqref{genArad})
\beq
\restr{A}{\Gamma}=A_{\rm rad}+\extd\varphi\,.\label{CSsplit}
\eeq
As in the Maxwell case, $\varphi$ is the edge mode, which is part of the global theory space from the start and clearly has to transform as $\delta_\alpha\varphi=\alpha$ (see the discussion in Secs.~\ref{sec_fieldrfs} and~\ref{sec:relational_obs}). This permits us to rewrite the pullback of the presymplectic potential to $\Gamma$ in the form\begin{align}
\theta &= \theta_{\rm rad} + \theta_{\rm gauge}\,,\\
\theta_{\rm rad} &= \delta A_{\rm rad} \wedge A_{\rm rad}\,, \\
\theta_{\rm gauge} &= \delta \extd \vphi \wedge A_{\rm rad} + \delta A_{\rm rad} \wedge \extd \vphi + \delta \extd \vphi \wedge \extd \vphi\,.
\end{align}

In order to implement the post-selection procedure of Sec.~\ref{subsec:foliation}, we now have to impose boundary conditions on the radiative data. It is clear that one can make $\omega_{\rm rad} =-\frac{1}{2}\delta A_{\rm rad}^+\wedge\delta A^-_{\rm rad}$ to vanish by imposing 
\beq
\delta A^+_{\rm rad} \heq 0 \qquad \text{or} \qquad \delta A^-_{\rm rad} \heq 0 \,, 
\eeq 
where
\beq
A^\pm_{\rm rad} = A^\pm - \left( \extd \vphi \pm \ast \extd \vphi \right)\,.
\eeq
To determine the correction term $\ell_{\rm corr}$ that will need to be included in order for the shifted presymplectic  potential $\theta_{\rm corr}$ to reduce to a corner term on-shell (see equation~\eqref{shifttheta}) across $\Gamma$, we firstly rewrite the radiative part as:
\beq
\theta_{\rm rad} = \frac{1}{2} \delta A^\mp_{\rm rad} \wedge A^\pm_{\rm rad} + \frac{1}{4} \delta\left( A^\pm_{\rm rad} \wedge A^\mp_{\rm rad} \right)\,.
\eeq 
The second term will have to be cancelled by a contribution $- \frac{1}{4} \left( A^\pm_{\rm rad} \wedge A^\mp_{\rm rad} \right)$ to $\ell_{\rm corr}$, for otherwise equation~\eqref{shifttheta} cannot be realized.

Next, let us turn to $\theta_{\rm gauge}$. As in Maxwell theory, this contains the edge mode contributions that will allow us to impose the gauge constraint $F = 0$ dynamically on $\Gamma$. However, the situation is slightly less trivial here as it requires a change of polarization. In Maxwell theory, we readily had $\theta_{\rm gauge} = \extd \star F \delta \vphi - \extd(\delta \vphi \star F)$; that is, the variation of the reference frame times a constraint, supplemented with a corner term. This is not so in Chern-Simons theory, as
\beq\label{eq:theta_gauge_Abelian_CS}
\theta_{\rm gauge} = - 2 F \delta \vphi + \extd (\delta \vphi A_{\rm rad}- \vphi \delta A_{\rm rad} + \delta \vphi \extd \vphi) + \delta(\vphi F)\,.  
\eeq
The first term includes the constraint $F \approx 0$, the second is a corner contribution, and the third\,---\,the difference to the Maxwell case\,---\,an exact field space form resulting from the change of polarization to radiative and edge mode degrees of freedom. Given that this last term vanishes on-shell, defining $\theta_\partial = - \delta \vphi A_{\rm rad} + \vphi \delta A_{\rm rad} - \delta \vphi \extd \vphi$ and $\ell_\partial = 0$ would be in compliance with the general algorithm summarized in Sec.~\ref{sec:summary_algo}. However, recalling the discussion at the end of Secs.~\ref{ssec_globsplit} and~\ref{ssec_genalg}, we will find it advantageous to cancel the exact form $\delta(\vphi F)$ even off-shell, by means of a non-vanishing $\ell_\partial := - \vphi F$, which in turn will contribute to $\ell_{\rm corr}$. All in all, we are thus led to defining: 
\beq
\ell_{\rm corr} := - \vphi F - \frac{1}{4}  A^{\mp}_{\rm rad} \wedge A^\pm_{\rm rad}\,.
\eeq

Recalling item (ii) from Sec.~\ref{ssec_globsplit}, the term $- \vphi F$ can be equivalently motivated by the desire to restore gauge-invariance of the subregion variational problem. Indeed, an important feature of Chern-Simons theory as compared to Maxwell theory is that $S_M$ fails to be gauge-invariant on its own. More precisely, for fictitious boundaries $S_M$ is kinematically not gauge-invariant, but on-shell it is because 
\beq
\delta_\alpha S_M = \int_{\Gamma\cup\Sigma_2\cup\Sigma_1}\alpha F\,.
\eeq
In principle, we thus do not need to add a piece to the action in order to restore gauge-invariance for a subregion delimited by fictitious boundaries.\footnote{In subsequent work \cite{CH2}, we shall also deal with physical boundaries on which the bulk equation of motion need not hold in the form $\restr{F}{\Gamma}=0$ on account of possible sources, in which case $S_M$ will not be gauge-invariant by itself on-shell.} However, it is convenient to have gauge-invariance built-in at the kinematical level, and therefore require the action to be gauge-invariant even off-shell. Including a term $- \vphi F$ into $\ell_{\rm corr}$ achieves just that. 

Note that the decomposition \eqref{eq:theta_gauge_Abelian_CS} is not unique as one can shift $\vphi F$ in the third term by an exact form $\extd k$ and compensate this change by a shift $-\extd \delta k$ of the second term. However, this translates into nothing else but the usual ambiguity in the definition of the boundary Lagrangian $\ell$ (which is unique up to integration by parts).

\medskip

Altogether, we are led to introducing the post-selection action (see equation~\eqref{Ssp}):
\begin{align}
S_{\rm ps} &= \int_\Gamma -\vphi F - \frac{1}{4}  A^{\mp}_{\rm rad} \wedge A^\pm_{\rm rad} + \lambda \wedge (A_{\rm rad}^\pm - A^\pm_0) \\
&= \int_\Gamma -\vphi F \pm \frac{1}{2}  \ast A_{\rm rad} \wedge A_{\rm rad} + \lambda \wedge (A_{\rm rad} \pm  \ast A_{\rm rad} - A^\pm_0)\,,
\end{align}
where $A^{\pm}_0$ is a background self-dual (resp.\ antiself-dual) form entering the boundary condition $A^{\pm} \heq A^\pm_0$, and $\lambda$ is a gauge-invariant $1$-form that we can furthermore assume to be antiself-dual (resp.\ self-dual). 
After evaluating the Lagrange multiplier on-shell ($\lambda \approx \frac{1}{2} A^{\mp}_{\rm rad}$), we obtain the following boundary Lagrangian (see equation~\eqref{bdryaction}):
\begin{align}
\ell^\pm &= -\vphi F + \frac{1}{2} A^\pm_0 \wedge A^{\mp}_{\rm rad} - \frac{1}{4}  A^{\pm}_{\rm rad} \wedge A^\mp_{\rm rad} \\
&= -\vphi F + \frac{1}{2} A^\pm_0 \wedge ( A_{\rm rad} \mp \ast A_{\rm rad} ) \pm \frac{1}{2}  A_{\rm rad} \wedge \ast  A_{\rm rad} \,.
\end{align}
In terms of the original variables $A$ and $\vphi$, this is simply:
\beq\label{eq:bdy_CS}
\ell^\pm = -\vphi \extd A + \frac{1}{2} A^\pm_0 \wedge ( A \mp \ast A - \extd \vphi \pm \ast \extd \vphi ) \pm \frac{1}{2}  A \wedge \ast  A \pm A \wedge \ast \extd \vphi \pm \frac{1}{2} \extd \vphi \wedge \ast \extd \vphi \,,
\eeq
from which we deduce that the frame field $\vphi$ inherits a non-trivial kinetic term $\pm \frac{1}{2} \extd \vphi \wedge \ast \extd \vphi$. By construction, this boundary Lagrangian allows us to impose the constraint $F = 0$ on $\Gamma$ together with the boundary condition $A_{\rm rad}^\pm = A^\pm_0$ as dynamical equations of motions.
Indeed, varying the action $S_{M\cup\Gamma}=\int_ML[A]+\int_\Gamma \ell^\pm_\Gamma$ gives
\begin{align}
\delta S_{M\cup\Gamma}&=\int_M2\delta A\wedge F+\int_{\Sigma_2-\Sigma_1}\delta A\wedge A \\
& \quad + \int_\Gamma\Big(\frac{1}{2} \delta A^\mp_{\rm rad}\wedge(A^\pm_{\rm rad}- A_0^\pm )- 2 F \delta\varphi + \extd (\delta \vphi A_{\rm rad} - \vphi \delta A_{\rm rad} + \delta \vphi \extd \vphi)\Big)\,,
\end{align}
which is stationary up to variation on $\Sigma_1$ and $\Sigma_2$ provided that: $F = 0$ in the bulk as well as on $\Gamma$, and $A^\pm_{\rm rad}= A_0^\pm$.

We note that our final expression \eqref{eq:bdy_CS} coincides with the boundary Lagrangian proposed in \cite{Mathieu:2019lgi}, up to an irrelevant integration by parts and in the special case of vanishing background field $A_0^\pm$. In this, our derivation justifies and generalizes this previous proposal. It is also closely related to the boundary action previously introduced in \cite{Geiller:2019bti}. Translated in our own notations, the authors of this paper proposed the boundary Lagrangian 
\beq\label{eq:CS_MarcPuttarak}
\tilde{\ell}^\pm =  - \vphi F + j \wedge A_{\rm rad} \pm \frac{1}{2}   j \wedge \ast j\,,
\eeq
in which an extra gauge-invariant and dynamical boundary $1$-form $j$ has been introduced. The equation of motion resulting from varying $j$ is $\ast j = \mp A_{\rm rad}$, and once coupled to the bulk action, the equation of motion resulting from varying $A$ is $j = A_{\rm rad}$. $A_{\rm rad}$ is therefore (anti)self-dual on-shell, and it turns out that we recover the same equations of motion for $(A, \vphi)$ as with $\ell^\pm$, again provided that $A^\pm_0=0$. In this sense, our proposal similarly extends the results of \cite{Geiller:2019bti} to non-vanishing $A^\pm_0$ (at least at the classical level, which is the main focus of the present work). In order to accommodate non-vanishing boundary conditions, the action \eqref{eq:CS_MarcPuttarak} can be augmented to: 
\beq
\tilde{\ell}_\Gamma^\pm = \int_\Gamma - \vphi F + j \wedge A_{\rm rad} + \frac{1}{2} A_0^\pm \wedge ( A_{\rm rad} \mp \ast A_{\rm rad} ) \pm \frac{1}{2}  j \wedge \ast j\,,
\eeq
which remains linear in $A_{\rm rad}$. By varying $A$ we now obtain $j = A_{\rm rad} - A^\pm_0$, while the variation of $j$ still yields $\ast j = \mp A_{\rm rad}$. It follows that $A^\pm_{\rm rad} = A^\pm_0$, as claimed.

\medskip

Let us conclude this subsection by disentangling gauge transformations from boundary symmetries and their charges. Computing $\omega_{\rm gauge} = \delta \theta_{\rm gauge}$, we find:
\beq
\omega_{\rm gauge} = - \delta \extd \vphi \wedge \delta \extd \vphi - 2 \delta \extd \vphi \wedge \delta A_{\rm rad} \approx - \extd \left( \delta \vphi \delta \extd \vphi + 2 \delta \vphi \delta A_{\rm rad} \right)\,.
\eeq
The (on-shell) consistency relation \eqref{eq:omega_bdy}, $\extd \omega_{\partial} \heq - \omega_{\rm gauge}$, can be solved by postulating a boundary presymplectic current
\beq
\omega_\partial := \delta \vphi \delta \extd \vphi + 2 \delta \vphi \delta A_{\rm rad} = 2 \delta \vphi \delta A - \delta \vphi \delta \extd \vphi\,.
\eeq
Vector fields $\fX_\alpha$ act on $\Omega_M = \int_\Sigma \omega + \int_{\partial \Sigma} \omega_{\partial}$ as
\begin{align}
\fX_\alpha \cdot \Omega_M &= - 2 \int_\Sigma \extd \alpha \wedge \delta A + \int_{\partial \Sigma} \left( 2 \alpha \delta A - 2 \delta \vphi \extd \alpha - \alpha \delta \extd \vphi + \delta \vphi \extd \alpha \right) \\
&= \int_\Sigma \left( - 2 \extd (\alpha \delta A) + 2 \alpha \delta F \right) + \int_{\partial \Sigma} \left( 2 \alpha \delta A - \extd(\alpha \delta \vphi) \right) \\
&= 2 \int_\Sigma \alpha \delta F \approx 0\,,
\end{align}
and are therefore gauge symmetries. Defining the constraints 
\beq
C[\alpha] = 2 \int_\Sigma \alpha F
\eeq
for field-independent $\alpha$, we find that they form an Abelian Poisson algebra
\beq
\{ C[\alpha], C[\beta]\}:= \fX_\beta \cdot \fX_\alpha \cdot \Omega_M =  0\,. 
\eeq

Next, we can introduce vector fields $\fY_\rho$ that only act on the edge reference frame as frame reorientations (see Sec.~\ref{ssec_edgeframereorient} for the corresponding general discussion): 
\beq
\Delta_\rho A:=\fY_\rho(A)=0 \qquad \mathrm{and} \qquad \Delta_\rho\varphi:=\fY_\rho(\varphi)= - \rho\,,
\eeq
so that, in particular, the frame-dressed observables vary according to $\Delta_\rho A_{\rm rad}:=\fY_\rho( A_{\rm rad} ) = \extd \rho$ and $\Delta_\rho A_{\rm rad}^\pm :=\fY_\rho( A^\pm_{\rm rad} ) = \extd \rho \pm \ast \extd \rho$. We then have $\fY_\rho\cdot \omega =0$, while
\beq
\fY_\rho\cdot \omega_{\partial} = - 2 \rho \delta A + \rho \delta \extd \vphi - \delta \vphi \extd \rho = - 2 \rho \delta A_{\rm rad} - \extd (\rho \delta \vphi)\,. 
\eeq
For a field-independent $\rho$ it follows that:
\beq\label{eq:charges_CS_global}
\fY_\rho \cdot \Omega_M = \delta Q[\rho] \,, \qquad \mathrm{where} \qquad Q[\rho] := -2 \int_{\partial \Sigma} \rho A_{\rm rad}\,,
\eeq
and the charges $Q$ obey a Ka\v{c}-Moody algebra:
\beq
\{ Q[\rho], Q[\sigma]\}:= \fY_\sigma \cdot \fY_\rho \cdot \Omega_M = - 2 \int_{\partial \Sigma} \rho \extd \sigma\,. \label{KMalgebra}
\eeq

We would find trivial charges ($\delta Q[\rho]\approx0$) if one were to naively impose a boundary condition on the full radiative connection $A_{\rm rad}$. By contrast, with the weaker (anti)-selfdual boundary condition $A_{\rm rad}^\pm = A_0^\pm$, $A_{\rm rad}^\mp$ is free to fluctuate, so that we can have non-trivial charges. More precisely, the regional covariant phase space is stable under $\fY_\rho$ provided that $A_{\rm rad}^\pm$ is left invariant. Such vector fields are labelled by functions $\rho$ on $\Gamma$, such that
\beq\label{eq:selfdual_sym}
\extd \rho \pm \ast \extd \rho = 0\,.
\eeq 
In the local coordinates $(t,x)$, this is nothing but $\partial_t \rho \pm \partial_x \rho = 0$, with general solution of the form $\rho = \rho(t \mp x)$. Crucially, and in sharp contrast with Maxwell theory, we therefore obtain an infinite-dimensional space of symmetries.\footnote{This difference holds more generally in comparison to Yang-Mills theories. As we will see below, an additional specificity of non-Abelian Yang-Mills theories is that the analogue of \eqref{eq:selfdual_sym} explicitly depends on the background fields defining the boundary condition. As a result, the dimension of the solution space itself depends on this background data.} This entails a non-trivial regional reduced phase space $\cP_M^{X_0}$ with propagating edge-mode-dressed degrees of freedom. For example, as in the case of Dirichlet and Robin boundary conditions in Maxwell theory, these may include observables relating $A$ in the bulk of $M$ to the edge frame via Wilson lines. Owing to the topological nature of the theory, such quasi-local observables would have been absent from the regional reduced phase space had one ignored edge modes in the first place.

Given that $\delta A_{\rm rad}^\pm = 0$, we can define the charges by
\beq
Q^\pm[\rho] = - \int_{\partial \Sigma} \rho A_{\rm rad}^{\mp}\,,
\eeq
so that $Q[\rho]=Q^+[\rho]+Q^-[\rho]$, where we have added an explicit superscript $\pm$ to distinguish our two sets of boundary conditions. On solutions to the boundary conditions $A^\pm_{\rm rad}=A^\pm_0$, the charges $Q^\mp[\rho]$ are no longer dynamical and thereby drop out of the Poisson-bracket relations in Eq.~\eqref{KMalgebra}. The surviving charges $Q^\pm[\rho]$ satisfy the Ka\v{c}-Moody commutation relations. Indeed, owing to the (anti)-selfduality condition \eqref{eq:selfdual_sym}: 
\beq\label{eq:sym_CS}
\{ Q^{\pm}[\rho], Q^\pm [\sigma]\}:= \fY_\sigma \cdot \fY_\rho \cdot \Omega_M = - \int_{\partial \Sigma} \rho (\extd \sigma \mp \ast \extd \sigma) = - 2 \int_{\partial \Sigma} \rho \extd \sigma\,.
\eeq
Finally, the vector fields $\fY_\rho$ parametrized by functions $\rho$ which fail to satisfy the stability condition \eqref{eq:selfdual_sym} are to be interpreted as meta-symmetries: owing to \eqref{eq:charges_CS_global}, they do generate symplectomorphisms, but between distinct regional covariant phase spaces. Note that, as a result, boundary gauge symmetries are excluded in this example.

\subsection{Yang-Mills theory}\label{ssec_YM}

As our last example, and in complement to Sec.~\ref{subsec:maxwell}, let us consider Yang-Mills theory for a general (compact and connected) matrix Lie group $G$. This will serve as a non-Abelian illustration of the post-selection and edge frame formalism. 

Yang-Mills theory is also one of the two main models originally studied in the pioneering work of Donnelly and Freidel on edge modes \cite{Donnelly:2016auv}. Where comparable, our technical results will be consistent with theirs. At the same time, our work extends their construction and analysis. For example,  1) by establishing the interpretation of edge modes as internalized external reference frames (see Sec.~\ref{sec_gaugesplit}), we are clarifying their overall physical meaning and that they are already included in the global theory to start with; 2) we are focusing on a spacetime region with a non-trivial time-like boundary, rather than a spatial Cauchy slice, which provides an alternative starting point for the construction of the regional presymplectic structure $\Omega_M$ (see Sec.~\ref{subsec:presympl}); 3) we demonstrate how to obtain the regional phase space and variational principle from the global ones via post-selection on edge-mode-dressed boundary conditions and clarify the status of edge modes in the context of different types of boundary conditions. In this regard, our work also complements the discussion in \cite{Riello:2021lfl,Riello:2020zbk,Gomes:2018dxs,Gomes:2019xto}, where the fate of edge modes in Yang-Mills theory in causal diamonds subject to Neumann boundary conditions was studied in a different formalism and we comment on that below.

The Lagrangian is defined as:
\beq
L = - \frac{1}{2} \Tr\left[ F \wedge \star F \right]
\eeq
where $F:= \extd A + A \wedge A$ is the curvature two-form of a $G$-connection $A$. We also recall that the covariant exterior derivative $\extd_A$ acts on a Lie-algebra valued form $\eta$ as $\extd_A \eta := \extd \eta + \left[ A, \eta \right]$.\footnote{Here, $[\cdot,\cdot]$ is the Lie bracket on Lie-algebra valued forms, which can be defined in terms of the wedge product as $[\eta_1 ,\eta_2]= \eta_1 \wedge \eta_2 -(-1)^{pq} \eta_2 \wedge \eta_1$, where $\eta_1$ is a $p$-form and $\eta_2$ is a $q$-form.} The variation of the curvature two-form can be nicely expressed as $\delta F = \extd_A \delta A$, from which we deduce:
\beq
\delta L = - \Tr\left[ \extd_A \delta A \wedge \star F \right] = - \Tr\left[ \extd_A \left(\delta A \wedge \star F \right)\right] - \Tr\left[  \delta A \wedge \extd_A \star F \right]  = - \extd \Tr\left[ \delta A \wedge \star F \right] - \Tr\left[  \delta A \wedge \extd_A \star F \right]\,.
\eeq
The bulk equations of motion and the presymplectic potential are therefore
\beq
\extd_A \star F \approx 0 \qquad \mathrm{and} \qquad \Theta = - \Tr\left[\delta A \wedge \star F \right]\,,
\eeq
from which we also read out the presymplectic current $\omega = \Tr\left[\delta A \wedge \delta \star F \right]$.

The radiative data relative to the reference frame $U$ on $\Gamma$ is (cf.\ equations~\eqref{genArad} and~\eqref{genFrad}):
\beq
A_{\rm rad} = U^{-1} A U + U^{-1} \extd U \qquad \mathrm{and} \qquad
(\star F)_{\rm rad} = U^{-1} \star F U\,.
\eeq
To determine the radiative presymplectic potential, we first remark that
\begin{align}
U \delta( U^{-1} \extd U) U^{-1} &= \extd ( \delta U U^{-1} ) \,, \\
U \delta( U^{-1} A U) U^{-1} &= \delta A + \left[ A, \delta U U^{-1}\right]\,,
\end{align}
and therefore
\beq
U \delta A_{\rm rad} U^{-1}= \delta A + \extd_A (\delta U U^{-1})\,. 
\eeq
It follows that
\beq
\theta_{\rm rad}:= - \Tr\left[\delta A_{\rm rad} \wedge (\star F)_{\rm rad} \right] = - \Tr\left[\delta A \wedge \star F \right] - \Tr\left[\extd_A (\delta U U^{-1} )  \wedge \star F \right]\,,
\eeq
or, in other words,
\beq
\theta_{\rm gauge} = \Tr\left[\extd_A (\delta U U^{-1} )  \wedge \star F \right] = \extd \Tr\left[ \delta U U^{-1}  \star F \right] - \Tr\left[\delta U U^{-1}  \extd_A \star F \right]\,.
\eeq
On-shell, the second term vanishes and $\theta_{\rm gauge}$ is therefore exact. 
We see in particular, as anticipated in Sec.~\ref{subsec:Maurer-Cartan}, that the (field-space) right-invariant Maurer-Cartan form $\delta U U^{-1}$ is the non-commutative analogue of the variation $\delta \vphi$ that was entering the gauge presymplectic structure of Maxwell and Abelian Chern-Simons theories. The main difference is that $\delta U U^{-1}$ is not field-space exact when $G$ is not Abelian, which leads to extra contributions to the gauge part of the presymplectic current (as compared to formula \eqref{eq:Maxwell_omega-gauge} in Maxwell theory):
\begin{align}
\omega_{\rm gauge} = \delta \theta_{\rm gauge} &=  \extd \Tr\left[ (\delta U U^{-1}) (\delta U U^{-1}) \star F \right] - \extd \Tr\left[ \delta U U^{-1}   \delta \star F \right] \nn \\
&\quad - \Tr\left[(\delta U U^{-1}) (\delta U U^{-1})  \extd_A \star F \right] +  \Tr\left[\delta U U^{-1}   \delta \extd_A \star F \right]\,.
\end{align}
When the equations of motion are satisfied on the boundary, $\restr{\extd_A \star F}{\Gamma} \approx 0$, $\omega_{\rm gauge}$ becomes exact:
\beq
\omega_{\rm gauge} \approx \extd \Tr\left[ (\delta U U^{-1}) (\delta U U^{-1}) \star F \right] - \extd \Tr\left[ \delta U U^{-1}   \delta \star F \right] \,.
\eeq
Our consistency condition \eqref{eq:omega_bdy} can therefore be solved by the following choice of boundary presymplectic current: 
\beq
\omega_\partial = \Tr\left[ \delta U U^{-1} \delta \star F \right] - \Tr\left[ (\delta U U^{-1}) (\delta U U^{-1}) \star F \right]\,.
\eeq

Let us explicitly check that the regional presymplectic structure $\Omega_M = \int_\Sigma \omega + \int_{\partial \Sigma} \omega_\partial$ is invariant under field-dependent gauge transformations, as argued around equation \eqref{eq:generalized_gauge}. Remember that the vector field $\fX_\alpha$ acts as $\fX_\alpha(A) = [\alpha , A] - \extd \alpha = - \extd_A \alpha$ and $\fX_\alpha(F)=[\alpha, F]$, where $\alpha$ is Lie-algebra-valued and possibly field-dependent. We can first compute
\begin{align}
\fX_\alpha \cdot \omega &= -\Tr\left[\extd_A \alpha \wedge \delta \star F \right] + \Tr\left[\delta A \wedge [\star F, \alpha] \right] \\
&= - \extd \Tr\left[ \alpha \delta \star F\right] + \Tr\left[\alpha \extd_A \delta \star F \right] + \Tr\left[\delta A \wedge [\star F, \alpha] \right] \nn \\
& = - \extd \Tr\left[ \alpha \delta \star F\right] + \Tr\left[\alpha  \delta \extd_A \star F \right] - \Tr\left[\alpha  [\delta A , \star F] \right] + \Tr\left[\delta A \wedge [\star F, \alpha] \right] \nn \\
& = - \extd \Tr\left[ \alpha \delta \star F\right] + \Tr\left[\alpha  \delta \extd_A \star F \right] \approx - \extd \Tr\left[ \alpha \delta \star F\right]\,, \nn
\end{align}
where we have used Leibniz's rule for $\extd_A$ in going from the first to the second line, and the commutation relation $[\delta , \extd_A]\eta = [\delta A, \eta]$ (for any Lie-algebra-valued form $\eta$) in going from the second to the third. Owing to the constraint $\delta \extd_A \star F \approx 0$,  $\fX_\alpha \cdot \omega$ is exact on-shell and reduces to a boundary term upon integration, as expected in a local gauge theory. Given that $\fX_\alpha \cdot \delta U U^{-1} = \alpha$, we also have
\begin{align}
\fX_\alpha \cdot \omega_\partial &= \Tr\left[ \alpha \delta \star F \right] + \Tr\left[\delta U U^{-1} [\star F, \alpha] \right] - \Tr\left[[\alpha , \delta U U^{-1}] \star F \right] = \Tr\left[ \alpha \delta \star F \right]\,.
\end{align}
As anticipated, we conclude that: 
\beq
\fX_\alpha \cdot \Omega_M = \int_\Sigma \Tr\left[\alpha  \delta \extd_A \star F \right] \approx 0\,,
\eeq
meaning that the vector field $\fX_\alpha$ is a gauge direction, even for a field-dependent $\alpha$. Specifically, invoking Cartan's magic formula, this entails invariance of the regional presymplectic structure, $\cL_{\fX_\alpha}\Omega_M\approx0$. In~\cite{Donnelly:2016auv}, such an invariance under \emph{field-dependent} gauge transformations was the primary postulate that led to the introduction of the edge field $U$, together with the boundary presymplectic structure $\omega_\partial$. From a conceptual point of view, our construction proceeds differently (see  Sec.~\ref{subsec:presympl}); in particular, invariance under field-dependent gauge transformation is an output, not an input of our procedure, but we recover the same results. For a field-independent $\alpha$, we can furthermore define the constraint generators
\beq\label{YMconstraints}
C[\alpha] := \int_\Sigma \Tr\left[\alpha  \extd_A \star F \right]\,, \qquad \delta C[\alpha] = \fX_\alpha \cdot \Omega_M \,. 
\eeq
They obey the following Poisson algebra:\footnote{Here, it is useful to remember that $g  \acts (\extd_A \star F) = g \extd_A \star F g^{-1}$.} 
\beq
\{ C[\alpha],C[\beta] \} := \fX_\beta \cdot \fX_\alpha \cdot \Omega_M = \int_\Sigma \Tr\left[\alpha  [\beta , \extd_A \star F ] \right] = \int_\Sigma \Tr\left[ [\alpha, \beta]  \extd_A \star F  \right] = C[[\alpha , \beta ]]\,.
\eeq
Hence, the constraints are first-class and define a homomorphism from $\mathfrak{g}$ to the constraint Poisson algebra.

On the other hand, the edge reference frame reorientations $U \to U g^{-1}$ (see Sec.~\ref{ssec_edgeframereorient}) are generated by vector fields $\fY_\rho$, defined as: 
\beq
\Delta_\rho A := \fY_\rho(A) = 0 \qquad \mathrm{and} \qquad  \Delta_\rho U U^{-1} := \fY_\rho \cdot \delta U U^{-1} = - U \rho U^{-1}\,.
\eeq
They act on the frame-dressed observables $A_{\rm rad}$ and $\star F_{\rm rad}$ like gauge transformations intrinsic to $\Gamma$, in the sense that:\footnote{Although being gauge-invariant, we can view $A_{\rm rad}$ as a vector potential for fields on $\Gamma$, which justifies the slight abuse of notation $\extd_{A_{\rm rad}}  \eta := \extd \eta + [A_{\rm rad} , \eta]$, for any Lie algebra valued form on $\Gamma$.}
\beq\label{YMrelobstrans}
\Delta_\rho A_{\rm rad}:=\fY_\rho (A_{\rm rad})= [\rho , A_{\rm rad}] - \extd \rho = - \extd_{A_{\rm rad}} \rho \,, \qquad  \Delta_\rho\left(\star F\right)_{\rm rad}:=\fY_\rho (\star F_{\rm rad}) = [\rho , \star F_{\rm rad}] \,.
\eeq
Note that these expressions are examples of the frame-reorientation induced relational observable transformations in equations~\eqref{ickweessochnich} and~\eqref{symactfield2}, however, in their infinitesimal form.
We then find that $\fY_\rho\cdot \omega = 0$, while 
\beq
\fY_\rho\cdot \omega_\partial = - \Tr[U \rho U^{-1} \delta \star F] + \Tr[ [U \rho U^{-1} , \delta U U^{-1}] \star F] = - \Tr[\rho \delta \star F_{\rm rad}]
\eeq 
Restricting to \emph{field-independent} $\rho$ we can define the charges 
\beq\label{eq:YM_charges}
Q[\rho] := - \int_{\partial \Sigma} \Tr\left[\rho \star F_{\rm rad} \right]\,, \qquad \delta Q[\rho] = \fY_\rho \cdot \Omega_M \,,
\eeq
which verify the algebra
\beq\label{eq:YM_charge_bracket}
\{ Q[\rho],Q[\sigma] \} := \fY_\sigma \cdot \fY_\rho \cdot \Omega_M =  - \int_{\partial \Sigma} \Tr\left[[\rho , \sigma ] \star F_{\rm rad} \right] = Q[[\rho , \sigma ]]\,.
\eeq
It is also clear that the charges are gauge-invariant and Poisson-commute with the constraints
\beq
\{Q[\rho],C[\alpha]\}=\fX_\alpha\cdot \fY_\rho\cdot\Omega_M=0\,.
\eeq

Even though the charges $Q$ and the constraints $C$ obey similar-looking algebras, it is important to realize that they are not isomorphic. Gauge transformations act in the whole spacetime region $M$, and are consequently labelled by Lie-algebra-valued fields ($\alpha$, $\beta$,...) on $M$. By contrast, the charges act solely on the edge field $U$, and are therefore labelled by Lie-algebra-valued fields ($\rho$, $\sigma$,...) on $\Gamma$.

\medskip

Let us now turn to the construction of boundary Lagrangians, following the algorithm put forward in Sec.~\ref{sec:splitting}. As before, we only consider boundary conditions which are covariant under local changes of inertial frames and linear in the canonical local Darboux coordinates of $\omega_{\rm rad}$. Similarly to \eqref{Maxdecomp}, the resulting generalized Robin boundary conditions can be parametrized by the following change of polarization:
\beq
X:=\mathbf{a}A_{\rm rad} + (-1)^{d-1} \,\mathbf{b} \ast \star F_{\rm rad} \,,\qquad\qquad Y:=\mathbf{c} A_{\rm rad}  + (-1)^{d-1}  \mathbf{d} \ast \star F_{\rm rad}\,,
\eeq
where the $\mathrm{SL}(2 , \mathbb{R})$ condition $\mathbf{a} \mathbf{d} - \mathbf{b} \mathbf{c} = 1$ ensures that
\beq
\omega_{\rm rad} = \Tr\left[ \delta A_{\rm rad} \wedge \delta \star F_{\rm rad} \right] = \Tr\left[\delta X \wedge \ast \delta Y \right]\,. 
\eeq 

In order to impose the boundary condition $X=X_0$, one should first reduce the on-shell presymplectic potential on $\Gamma$ to a corner term as in equation~\eqref{shifttheta}. This is the role of the correction term $\ell_{\rm corr}$, which by construction must verify (cf.\ equations~\eqref{ellcorr0} and~\eqref{Maxellcor}):
\begin{align}
\delta \ell_{\rm corr} &\approx - \Tr\left[ \delta X \wedge \ast Y \right] - \theta_{\rm rad} - \theta_{\rm gauge} - \extd C \\
&\approx - \Tr\left[ \delta X \wedge \ast Y \right] + \Tr\left[ \delta A_{\rm rad} \wedge \star F_{\rm rad} \right] - \extd\left( C + \Tr\left[ \delta U U^{-1}  \star F \right] \right) \\
& = \frac{1}{2}\delta \left( \Tr\left[ A_{\rm rad} \wedge \star F_{\rm rad} \right] - \Tr\left[ X \wedge \ast Y \right]\right) - \extd\left( C + \Tr\left[ \delta U U^{-1}  \star F \right] \right)
\end{align}
This can be satisfied by defining 
\beq
\ell_{\rm corr} := \frac{1}{2} \left( \Tr\left[ A_{\rm rad} \wedge \star F_{\rm rad} \right] - \Tr\left[  X \wedge \ast Y \right]\right) \qquad \mathrm{and} \qquad C := - \Tr\left[ \delta U U^{-1}  \star F \right]\,.
\eeq
The second step in our algorithm consists in a dynamical imposition of the boundary condition $X= X_0$, by means of the post-selection action \eqref{Ssp}: 
\beq
S_{\rm ps} = \int_\Gamma \ell_{\rm corr} + \Tr\left[ \lambda \wedge (X- X_0)  \right]\,.
\eeq
After solving for $\lambda$, we obtain the following boundary Lagrangian for \textbf{Robin boundary conditions}:
\beq
\ell = - \Tr\left[ X_0 \wedge \ast Y \right] + \frac{1}{2} \left( \Tr\left[ A_{\rm rad} \wedge \star F_{\rm rad}  \right] + \Tr\left[ X \wedge \ast Y \right] \right) \,.
\eeq 
This is a straightforward generalization of the boundary Lagrangian \eqref{eq:Maxwell_l} of Maxwell theory, the important difference lying in the fact that $\star F$ is not gauge invariant in the non-Abelian case and therefore needs to be approppriately dressed. As in the Maxwell case, variation of this Lagrangian yields the boundary conditions as boundary equations of motion, as well as the bulk equations of motions (incl.\ on $\Gamma$).

We can impose a {\bf Dirichlet boundary condition} $A_{\rm rad} = A_0$ by specializing to $\mathbf{a} = \mathbf{d} = 1$ and $\mathbf{b} = \mathbf{c} = 0$, which leads to a vanishing generating function $\ell_{\rm corr}=0$ and (cf.\ equation~\eqref{MaxDirichlet} in Maxwell theory)
\begin{align}
\ell &= \Tr\left[ (A_{\rm rad} - A_0 ) \wedge \star F_{\rm rad} \right]  \\
&= \Tr\left[ (A - U A_0 U^{-1}) \wedge \star F \right] + \Tr\left[ \extd U U^{-1} \wedge \star F \right]\,.
\end{align}
Choosing $\mathbf{a} = \mathbf{d} = 0$ and $\mathbf{b} = - \mathbf{c} = 1$ instead, yields the {\bf Neumann boundary condition} $\star F_{\rm rad} = f_0 := \ast X_0$ (cf.\ equation~\eqref{MaxNeumannac} in Maxwell theory):
\begin{align}
\ell &= \Tr\left[ X_0 \wedge \ast A_{\rm rad} \right] + \frac{1}{2} \left( \Tr\left[ A_{\rm rad} \wedge \star F_{\rm rad}  \right] + \Tr\left[ (-1)^{d} \ast \star F_{\rm rad} \wedge \ast A_{\rm rad} \right] \right) \\
&= \Tr\left[ A_{\rm rad} \wedge f_0 \right] = \Tr\left[ U^{-1} A U \wedge f_0 \right] + \Tr\left[ U^{-1} \extd U \wedge f_0 \right]\,.
\end{align}

\medskip

We conclude by analyzing the physical status of the frame reorientations (symmetries) $\fY_\rho$ and edge modes, which varies depending on the type of boundary condition being imposed, similar to the Maxwell case at the end of Sec.~\ref{subsec:maxwell}. Ignoring mixed types of boundary conditions for simplicity, we can distinguish two cases.

Suppose first that $\mathbf{a}=0$ everywhere on $\Gamma$. The boundary condition is then of the \textbf{Neumann type} and reduces to $\star F_{\rm rad} = f_0$ for some background function $f_0$. Owing to equation~\eqref{YMrelobstrans}, the vector field $\fY_\rho$ leaves the boundary condition invariant if and only if $[\rho , f_0] = 0$. However, by virtue of \eqref{eq:YM_charges}, any such vector field corresponds to a gauge direction of $\Omega_M$. Indeed, in analogy to the edge constraints \eqref{edgeconstraint} of Maxwell theory, we now obtain the first-class edge constraints
\beq 
C_\partial[\rho]:=-\int_{\partial\Sigma}\Tr\left[\rho \left(\star F_{\rm rad}-f_0\right) \right]
\eeq
in addition to the standard bulk constraints in equation~\eqref{YMconstraints}. 
The condition $[\rho , f_0] = 0$ ensures that the Poisson brackets of such constraints does vanish on $\cS_M^{X_0}$:
\beq\label{eq:YM_boundary_gauge_alg}
\{ C_\partial [\rho] , C_\partial [\sigma]\} = - \int_{\partial \Sigma} \Tr[[\rho, \sigma] \star F_{\rm rad}] = \int_{\partial \Sigma} \Tr[\sigma [\rho, \star F_{\rm rad}]]  \heq \int_{\partial \Sigma} \Tr[\sigma [\rho, f_0]] = 0\,,
\eeq
that is, they constitute a first-class algebra. By contrast, any $C_\partial [\rho]$ with $[\rho, f_0] \neq 0$ is second-class,\footnote{This can be observed by computing its Poisson bracket with other constraints $C_\partial [\sigma]$ with $[\sigma, f_0]\neq 0$.} and therefore generates a frame reorientation $\fY_\rho$ that does not preserve the boundary conditions. Such transformations constitute meta-symmetries that induce a change of subregion theory $\cS_M^{X_0}$ within the foliation \eqref{eq:foliation} of the global solution space $\cS$. This is in contrast to the case of Neumann boundary conditions in Maxwell theory, where, owing to the Abelian character of the group, no such meta-symmetry exists. As in the Maxwell case, however, Neumann boundary conditions are not compatible with proper symmetries of the regional solution space $\cS_M^{X_0}$.
We conclude that edge modes are somewhat redundant with such boundary conditions: even though they play a conceptually enlightening role at intermediate steps of our construction and are crucial in the formulation of the boundary value problem through post-selection, they do not manifest themselves at the level of the regional physical phase space $\cP_M^{X_0}$. Indeed, as emphasized in the Maxwell case, Neumann boundary conditions correspond to reflecting boundary conditions, in which case $M$ and its complement $\bar M$ give rise to largely independent regional dynamics. Incidentally, alternative constructions without edge modes have been proposed in the literature for this particular class of boundary conditions (see again \cite{Gomes:2018dxs, Gomes:2019xto, Riello:2020zbk, Riello:2021lfl}, and in particular \cite{Riello:2021lfl} for a nice summary).

Suppose instead that $\mathbf{a} \neq 0$ everywhere on $\Gamma$, so that we are dealing either with \textbf{Dirichlet} or \textbf{Robin boundary conditions}. We can then choose $\mathbf{a} =1$ without any loss of generality. In order for the vector field $\fY_\rho$ to leave the boundary condition $X = X_0$ invariant, the function $\rho$ must verify:
\beq
\Delta_\rho X = [\rho , A_{\rm rad}] - \extd \rho + (-1)^{d-1} \mathbf{b} \ast [\rho , \star F_{\rm rad}] = - \extd_{X_0} \rho =0 \,. \label{YMbcpres}
\eeq  
This generalizes the condition $\extd \rho = 0$ we already came across for Maxwell theory with boundary condition on $A_{\rm rad}$ (see Sec.~\ref{subsec:presympl} and the end of Sec.~\ref{subsec:maxwell}). In view of \eqref{eq:YM_charges}, $\delta Q[\rho]$ does not generally vanish since $\star F_{\rm rad}$ is free to fluctuate. We conclude that the charges $Q[\rho]$ such that $\extd_{X_0} \rho = 0$ generate a non-trivial algebra of symmetries, with commutation relations \eqref{eq:YM_charge_bracket}. However, equation~\eqref{YMbcpres} being a first order partial differential equation for $\rho$, the dimension of this algebra can at most be equal to $\dim\mathfrak{g}$,\footnote{A more formal justification of this statement would invoke the Cauchy-Lipschitz theorem.} and for non-Abelian groups, it depends on the choice of background field $X_0$.\footnote{For a generic $X_0$, one actually expects $\extd_{X_0} \rho = 0$ to have no solution.} For  fixed Dirichlet or Robin boundary conditions, the remaining frame reorientations $\fY_\rho$ will violate equation~\eqref{YMbcpres} and thereby correspond to meta-symmetries.

The situation in non-Abelian Yang-Mills theory is therefore very similar to the Abelian Maxwell theory case, the only difference being that here the dimension of the symmetry algebra can be anything between zero and $\dim\mathfrak{g}$, while in Maxwell theory it is always of dimension $\dim\mathfrak{u}(1)=1$. When there are solutions to equation~\eqref{YMbcpres} and the algebra is of non-trivial dimension, we can invoke the same argumentation as at the end of Sec.~\ref{subsec:maxwell} to conclude that edge modes are of direct physical significance in the regional reduced phase space $\cP_M^{X_0}$.\footnote{In the language of \cite{Gomes:2019xto}, the existence of \emph{stabilizers} (that is, of transformations $\fY_\rho$ such that \eqref{YMbcpres} holds) leads to a stratification of the reduced phase space generated by the symmetry charges.} This phase space will contain edge-frame-dressed observables, including non-local ones that relate bulk properties of $M$ to the edge frame $U$ via Wilson lines. By contrast, when the symmetry algebra happens to be zero-dimensional, edge modes do not manifest themselves at the level of the regional reduced phase space; the frame-dressed observables rather parametrize transversal directions to the regional solution space $\cS_M^{X_0}$ in the foliation of the global space of solutions $\cS$.

Our conclusion that edge modes are physically significant in the presence of Dirichlet or Robin boundary conditions verifying equation \eqref{YMbcpres} is consistent with the findings of \cite{Gomes:2019xto, Gomes:2019xhu}. There, it was shown without invoking edge modes that an equation analogous to \eqref{YMbcpres} can be understood as encoding a fundamental ambiguity in the gluing properties of the system. That is, whenever this equation admits non-trivial solutions, it turns out not to be possible to reconstruct a unique state for $\Sigma\cup \bar \Sigma$ from the knowledge of the local states characterizing the gauge-invariant data with support respectively in $\Sigma$ and $\bar\Sigma$. The physical symmetries generated by the analog of $Q[\rho]$ in  \cite{Gomes:2019xto, Gomes:2019xhu} capture the missing holistic properties of the global system, and once specified, restore unicity of the gluing procedure. In our construction of edge modes as reference frames, this holistic nature of the symmetry charges $Q[\rho]$ is explicit from the outset: $U$ is non-locally defined and is precisely introduced to account for the gauge-invariant observables of the system which have support on both $M$ and $\bar M$. However, it is enlightening to note that the authors of \cite{Gomes:2019xto, Gomes:2019xhu} arrived at a similar conclusion without introducing edge modes \emph{per se}, but instead relied on a careful analysis of the gluing properties of partial Cauchy slices.

Altogether, this shows that edge modes, in the sense of \cite{Donnelly:2016auv}, do not always manifest themselves at the physical level in Yang-Mills theory; and when they do, they add a finite number of parameters to an otherwise infinite-dimensional regional reduced phase space. In this regard, Yang-Mills theory is dramatically different from (Abelian) Chern-Simons theory: in the latter case, frame reorientations were giving rise to an infinite-dimensional algebra of edge symmetries \eqref{eq:sym_CS}, owing to the fact that the analog of \eqref{YMbcpres}, equation~\eqref{eq:selfdual_sym}, admits infinitely many solutions (irrespectively of the background). In this respect, our conclusions seem consistent with at least some of the views on Yang-Mills edge modes laid out in \cite{Gomes:2018dxs, Gomes:2019xto, Riello:2020zbk, Riello:2021lfl} (see also \cite{Gomes:2019xhu, Gomes:2019otw} for a philosophical discussion), and it would be interesting to compare the two frameworks more closely. Notwithstanding, also in Yang-Mills and Maxwell theory, edge modes play a conceptually illuminating role and are crucial for consistently formulating the regional boundary value problem, as well as the foliation of distinct subregion theories. Ultimately, they provide the proper tools to relate $M$ to its complement $\bar M$.

\section{Conclusion and outlook}\label{sec:conclu}

In this paper, we have provided a self-contained and detailed analysis of edge modes in gauge field theory, which we believe clarifies their physical nature. By embedding a spacetime region $M$ with time-like boundary $\Gamma$ into a global spacetime $M\cup \bar M$, we were able to view the regional dynamics in $M$ as the result of a post-selection procedure on the global solution space. This also constitutes a connection between covariant phase space constructions for global spacetimes \cite{Lee:1990nz} and bounded subregions as e.g.\ in \cite{Harlow:2019yfa,Margalef-Bentabol:2020teu,Geiller:2019bti,Chandrasekaran:2020wwn}. This global perspective allowed us to make precise sense of edge modes as dynamical reference frames, and thereby elucidate their interpretation. Our main conclusions are the following.
\begin{itemize}
    \item Edge modes can be systematically and consistently interpreted as dynamical reference frames, in the same sense in which they have appeared in the recent quantum reference frame literature \cite{Giacomini:2017zju,Vanrietvelde:2018dit,Vanrietvelde:2018pgb,Hohn:2018iwn,Hohn:2018toe,Hoehn:2021wet,Hohn:2019cfk,Hoehn:2020epv,Krumm:2020fws,Hoehn:2021flk,delaHamette:2020dyi,Castro-Ruiz:2019nnl,Giacomini:2020ahk,Ballesteros:2020lgl}. Here, they are in general necessary to specify gauge-invariant boundary conditions on the time-like boundary $\Gamma$ of a finite region $M$. They are non-local functionals of the global configuration fields (defined e.g.\ via systems of Wilson lines with support in the complementary region $\bar M$), which however materialize themselves as local fields on $\Gamma$. See Sec.~\ref{sec_gaugesplit}. 
    
    \item Hence, edge modes are a direct manifestation of the relational character of gauge field degrees of freedom \cite{Rovelli:2013fga,Rovelli:2020mpk}. Indeed, frame-dressed observables on $\Gamma$ are gauge-invariant functionals which relate the region $M$ to its complement $\bar M$, and that can be precisely interpreted as relational observables \cite{Rovelli:1989jn,Rovelli:1990jm,Rovelli:1990pi,Dittrich:2004cb,Dittrich:2005kc,rovelliQuantumGravity2004,Rovelli:2001bz} in a covariant phase space setting (see Sec.~\ref{sec:relational_obs}).
    
    \item Given a suitable set of gauge-invariant boundary conditions on $\Gamma$ (they can be of Dirichlet, Neumann, Robin or mixed type), we have developed a systematic algorithm allowing one to induce a consistent variational principle for subregion $M$ from the global one on $M\cup\bar M$. This includes the definition of a presymplectic form $\Omega_M = \int_\Sigma \omega + \int_{\partial \Sigma} \omega_\partial$, and more generally, of an action $S_{M\cup \Gamma} = \int_M L + \int_{\Gamma} \ell$. The presymplectic structure is made independent from the choice of partial Cauchy slice $\Sigma$ thanks to the inclusion of a corner term verifying the on-shell consistency condition $\omega_\partial + \extd \omega_{\rm gauge} \heq 0$ (see Sec.~\ref{sec:geometry}). The boundary Lagrangian form $\ell$ can be motivated by similar arguments at the off-shell level (see Sec.~\ref{sec:splitting}).
    
    \item While edge modes are always useful concepts in the construction of the variational problem for subregion $M$, they may sometimes drop out from the resulting on-shell presymplectic structure. However, this only happens for very specific choices of boundary conditions (such as a Neumann boundary condition on $\star F$ in Maxwell theory). For generic boundary conditions, the edge reference frame does contribute to the regional presymplectic structure.
    
    \item As was originally emphasized in \cite{Donnelly:2016auv}, symmetries need to be carefully distinguished from gauge transformations. As we have shown, symmetry transformations of edge degrees of freedom can be interpreted as reference frame reorientations, which makes their physical character all the more clear.  At the level of the subregion $M$, the inclusion of boundary conditions necessitates a refinement of this dichotomy. Frame reorientations then split into three further subcategories: proper boundary symmetries, which leave the regional field space (in particular, the boundary conditions) invariant and carry non-trivial charges; boundary gauge symmetries, which also leave the regional field space invariant, but whose charges become trivial on-shell of the boundary conditions; and finally meta-symmetries, which, by changing the boundary conditions themselves, give rise to symplectomorphisms between distinct regional field spaces.     
    \item Our formalism is in principle applicable to any gauge field theory, as examplified in Sec.~\ref{sec:examples} (see also Sec.~\ref{sec:mechanical} for a mechanical model featuring both gauge symmetries and edge modes). In particular, we do not see any major difference in the physical interpretation of Yang-Mills, Maxwell and Chern-Simons edge modes: they can all be derived in a systematic way by application of our algorithm. However, in contrast to Yang-Mills theories (including Maxwell), Chern-Simons theories do have the peculiarity of supporting an infinite-dimensional algebra of boundary symmetries, even after suitable boundary conditions have been imposed on $\Gamma$ (see Sec.~\ref{subsec:CS} for an illustration of this claim in the Abelian context). In that, the dynamical role of edge modes is greatly enhanced in such theories.
\end{itemize}

The present article has been conceived as a first in a series of upcoming publications \cite{CH2}. In complement to splitting post-selection, we plan to examine the reverse process of gluing in the language of dynamical reference frames. It will also be important to test our formalism on other interesting gauge systems. An obvious example to consider, that was left out of the present work for simplicity, is non-Abelian Chern-Simons theory. 
Finally, beyond the definition of consistent variational principles, our main objective is to explore the statistical (and, ultimately, quantum) properties of bounded subregions in field theory, from the point of view of universal fluctuation relations in non-equilibrium thermodynamics.

Another and related objective we leave for the future is to investigate the nature of edge modes in gravitational systems from the reference frame perspective advocated in the present paper. In the context of full Einstein gravity, it will be particularly interesting to revisit recent proposals such as \cite{Freidel:2020xyx,Freidel:2019ees, Chandrasekaran:2020wwn, Donnelly:2020xgu,Wieland:2020gno,Wieland:2021vef,Wieland:2017zkf}. Finally, in order to test our construction in the realm of quantum gravity, it will also be enlightening to investigate symmetry-reduced models which are reasonably well understood at the quantum level (such as Jackiw-Teitelboim gravity).  

\section*{Acknowledgements}

We thank Laurent Freidel, Marc Geiller, Henrique Gomes, Aldo Riello, and Nicholas Teh for helpful discussions, and Josh Kirklin for comments on the manuscript. SC is supported by a Radboud Excellence Fellowship from Radboud University in Nijmegen, the Netherlands. PH is grateful for support through an `It from Qubit' fellowship of the Simons Foundation in the initial stages of this work and from the Foundational Questions Institute under grant number FQXi-RFP-1801A. He furthermore thanks Perimeter Institute for support and hospitality during a collaborative visit for the project.  

\noindent This project was initiated at Perimeter Institute and University College London. Research at Perimeter Institute is supported, in part, by the Government of Canada through the Department of Innovation, Science and Economic Development Canada, and by the Province of Ontario through the Ministry of Colleges and Universities. This work was also supported in part by funding from Okinawa Institute of Science and Technology Graduate University.

\bibliographystyle{JHEP}
\addcontentsline{toc}{section}{References}

\bibliography{STD}


\end{document}